\documentclass[]{aa}

\usepackage{bm}
\usepackage{amsmath}
\usepackage{xcolor}
\usepackage{graphicx}
\usepackage[outdir=./]{epstopdf}
\usepackage{hyperref}
\usepackage[all]{hypcap} 
\usepackage{dblfloatfix}    

\begin{document}
	
	\title{Transverse kink oscillations of inhomogeneous prominence threads: numerical analysis and H$\alpha$ forward modelling}
	
	\author{David Martínez-Gómez\inst{\ref{inst1},\ref{inst4}}
		\and Roberto Soler\inst{\ref{inst2},\ref{inst3}}
		\and Jaume Terradas\inst{\ref{inst2},\ref{inst3}}
		\and Elena Khomenko\inst{\ref{inst1},\ref{inst4}}}
	
	\institute{Instituto de Astrofísica de Canarias, 38205 La Laguna, Tenerife, Spain; \label{inst1}
		\and Departament de Física, Universitat de les Illes Balears, 07122, Palma de Mallorca, Spain\label{inst2}
		\and Institut d'Aplicacions Computacionals de Codi Comunitari (IAC3), Universitat de les Illes Balears, 07122, Palma de Mallorca, Spain\label{inst3}
		\and Departamento de Astrofísica, Universidad de La Laguna, 38205 La Laguna, Tenerife, Spain\label{inst4} \\
		\email{dmartinez@iac.es}
	}
	
	\abstract{Prominence threads are very long and thin flux tubes which are partially filled with cold plasma. Observations have shown that transverse oscillations are frequent in these solar structures. The observations are usually interpreted as the fundamental kink mode, while the detection of the first harmonic remains elusive.}{The properties of oscillations in threads are greatly affected by the density distribution along the flux tube. Here, we aim to study how the density inhomogeneity in the longitudinal and radial directions modify the periods and damping times of kink oscillations, and how this effect would be reflected in the observations.}{We solve the ideal magnetohydrodynamics equations through two different methods: a) performing 3D numerical simulations, and b) solving a 2D generalised eigenvalue problem. We study the dependence of the periods, damping times and amplitudes of transverse kink oscillations on the ratio between the densities at the centre and at the ends of the tube, and also on the average density. We apply forward modelling on our 3D simulations to compute synthetic H$\alpha$ profiles.}{We confirm that the ratio of the period of the fundamental oscillation mode to the period of the first harmonic increases as the ratio of the central density to the footpoint density is increased or as the averaged density of the tube is decreased. We find that the damping times due to resonant absorption decrease as the central to footpoint density ratio increases. Contrary to the case of longitudinally homogeneous tubes, we find that the damping time to period ratio also increases as the density ratio is increased or the average density is reduced. We present snapshots and time-distance diagrams of the emission in the H$\alpha$ line.}{The results presented here have implications for the field of prominence seismology. While the H$\alpha$ emission can be used to detect the fundamental mode, the first harmonic is barely detectable in H$\alpha$. This may explain the lack of detections of the first harmonic. A combination of different spectral lines is required to get information about the period ratio and to use it to infer physical properties of the threads.}
	
	
	\titlerunning{Inhomogeneous flux tubes}
	\authorrunning{Martínez-Gómez et al.}
	
	\maketitle

\section{Introduction}
	Solar prominences are inhomogeneous condensations of cool plasma in the corona \citep[e.g.,][]{2005SoPh..226..239L,2008ApJ...676L..89B}. The fine structure of prominences consists of very long and thin magnetic flux tubes partially filled with cold plasma (in comparison with the hot coronal plasma), which usually are referred to as threads or fibrils \citep{2007Sci...318.1577O}. The typical length of these magnetic flux tubes is of the order of $10^{5} \ \rm{km}$, with a radius between $50 \ \rm{km}$ and $500 \ \rm{km}$. The length of the region of the tube filled with cold plasma inferred from H$\alpha$ observations is in the range of $3000 \ \rm{km}$ to $28000 \ \rm{km}$ \citep{2008AdSpR..42..803L,2011SSRv..158..237L}.
	
	Observations have revealed the presence of transverse oscillations in prominence threads, with typical periods from $1$ to $20$ minutes \citep{2007SoPh..246...65L,2009ApJ...704..870L,2009A&A...499..595N}. These oscillations have usually been interpreted as magnetohydrodynamic (MHD) kink waves \citep[see, e.g.,][]{2008ApJ...678L.153T,2010ApJ...722.1778S,2011A&A...533A..60A}. Their properties have been used to perform prominence seismology \citep{2012ASPC..455..211A,2014IAUS..300...30B,2018LRSP...15....3A} with the intention of inferring information about the physical parameters of the plasma and/or the magnetic field \citep{2019A&A...622A..88M}.
	
	For instance, the comparison of the periods of the fundamental mode, $P_{0}$, and the corresponding first overtone, $P_{1}$, can be used to obtain information about the longitudinal distribution of density. The ratio of the damping time to the period has been commonly used to infer the radial distribution and determine which physical mechanism is responsible for that attenuation. In a longitudinally homogeneous tube, the period ratio is $P_{0} / P_{1} = 2$ \citep{1983SoPh...88..179E} but it differs from $2$ if a longitudinal inhomogeneity is considered. \cite{2005ApJ...624L..57A,2005A&A...430.1109A} showed that for the case of coronal loops, which are denser at the footpoints than at the apex, the period ratio is smaller than 2. Later, \citet{2010ApJ...725.1742D} showed that for a prominence thread, which is denser at the centre of the tube, the period ratio is always larger than $2$, a result that was confirmed by, e.g., \citet{2011A&A...533A..60A} and \citet{2011A&A...531A.167S}. The relation $P_{0} / P_{1}$ has been used to show that the ratio between the average density of the tube and the density at its centre is very small, which implies that there are very strong gradients of density along the threads \citep{2015A&A...575A.123S}. Regarding the damping time to period ratio, \citet{2009ApJ...707..662S,2014IAUS..300...48S} analysed several possible damping mechanisms and determined that the process that better describes the data from observations of prominence threads is resonant absorption \citep{1978ApJ...226..650I,2002ApJ...577..475R,2008ApJ...682L.141A,2010ApJ...722.1778S}. This phenomenon occurs when there is a smooth variation between the internal and external densities of the flux tube and depends on the thickness of this radial transition layer and on the density contrast between the internal and external plasmas.
	
	The previously mentioned works usually focused on the density inhomogeneity of the threads in one direction only, namely the longitudinal or the radial one. In that way, analytical expressions could be provided for the periods and damping times of oscillations in tubes with longitudinal uniform density or represented by piecewise constant models. On the contrary, the consideration of the inhomogeneity along those two directions at the same time requires a numerical treatment \citep{2005A&A...441..361A,2011A&A...533A..60A}. 
	
	Here, we use a model of a prominence thread in which the density follows a Lorentzian profile along the longitudinal direction. The main parameter that describes this profile is the longitudinal density ratio, $\chi$, which is defined as the ratio between the densities at the centre and at the ends of the tube. For prominence threads, this parameter takes values larger than $1$. In the radial direction there is a smooth transition layer that connects the internal plasma with the external plasma. The variation of density in this radial transition layer is given by a sinusoidal profile. Then, we study the properties of transverse kink oscillations by applying two different methods. In the first place, we use the numerical code MoLMHD \citep{2016ApJ...820..125T,2018ApJ...853...35T}, that solves the non-linear ideal MHD equations, to run 3D simulations and compute the full temporal evolution of the flux tube. Then, we compare the obtained results with the solutions from a 2D generalised eigenvalue problem \citep{2015A&A...575A.123S}. Numerical simulations are very computationally costly but they have the advantage that they capture the non-linear dynamics of this kind of oscillations. In contrast, the eigenvalue problem is restricted to the linear regime but it is much faster and it is more appropriate for performing a parametric analysis of the periods and damping times. Therefore, both methods complement each other.
	
	We study how the periods, damping times and amplitudes of the fundamental and first harmonic modes vary with the longitudinal density ratio and with the average density. We compute the period ratio $P_{0} / P_{1}$ and the damping to period ratio, and compare the obtained results with those corresponding to tubes with longitudinally uniform profiles of density.
	
	After this analysis, we go a step further in our numerical research and use the results from the 3D simulations to perform forward modelling and provide synthetic profiles in the H$\alpha$ line. We use the approximate method described in \citet{2015A&A...579A..16H}. We compute the H$\alpha$ intensity along two different lines of sight (parallel and perpendicular to the direction of oscillation) and provide snapshots that show the longitudinal and radial distribution of the density of prominence threads. We discuss the observational signatures of the transverse oscillations and how the strong variation of density along the thread may have implications for prominence seismology. 

	The synthetic H$\alpha$ profiles do not only provide information about the amplitudes, periods and damping times of the kink oscillations. They also include evidences of non-linear processes that are taking place in the threads. For instance, large velocity gradients at the boundaries between the internal and external layers of the flux tube trigger shear instabilities, like the Kelvin-Helmholtz instability \citep[KHI,][]{1961hhs..book.....C,1984A&A...131..283B,2008ApJ...687L.115T}. As a consequence, large deformations appear at those boundaries as the instability develops \citep[see, e.g.,][]{2014ApJ...787L..22A,2015A&A...582A.117M,2018ApJ...853...35T}.
	We discuss how the variations in intensity of the H$\alpha$ profiles are related to these non-linear effects.

	The present work is organised as follows. In Section \ref{sec:setup} we describe the model we use to represent an inhomogeneous flux tube, the equations that govern its dynamics and the methods applied to solve those equations. In Section \ref{sec:results} we analyse the characteristics of the transverse oscillations, comparing the cases of uniform and non-uniform tubes. Then, in Section \ref{sec:fomo} we perform forward modelling to compute synthetic H$\alpha$ profiles and show the observational signatures of several linear and non-linear features of the oscillations. Finally, our conclusions and a brief discussion on possible improvements are presented in Section \ref{sec:concl}.
	
\section{Model and numerical setup} \label{sec:setup}

\subsection{Flux tube model} \label{sec:model}
	\begin{figure}
		\centering
		\resizebox{\hsize}{0.5\vsize}{\includegraphics[]{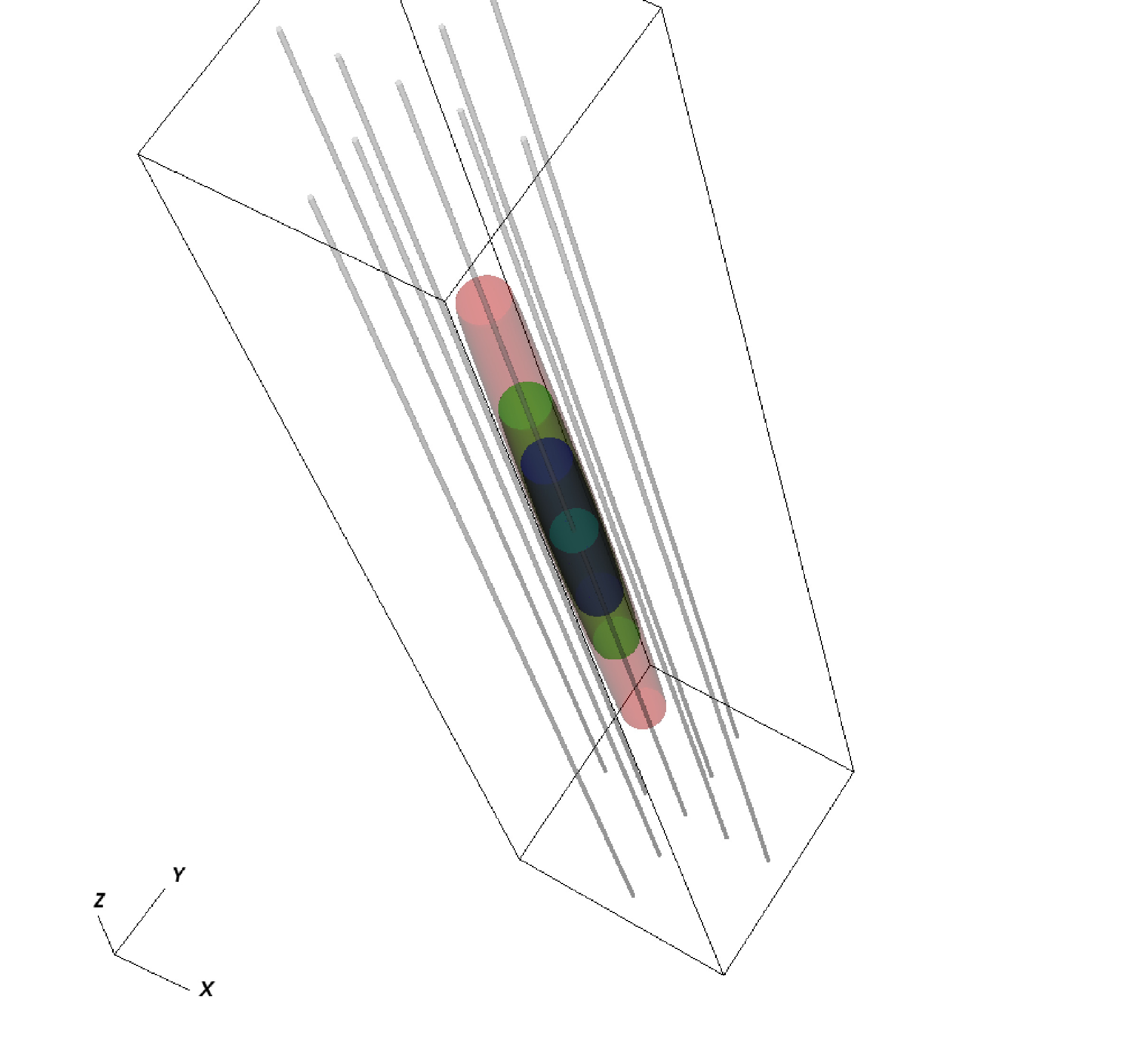}}
		\caption{Sketch of the equilibrium state for a thread with a density ratio of $\chi = 10$. Density contours for $\rho = 25 \rho_{\rm{ex}}$ (red), $\rho = 50 \rho_{\rm{ex}}$ (green), $\rho = 75 \rho_{\rm{ex}}$ (dark blue), and $\rho = 100 \rho_{\rm{ex}}$ (light blue). Magnetic field lines, with line-tying conditions applied at the top and bottom planes, are shown in grey colour.}
		\label{fig:sketch}
	\end{figure}

	We represent prominence threads as straight flux tubes with inhomogeneous density embedded in a hotter and lighter background (which represents the solar corona). In the longitudinal direction the density follows a Lorentzian profile:
	\begin{equation} \label{eq:lorentz}
		\rho_{\rm{i}}(z) = \frac{\rho_{\rm{i,0}}}{1+4 \left(\chi-1\right)z^2/L_{z}^2},
	\end{equation}
	where $\rho_{\rm{i,0}}$ is the density at $z=0$, $L_{z}$ is the length of the tube, and $\chi = \rho_{\rm{i,0}} / \rho_{\rm{i}}(z = L_{z}/2)$ is the longitudinal density ratio. The average density of this Lorentzian profile is given by
	\begin{equation} \label{eq:rho_av}
		\langle \rho_{\rm{i}}\rangle = \frac{1}{L_{z}} \int_{-L_{z}/2}^{L_{z}/2} \rho_{\rm{i}}(z) \rm{d}z = \rho_{\rm{i,0}} \frac{\arctan \sqrt{\chi-1}}{\sqrt{\chi-1}}.
	\end{equation}
	According to this formula, the value of $\langle \rho_{i}\rangle$ decreases as $\chi$ increases.

	In the transverse direction to the tube, the density is radially uniform inside the tube with a smooth transition between the internal and external values of density, $\rho_{\rm{in}}$ and $\rho_{\rm{ex}}$, respectively. In this transition layer the density varies as
	\begin{small}
	\begin{equation} \label{eq:rho_layer}
		\rho(r,z) = \frac{\rho_{\rm{i}}(z)}{2} \left[1 + \frac{\rho_{\rm{ex}}}{\rho_{\rm{i}}(z)}- \left(1-\frac{\rho_{\rm{ex}}}{\rho_{\rm{i}}(z)}\right) \sin \left(\pi \frac{r - R}{l} \right) \right],
	\end{equation}
	\end{small}
	with $r = \sqrt{x^2+y^2}$, $R$ the radius of the tube and $l$ the width of the transition layer. In this work, the relations between the internal and external densities and between the parameters $l$ and $R$ are set to $\rho_{\rm{i,0}} = 100 \rho_{\rm{ex}}$ (where $\rho_{\rm{ex}} = \rho_{\rm{ref}}$ and $\rho_{\rm{ref}}$ is the reference value of the coronal density) and $l/R=0.3$, respectively.

\begin{figure}
	\centering
	\resizebox{\hsize}{!}{\includegraphics[width=0.8\hsize]{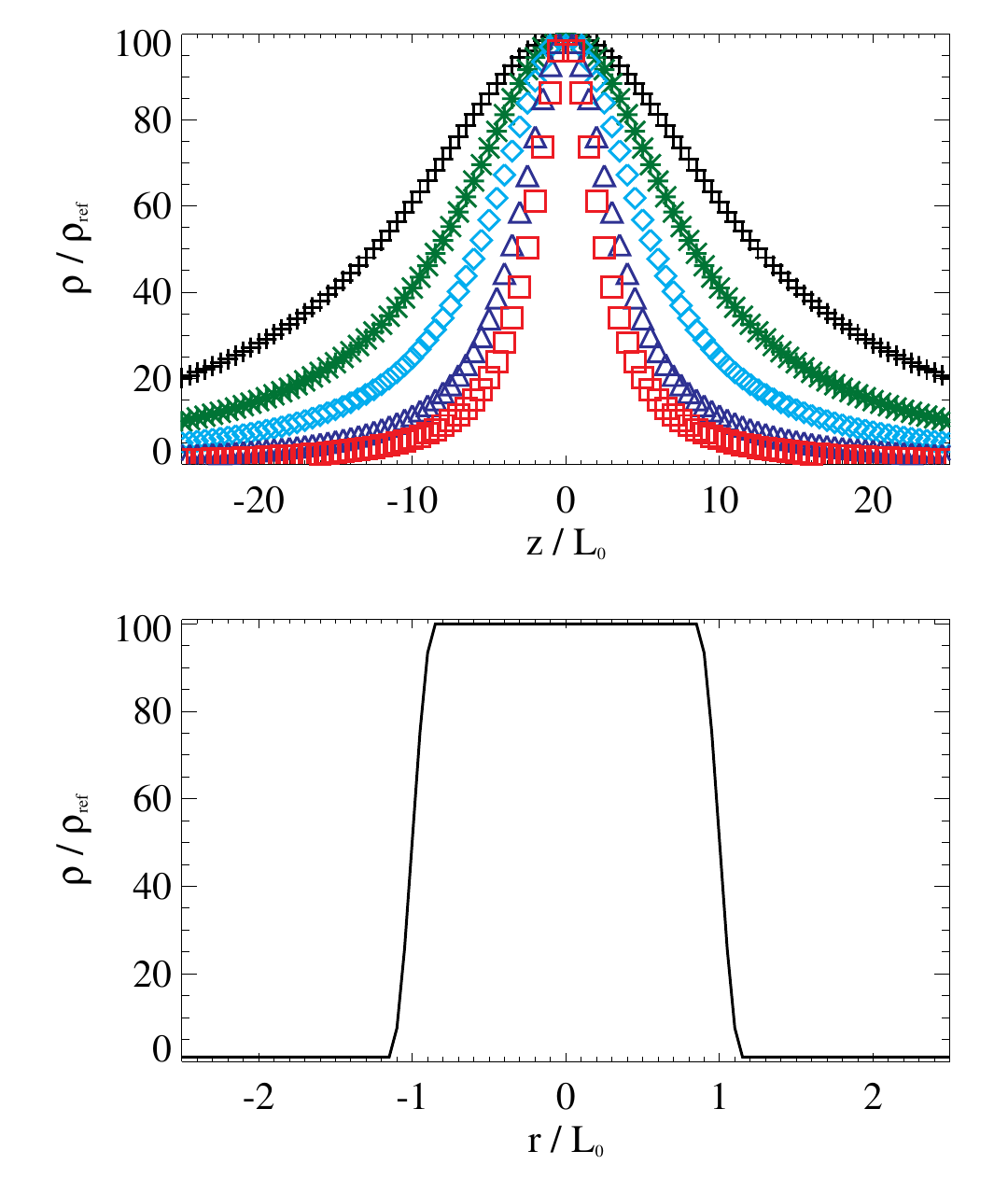}}
	\caption{Top panel: vertical profiles of density for several values of the density ratio: $\chi = 5$ (black), $\chi = 10$ (green), $\chi = 20$ (light blue), $\chi = 50$ (dark blue), and $\chi = 100$ (red). Bottom panel: radial profile of density at $z = 0$.}
	\label{fig:profiles}
\end{figure}

	The flux tube model is displayed in Figures \ref{fig:sketch} and \ref{fig:profiles}. A 3D sketch of an inhomogeneous thread with $\chi = 10$ is shown in the former: the red, green and blue surfaces correspond to different density contours, and the vertical grey lines show the magnetic field lines.	Figure \ref{fig:profiles} displays the vertical distribution of density for different values of the density ratio (top panel) and its radial profile at the tube's mid-height (bottom panel).

	To perform the simulations we consider a numerical domain of $-5 < x/L_{0} < 5$, $-5 < y/L_{0} < 5$ and $-L_{z} / 2 < z/L_{0} < L_{z}/2$, where $L_{0} = 1 \ \rm{Mm}$ is the reference length of the system and $L_{z} = 50 R$, with $R = L_{0}$. We note that the considered ratio $L_{z} / R$ produces a tube that is thicker than what is expected for a prominence thread. For instance, the observations performed by \citet{2007Sci...318.1577O}, \citet{2008AdSpR..42..803L}, and \citet{2011SSRv..158..237L} provided much larger values of the ratio $L_{z} / R$, ranging from $400$ to $2000$. The reason why we chose this small ratio is related to the computing time of the simulations: the oscillation period of a flux tube is proportional to its length, as shown by \citet{1983SoPh...88..179E}, so the study of transverse oscillations of tubes with much larger values of $L_{z} / R$ requires much longer computing times. The investigation we present in this paper required performing a considerable number of simulations, so it would have been impractical to use tubes with much higher values of $L_{z} / R$. Nevertheless, we expect that this choice does not affect the general validity of our results, as they mainly depend on the longitudinal variation of the tube properties and not on its actual length or radius. In addition, we consider the presence of a uniform longitudinal magnetic field of $B_{0} = 10 \ \rm{G}$ which permeates the whole domain and, in combination with the reference value of the coronal density $\rho_{\rm{ref}} = 3.656 \times 10^{-13} \ \rm{kg \ m^{-3}}$, gives a reference Alfvén speed of $c_{\rm{A,ref}} = B_{0} /\sqrt{\mu \rho_{\rm{ref}}} \approx 1494 \ \rm{km \ s^{-1}}$ and a reference time $t_{0} = L_{0} / c_{\rm{A,ref}} \approx 0.67 \ \rm{s}$.
	
	The pressure is uniform in the whole domain, which implies that the system is initially in mechanical equilibrium. We use a value of $P = 0.005 \ \rm{Pa}$, which corresponds to a plasma beta of $\beta = 2 \mu_{0} P / B_{0}^2 \approx 0.013$. In this low-$\beta$ limit, pressure has a negligible effect on the linear regime of transverse oscillations. However, it becomes relevant in the non-linear regime. \citet{1971JGR....76.5155H}, \citet{1994JGR....9921291R,1995GeoRL..22.1741R}, \citet{1995PhPl....2..501T} and \citet{2004ApJ...610..523T} demonstrated that the ponderomotive force causes longitudinal flows that tend to accumulate the mass at the centre of the tube. In a pressureless plasma this accumulation would go on without bound. On the contrary, the effect of pressure is to prevent this unlimited growth and turn it into an oscillatory behaviour.

	We apply line-tying boundary conditions at the ends of the flux tube to represent the field lines anchoring in the photosphere, meaning that the three components of the velocity are set to zero and the normal component of the magnetic field is fixed, while the normal derivatives of the rest of variables are equal to zero. For the lateral planes, closed boundary conditions are chosen. We note that these conditions may produce unwanted reflections of perturbations but we have set the lateral boundaries far enough from the numerical domain of interest so they have a negligible influence on the evolution of the flux tube.

\subsection{Methods and equations}
	On the one hand, we perform 3D simulations using the numerical code MoLMHD \citep{2016ApJ...820..125T,2018ApJ...853...35T}, which solves the non-linear ideal MHD equations:
	\begin{equation} \label{eq:cont}
		\frac{\partial \rho}{\partial t}+\nabla \cdot \left(\rho \bm{V}\right)=0,
	\end{equation}
	\begin{equation} \label{eq:mom}
		\frac{\partial \left(\rho \bm{V}\right)}{\partial t}+\nabla \cdot \left(\rho \bm{VV}+P\mathbb{I}-\frac{\bm{BB}}{\mu}+\frac{\bm{B}^{2}}{2\mu}\mathbb{I}\right)=0,
	\end{equation}
	\begin{equation} \label{eq:induction}
		\frac{\partial \bm{B}}{\partial t}=\nabla \times \left(\bm{V} \times \bm{B}\right)
	\end{equation}
	\begin{equation} \label{eq:pres}
		\frac{\partial P}{\partial t}+\nabla \cdot \left(\gamma P \bm{V}\right)=\left(\gamma - 1 \right)\bm{V} \cdot \nabla P,
	\end{equation}
	where $\rho$, $\bm{V}$, $\bm{B}$ and $P$ represent the density, velocity, magnetic field and pressure, respectively, $\mathbb{I}$ is the unit tensor, $\mu$ is the magnetic permeability and $\gamma$ is the adiabatic index. These equations are discretised using a combination of a sixth-order central finite-differences scheme and a WENO scheme (see details in \citet{2015ApJ...799...94T}) and are advanced in time through a 3rd order TVD Runge-Kutta method \citep{1983JCoPh..49..357H}.
	
	On the other hand, we solve an updated version of the 2D generalised eigenvalue problem described in \citet{2015A&A...575A.123S}. That work did not take into account the presence of the transition layer between the internal and external densities but considered the existence of a density jump. Therefore, it could not predict the damping of the oscillations associated with the phenomenon of resonant absorption. For the present work, the effect of transverse inhomogeneity has been included in the solution of the eigenvalue problem. For this method, we linearise the basic MHD Equations (\ref{eq:cont}) -- (\ref{eq:pres}). We only consider transverse oscillations, which are accurately described in the $\beta = 0$ approximation. Therefore, we drop the effect of gas pressure for simplicity. Unlike in \citet{2015A&A...575A.123S}, the considered model is transversely non-uniform and the ideal global mode perturbations are singular (resonant) at the specific radial position where the global mode frequency matches the local Alfvén frequency. In order to avoid the ideal singularity, which is not tractable numerically, here we need to add a small diffusive term in the induction equation. This term does not alter the overall behaviour of the perturbations, neither it modifies the eigenfrequencies, as long as it is sufficiently small \citep[see e.g.,][]{2011A&A...533A..60A}. So, the relevant linearised equations for the computation of eigenvalues are
	\begin{equation} \label{eq:mom_lin}
		\rho_{0} \frac{\partial \bm{v}}{\partial t} = \frac{1}{\mu} \left(\nabla \times \bm{b} \right) \times \bm{B_{0}},
	\end{equation}
	
	\begin{equation} \label{eq:ind_lin}
		\frac{\partial \bm{b}}{\partial t} = \nabla \times \left(\bm{v} \times \bm{B_{0}} \right) - \eta \nabla \times \nabla \times \bm{b},
	\end{equation}
	where $\bm{v} = \left(v_{r}, v_{\phi},0 \right)$ and $\bm{b} = \left(b_{r},b_{\phi},b_{z} \right)$ are the velocity and magnetic field perturbations, respectively, expressed in a cylindrical coordinate system with the $z$-axis aligned with the flux tube axis, $\rho_{0}$ and $\bm{B_{0}} = (0,0,B_{0})$ are the background density and magnetic field, respectively, and $\eta$ is the coefficient of magnetic diffusion (assumed uniform). We put the perturbations proportional to $\exp \left( i m \phi \right)$, where $m$ is the azimuthal wavenumber, and set $m = 1$ corresponding to kink modes. In addition, we express the temporal dependence of the perturbations as $\exp \left(-i \omega t \right)$, where $\omega$ is the complex eigenfrequency. The period of the oscillation is $P = 2\pi/\text{Re}(\omega)$ and the exponential damping time is $\tau_{\rm{D}} = -1 /\text{Im}(\omega)$. Then, Eqs. (\ref{eq:mom_lin}) -- (\ref{eq:ind_lin}) are expanded as
	\begin{equation} \label{eq:vr_lin}
		\omega v_{r} = i \frac{B_{0}}{\mu \rho_{0}} \left( \frac{\partial b_{r}}{\partial z} - \frac{\partial b_{z}}{\partial r} \right),
	\end{equation}
	
	\begin{equation} \label{eq:vphi_lin}
		\omega v_{\phi} = \frac{B_{0}}{\mu \rho_{0}} \left(\frac{b_{z}}{r} + i \frac{\partial b_{\phi}}{\partial z} \right),
	\end{equation}
	
	\begin{eqnarray} \label{eq:br_lin}
		\omega b_{r} &=& i B_{0} \frac{\partial v_{r}}{\partial z} - i \eta \bigg(\frac{b_{r}}{r^{2}} + i\frac{1}{r}\frac{\partial b_{\phi}}{\partial r} + i \frac{b_{\phi}}{r^{2}} \nonumber \\
		&-& \frac{\partial^{2} b_{r}}{\partial z^{2}} + \frac{\partial^{2} b_{z}}{\partial r \partial z} \bigg),
	\end{eqnarray}
	
	\begin{eqnarray} \label{eq:bphi_lin}
		\omega b_{\phi} &=& i B_{0} \frac{\partial v_{\phi}}{\partial z} + i \eta \bigg(\frac{\partial^{2} b_{\phi}}{\partial r^{2}} + \frac{1}{r} \frac{\partial b_{\phi}}{\partial r} - \frac{b_{\phi}}{r^{2}} \nonumber \\ 
		&-& i \frac{1}{r} \frac{\partial b_{r}}{\partial r} + i \frac{b_{r}}{r^{2}} - i \frac{1}{r} \frac{\partial b_{z}}{\partial z} + \frac{\partial^{2} b_{\phi}}{\partial z^{2}}\bigg),
	\end{eqnarray}
	
	\begin{eqnarray} \label{eq:bz_lin}
		\omega b_{z} &=& -i B_{0} \bigg(\frac{\partial v_{r}}{\partial r} + \frac{v_{r}}{r} + i \frac{v_{\phi}}{r} \bigg) + i \eta \bigg(\frac{\partial^{2} b_{z}}{\partial r^{2}} + \frac{1}{r}\frac{\partial b_{z}}{\partial r} \nonumber \\
		&-& \frac{b_{z}}{r^{2}} - \frac{\partial^{2} \partial b_{r}}{\partial r \partial z} - \frac{1}{r}\frac{\partial b_{r}}{\partial z} - i \frac{1}{r} \frac{\partial b_{\phi}}{\partial z} \bigg).
	\end{eqnarray}
	
	Following \citet{2015A&A...575A.123S}, trapped kink oscillations must satisfy the boundary conditions
	\begin{equation} \label{eq:boundary1}
		\frac{\partial v_{r}}{\partial r} = \frac{\partial v_{\phi}}{\partial r} = \frac{\partial b_{r}}{\partial r} = \frac{\partial b_{\phi}}{\partial r} = 0, \ \ b_{z} = 0, \ \ \text{at} \ r = 0,
	\end{equation}
	
	\begin{equation} \label{eq:boundary2}
		v_{r} = v_{\phi} = b_{r} = b_{\phi} = b_{z} = 0, \ \ \text{at} \ r = r_{\rm{max}},
	\end{equation}
	
	\begin{equation} \label{eq:boundary3}
		v_{r} = v_{\phi} = b_{z} = 0, \ \frac{\partial b_{r}}{\partial z} = \frac{\partial b_{\phi}}{\partial z} = 0, \ \ \text{at} \ z = \pm L_{z}/2,
	\end{equation}
	where $r = 0$ is the centre of the tube and $r = r_{\rm{max}}$ is a radial position sufficiently far away from the tube. Ideally, $r_{\rm{max}} \to \infty$, while here we consider $r_{\rm{max}} = 20 R$ for practical purposes. Equations (\ref{eq:vr_lin}) -- (\ref{eq:bz_lin}) together with the above boundary conditions define a 2D generalised eigenvalue problem, where $\omega$ is the eigenvalue and the perturbations form the vector of eigenfunctions. The eigenvalue problem is solved with a more efficient numerical method than that used in \citet{2015A&A...575A.123S}. The scheme is implemented in a \textit{Wolfram Mathematica} routine, where the spatial integration of the equations in the $r$ and $z$ directions is done with finite elements in a 2D structured mesh with a resolution of $400 \times 40$ cells. The mesh resolution is non-uniform in the radial direction, as a much fine resolution is needed in the non-uniform transitional layer to correctly describe the spatial scales associated with the resonant behaviour of the perturbations. The routine returns the closest eigenvalue to an initially provided guess and can be run iteratively to compute the solution as a function of a specific parameter of the model.
	
\subsection{Initial perturbation}
	Transverse kink oscillations in flux tubes have the azimuthal wavenumber $m = 1$ \citep{1983SoPh...88..179E,2005LRSP....2....3N}. To trigger this kind of oscillations in our simulations, we apply the following perturbation to the $x$-component of velocity:
	\begin{equation} \label{eq:pert}
		V_{x}(t=0) = V_{0} f(r) \cos \left(n k_{z} z + \frac{\left(n-1 \right)}{2} \pi \right),
	\end{equation}
	where
	\begin{equation}
		f(r) = \frac{\rho(r,z=0)-\rho_{\rm{ex}}}{\rho_{\rm{i,0}}-\rho_{\rm{ex}}},
	\end{equation}
	$k_{z}=\pi / L_{z}$ and $n$ is the longitudinal oscillation mode. The amplitude of the perturbation is expressed in terms of the internal Alfvén speed, $V_{0} = \alpha c_{\rm{A,in}}$, where $c_{\rm{A,in}} = B_{0} / \sqrt{\mu \rho_{i,0}}$ and $\alpha$ is a dimensionless factor that controls the strength of the perturbation.

	The factor $f(r)$ is used to ensure that the perturbation is applied to the tube and the transition layer but not to the external background plasma.

	If $n = 1$, the perturbation corresponds to the fundamental longitudinal mode of oscillation; $n = 2$ corresponds to the first longitudinal overtone. In a tube with a uniform density, the periods of these modes in the thin tube approximation are given by \citep{1983SoPh...88..179E}
	\begin{eqnarray} \label{eq:period}
		P_{0} = \frac{2 L_{z}}{c_{\rm{A,in}}} \sqrt{\frac{1+\rho_{\rm{ex}}/\rho_{\rm{i,0}}}{2}} \ \ \text{and} \ \ P_{1} = \frac{P_{0}}{2},
	\end{eqnarray}
	respectively. In the following sections we check how the inhomogeneity in density modifies the oscillation periods of these two modes.
	
	We note here that the longitudinal dependence given in Eq. (\ref{eq:pert}) corresponds to an eigenfunction of kink oscillations in longitudinally uniform tubes but not for the case with a longitudinal inhomogeneity \citep{2010ApJ...722.1778S,2011A&A...533A..60A,2015A&A...575A.123S}. Therefore, this perturbation will trigger additional longitudinal harmonics. However, we expect that the contribution of these modes has a negligible effect on the evolution of the transverse oscillations we are interested in analysing.
	
\section{Analysis of numerical results} \label{sec:results}
	In this section we present the results of our numerical study of oscillations of inhomogeneous threads and perform comparisons with the case of longitudinally homogeneous threads. All the simulations analysed here used a numerical domain of $200 \times 200 \times 100$ points, which corresponds to a spatial resolution of $50 \ \rm{km}$ in the $x$ and $y$ directions, and of $500 \ \rm{km}$ in the vertical direction.
	
\subsection{Example simulations} \label{sec:ref_sims}
	\begin{figure*}
		\centering
		\includegraphics[width=4cm,height=4cm]{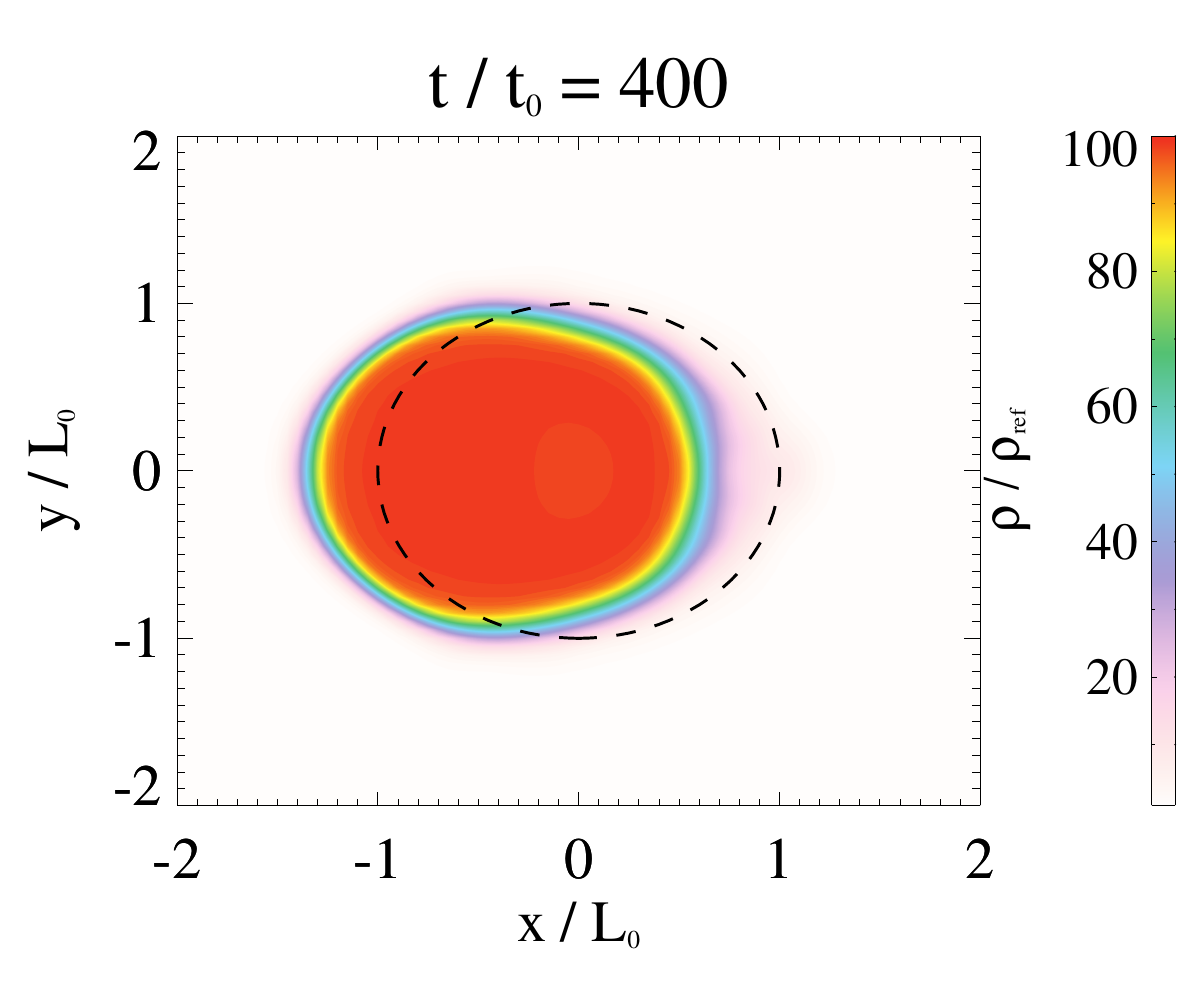}
		\includegraphics[width=4cm,height=4cm]{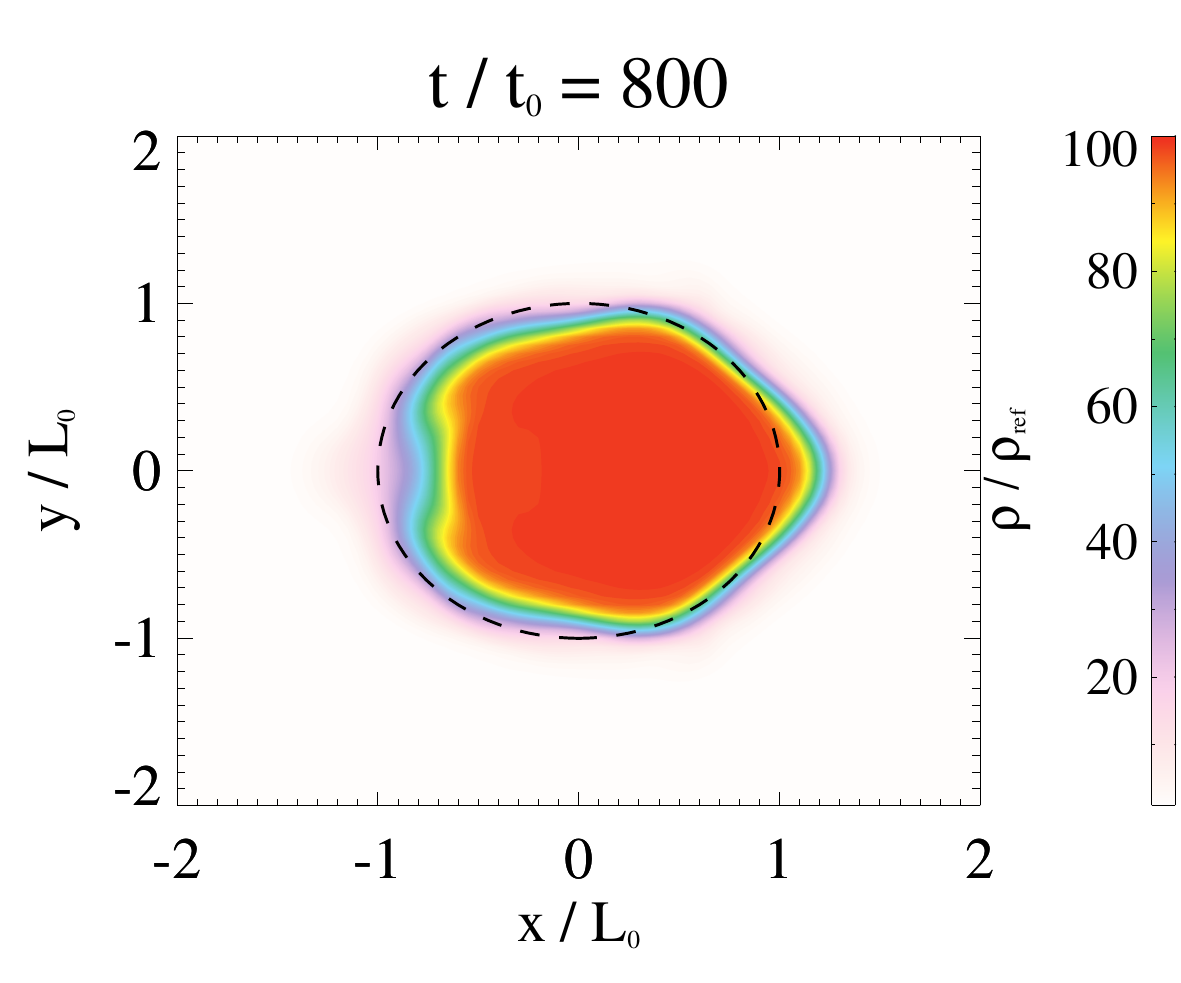}
		\includegraphics[width=4cm,height=4cm]{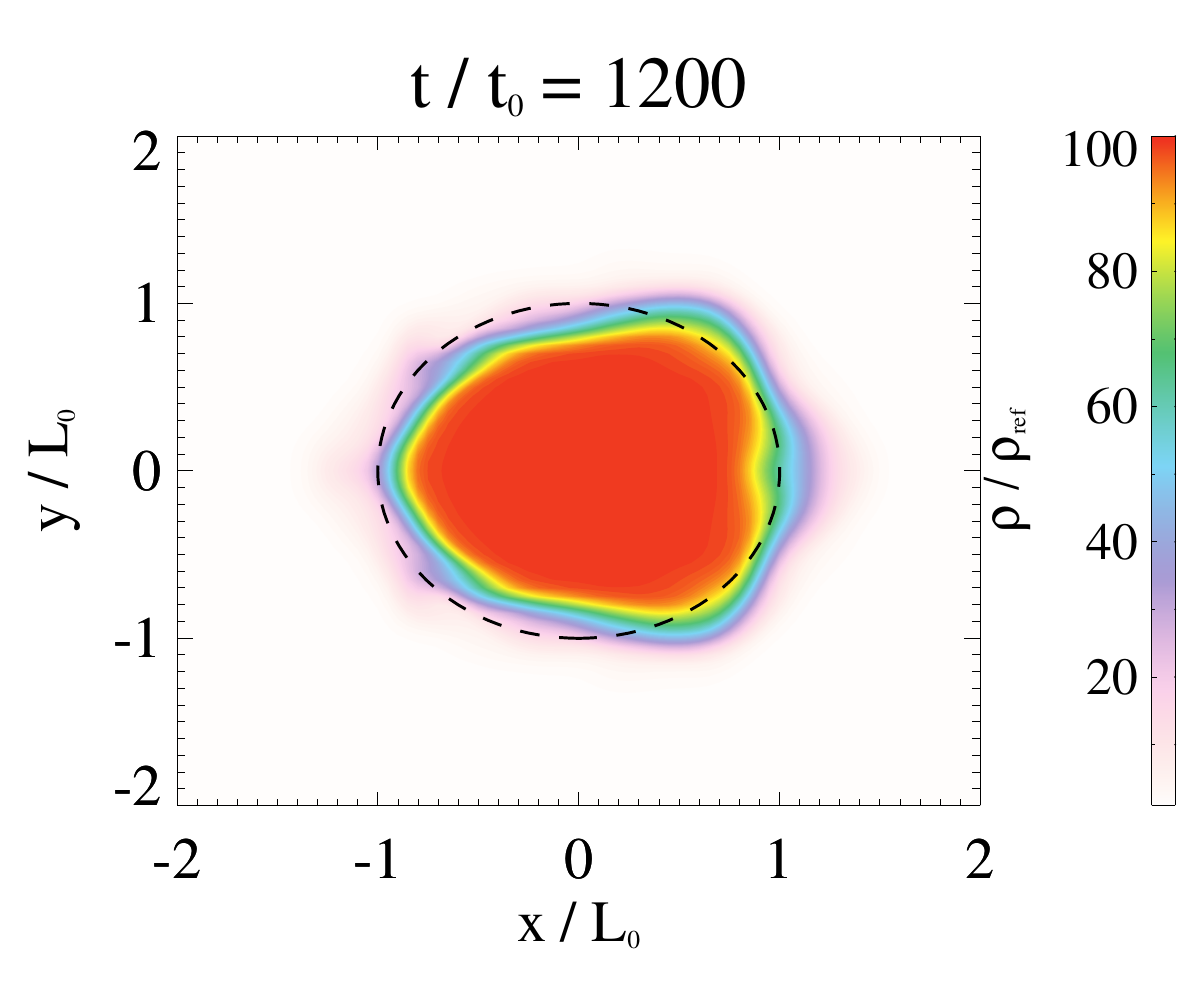}
		\includegraphics[width=4cm,height=4cm]{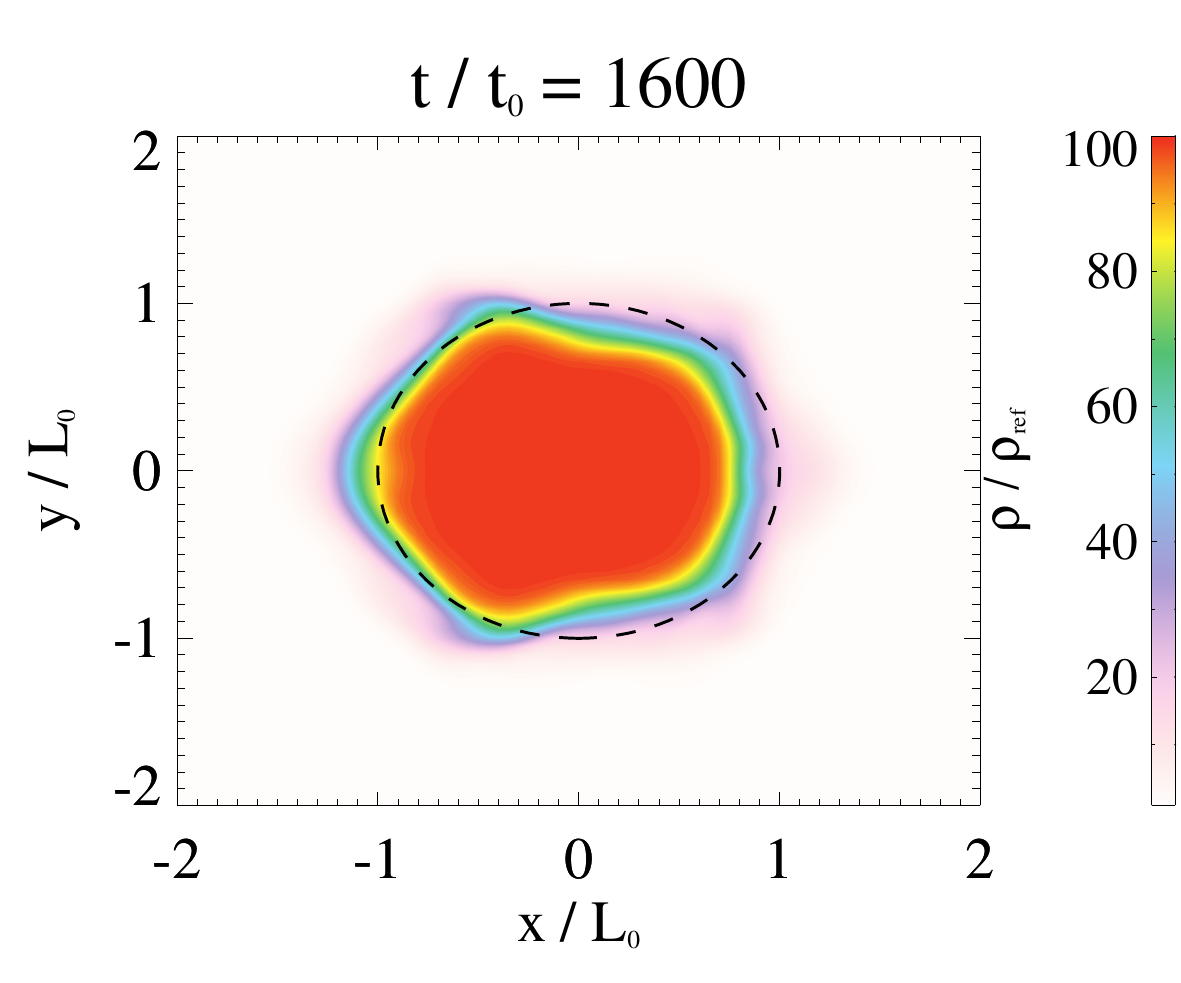} \\
		\includegraphics[width=4cm,height=4cm]{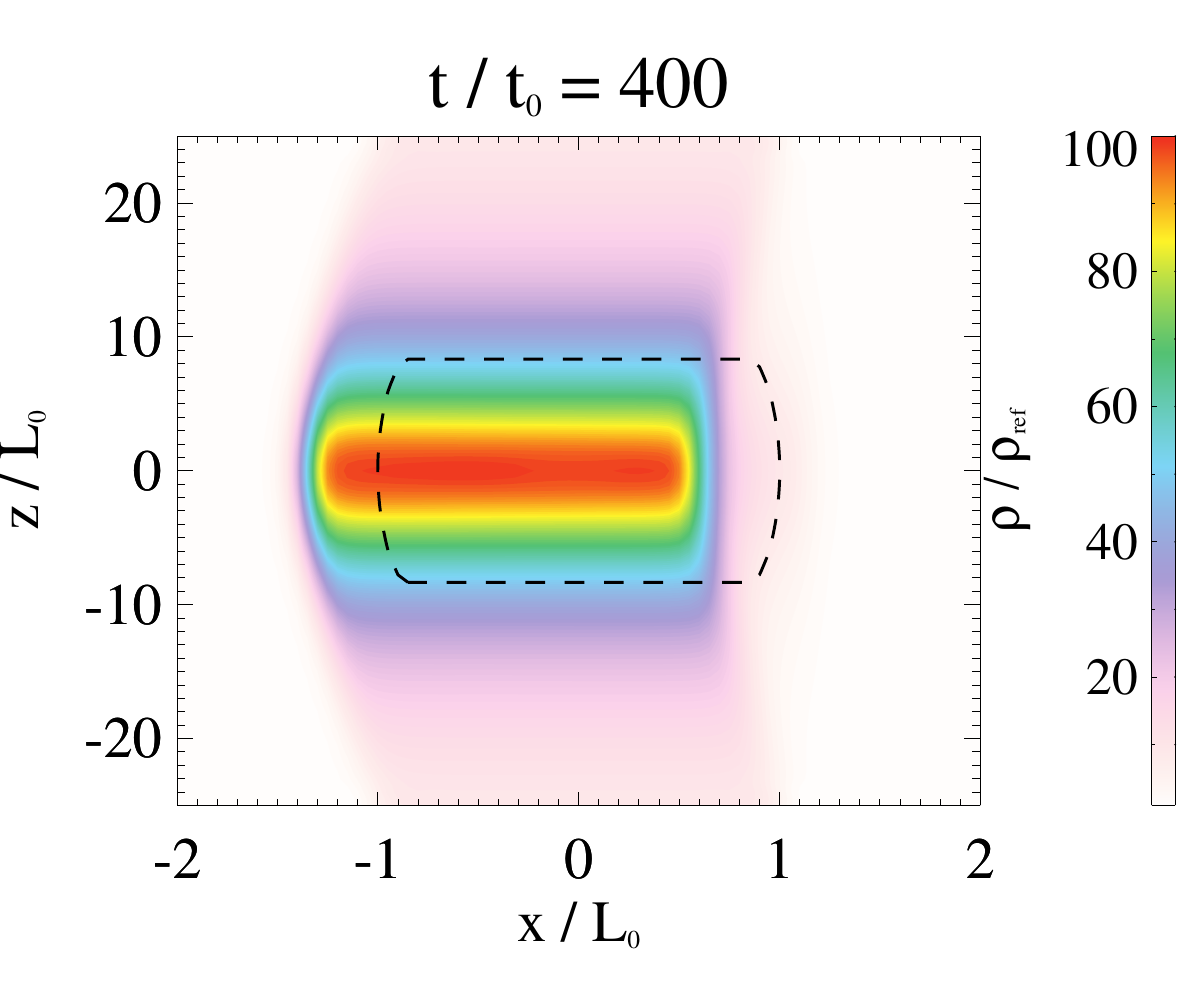}
		\includegraphics[width=4cm,height=4cm]{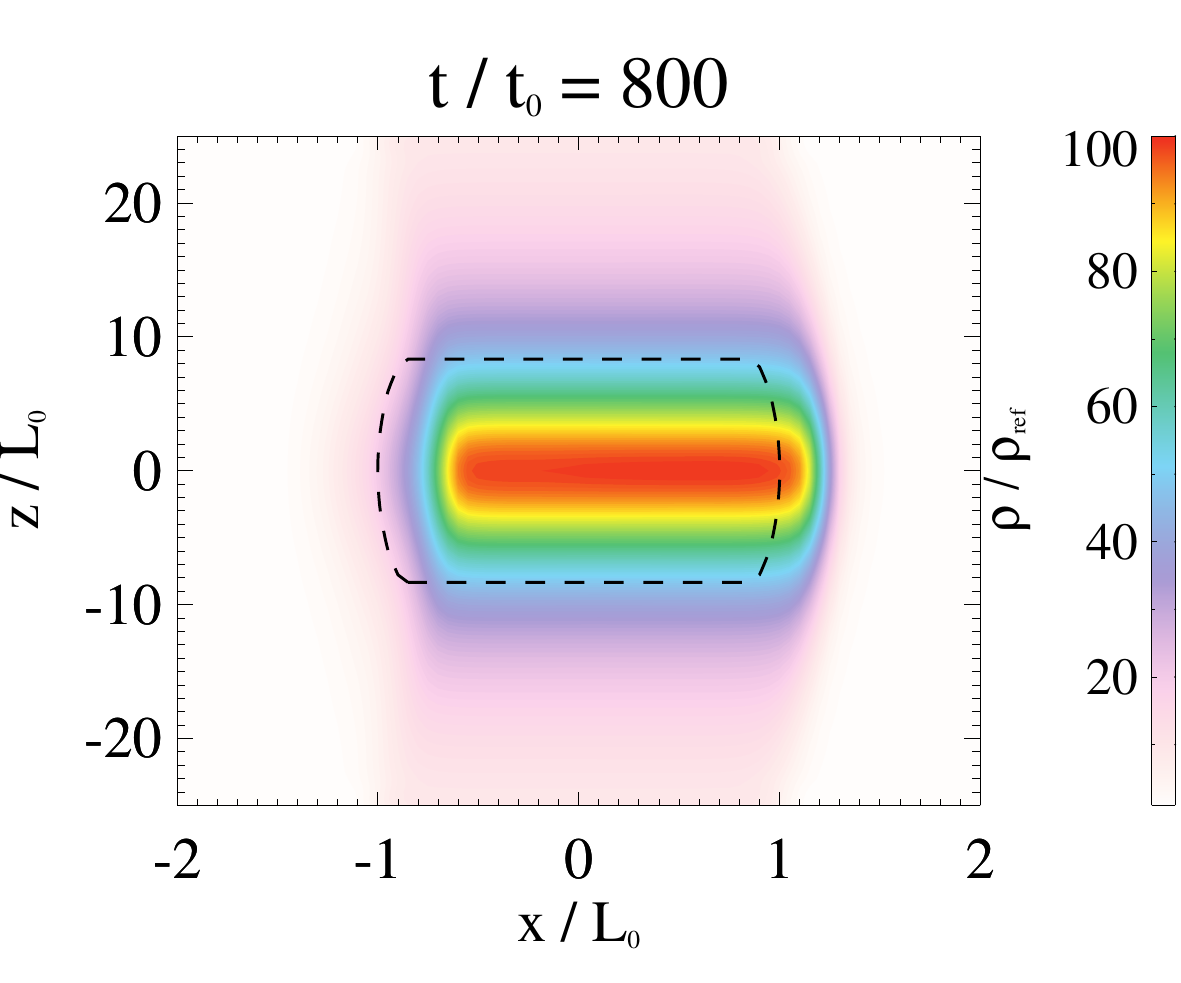}
		\includegraphics[width=4cm,height=4cm]{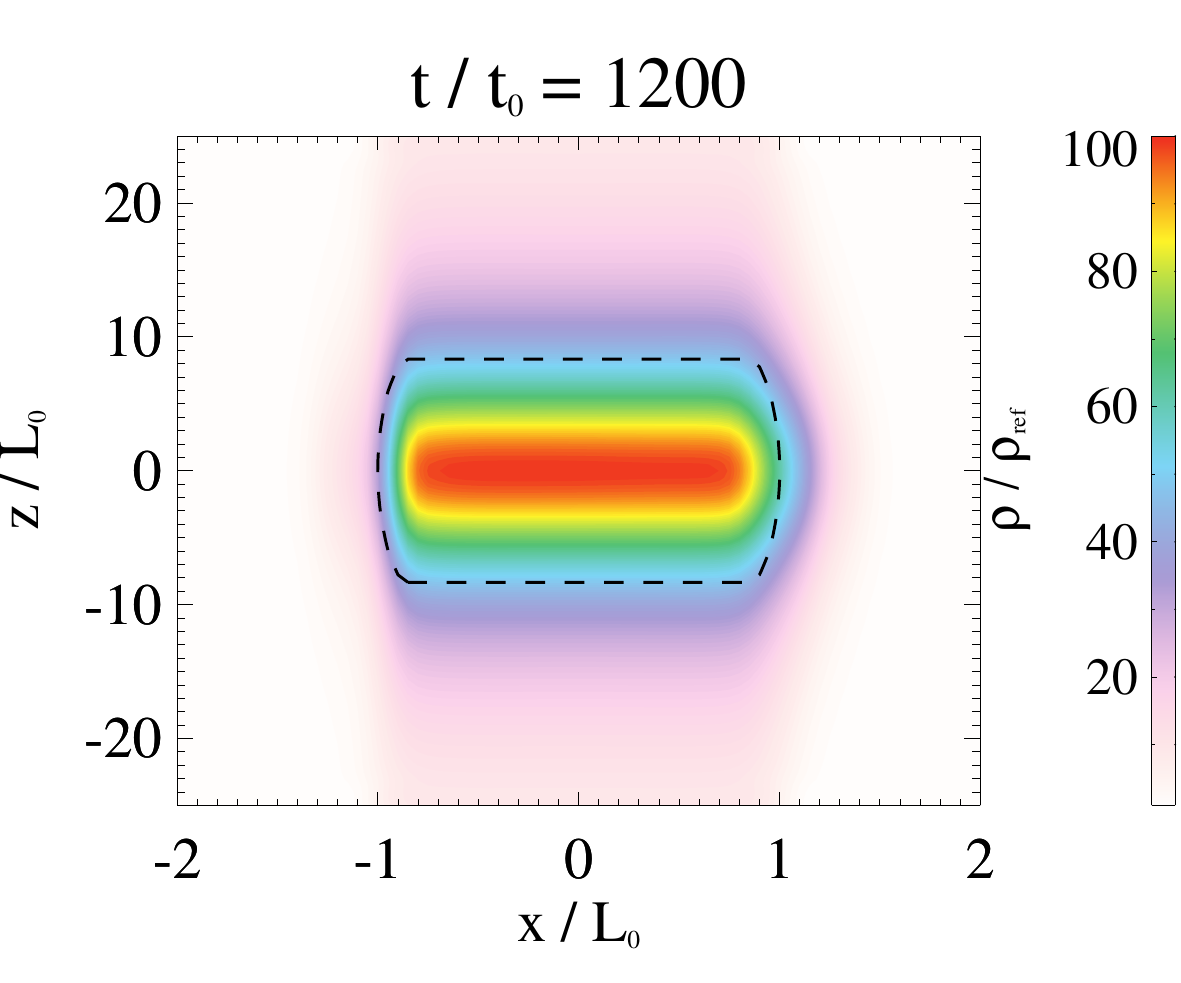}
		\includegraphics[width=4cm,height=4cm]{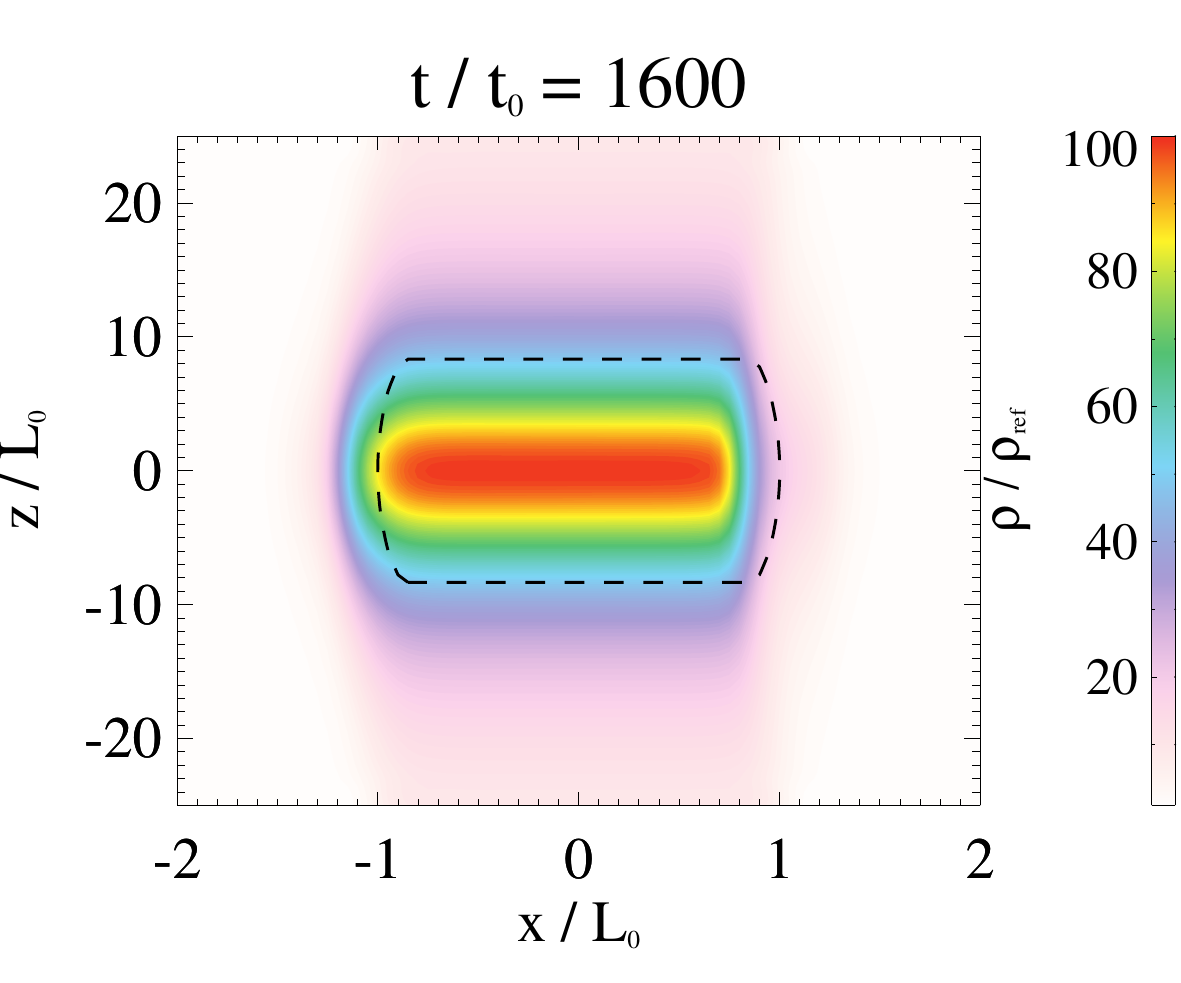}
		\caption{Snaphots of a simulation with $\chi = 10$, $V_{0} = 0.05 c_{\rm{A,in}}$, $n = 1$, and a numerical domain of $200 \times 200 \times 100$ points. Top: density colour maps at $z=0$. Bottom: density colour maps at $y = 0$. Dashed lines represent the density contour $\rho = 50 \rho_{\rm{ref}}$ at $t = 0$.}
		\label{fig:rho_snapshots}
	\end{figure*}

    \begin{figure*}
    	\centering
    	\includegraphics[width=4cm,height=4cm]{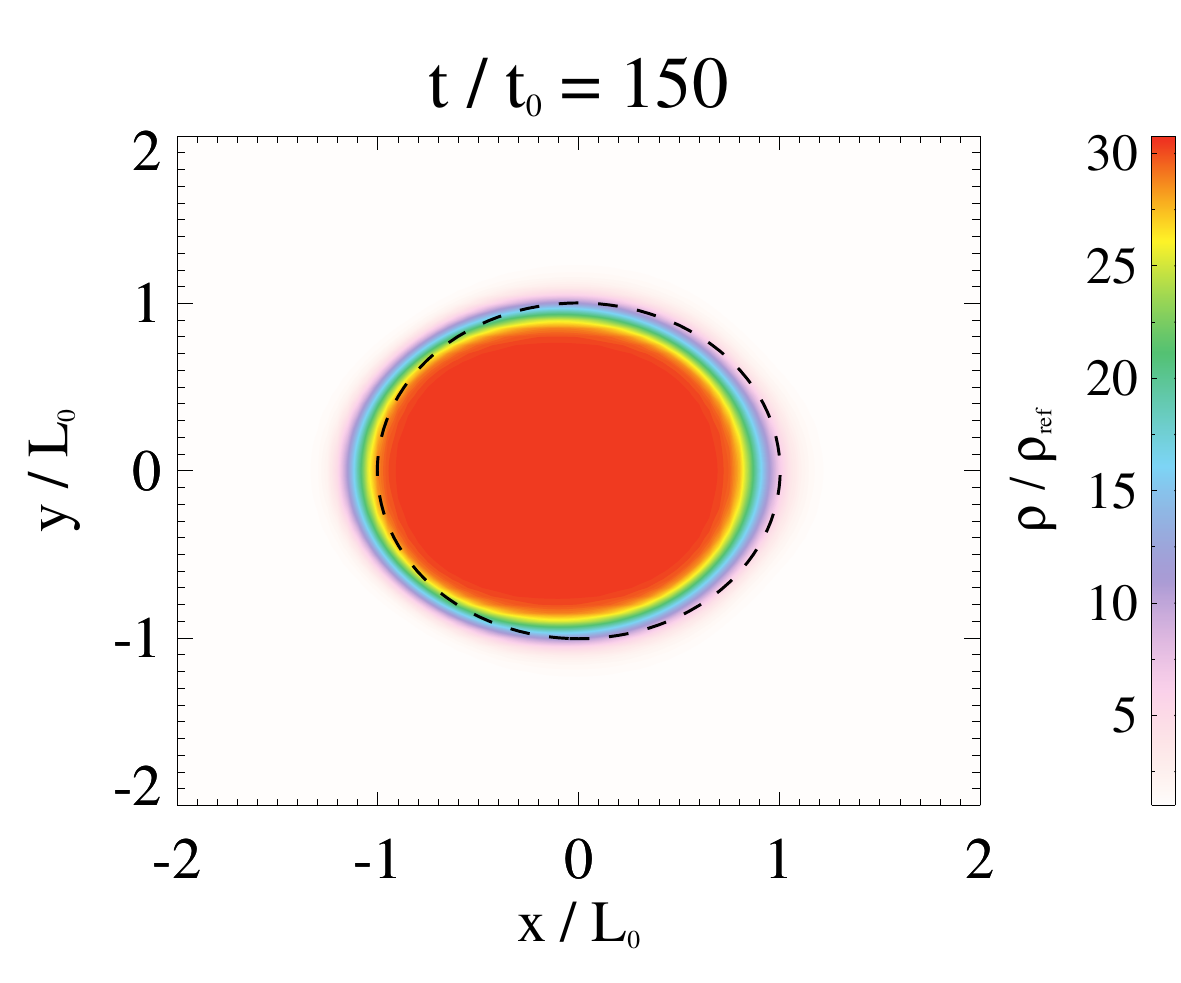}
    	\includegraphics[width=4cm,height=4cm]{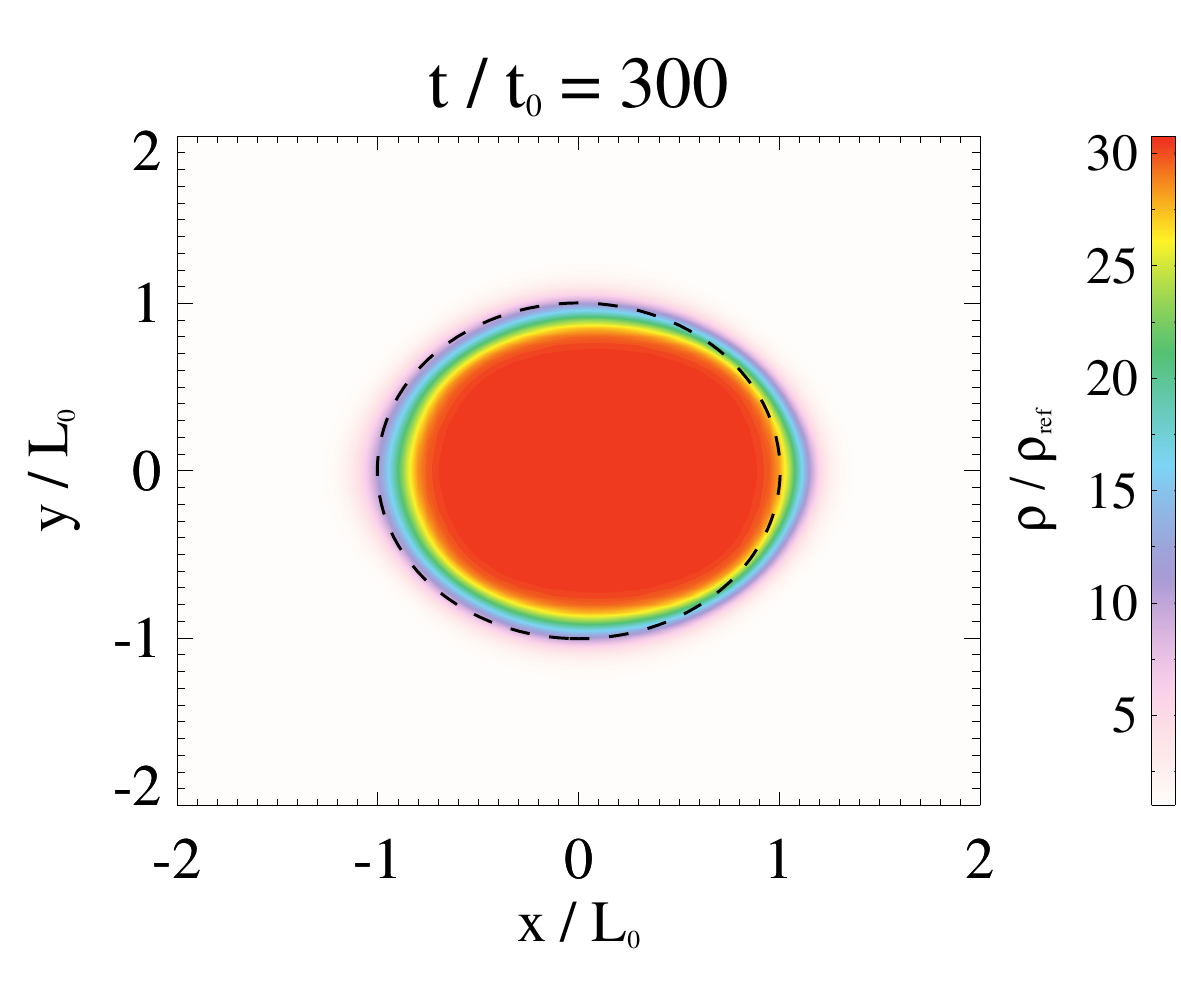}
    	\includegraphics[width=4cm,height=4cm]{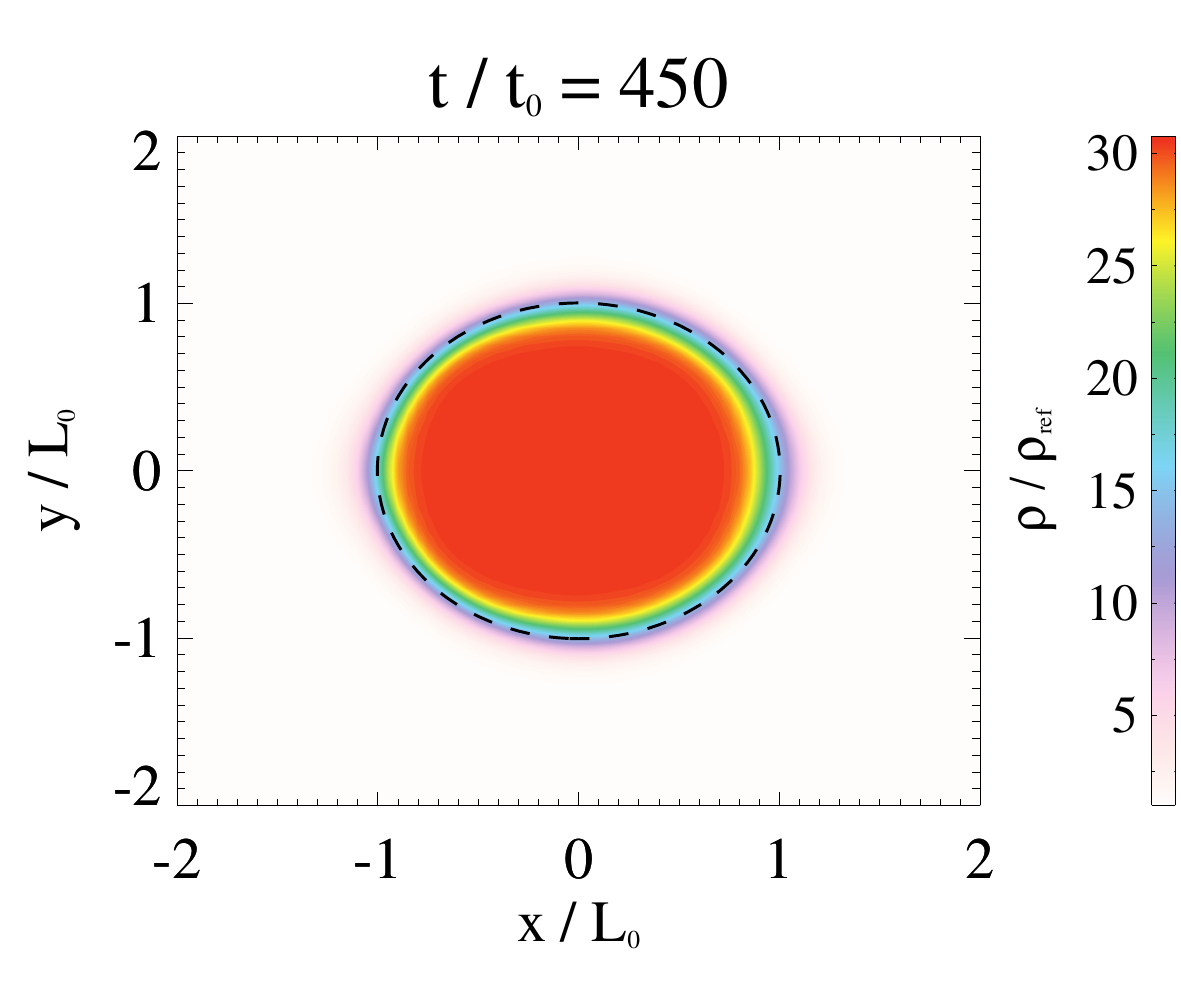}
    	\includegraphics[width=4cm,height=4cm]{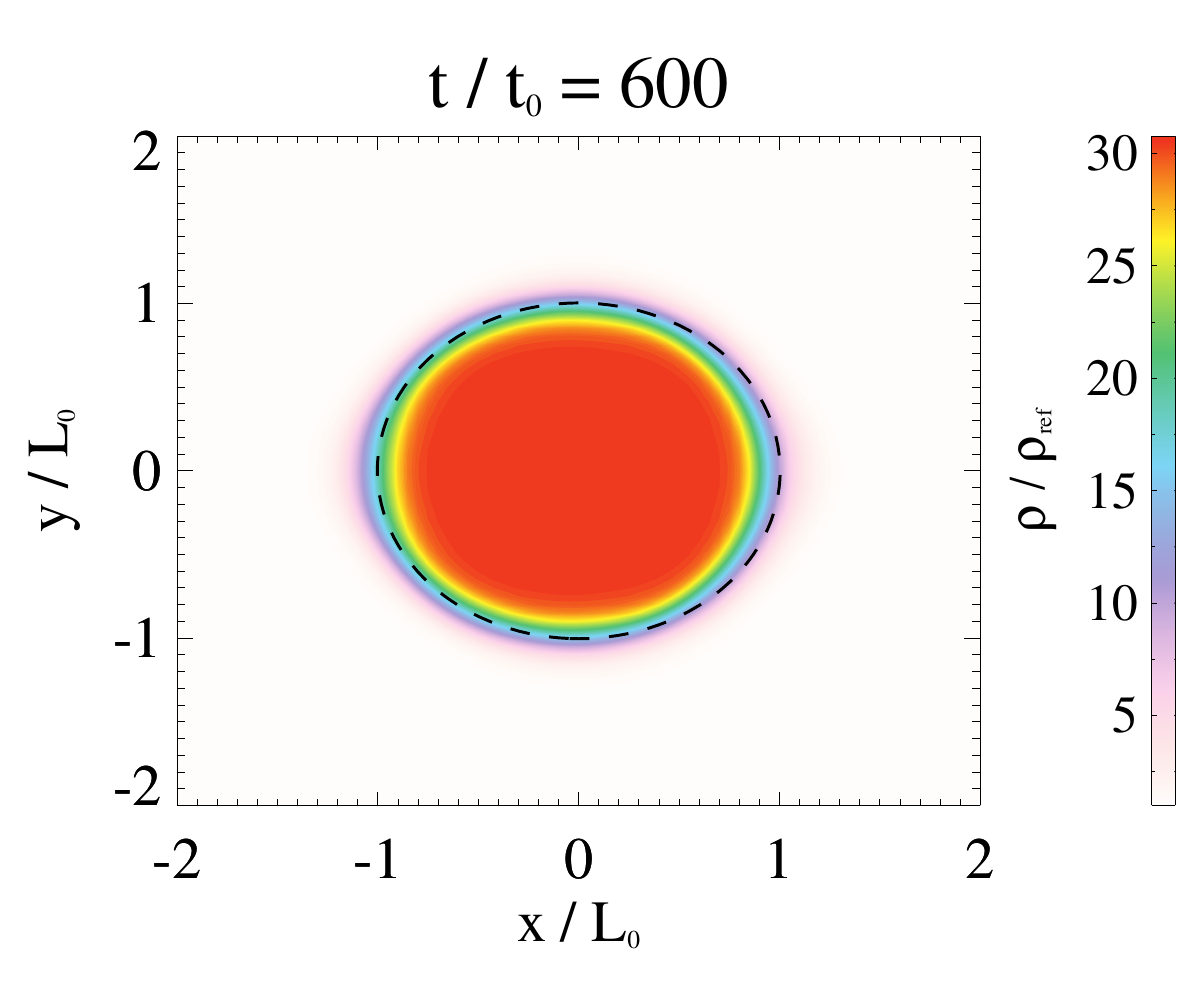} \\
    	\includegraphics[width=4cm,height=4cm]{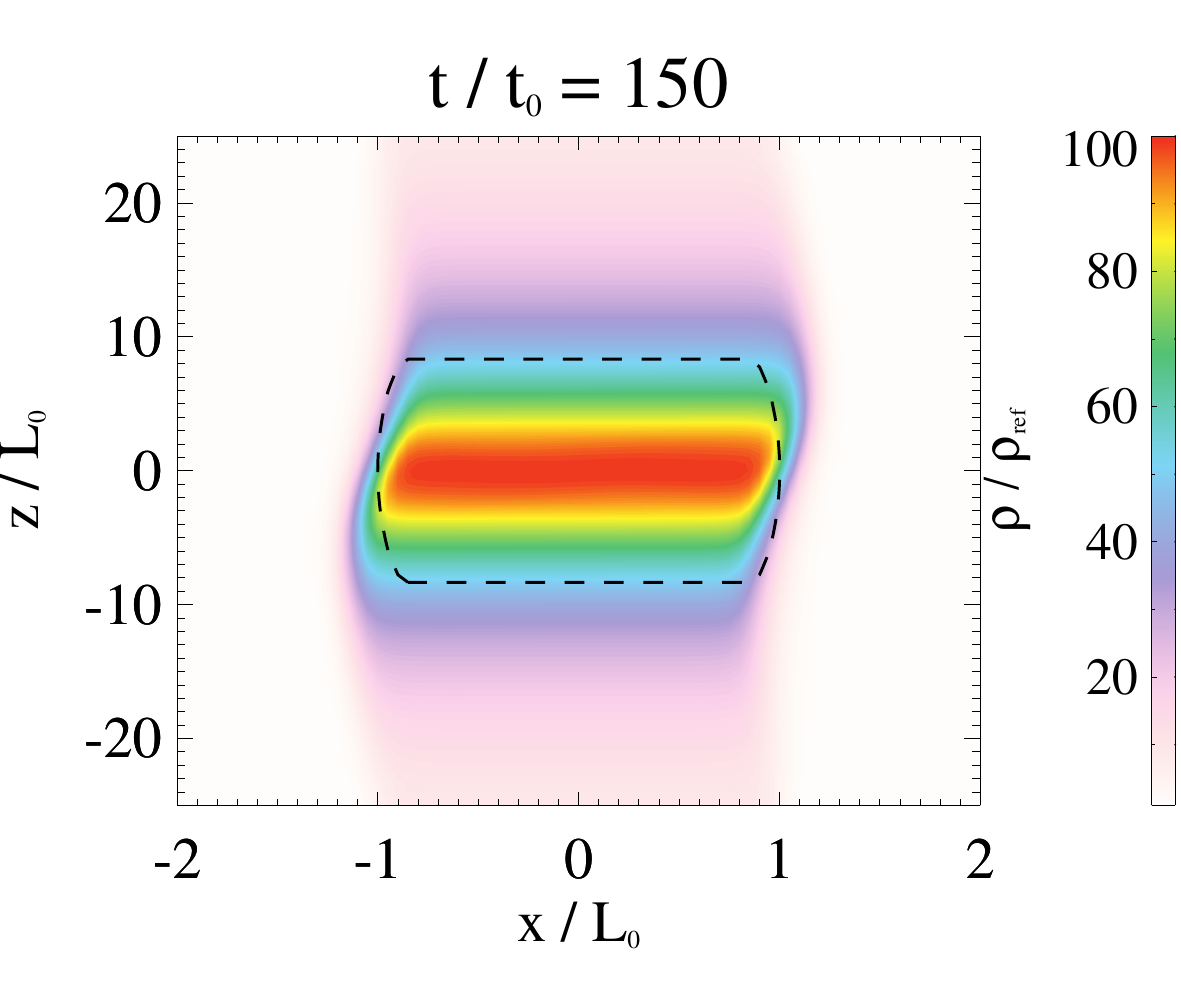}
    	\includegraphics[width=4cm,height=4cm]{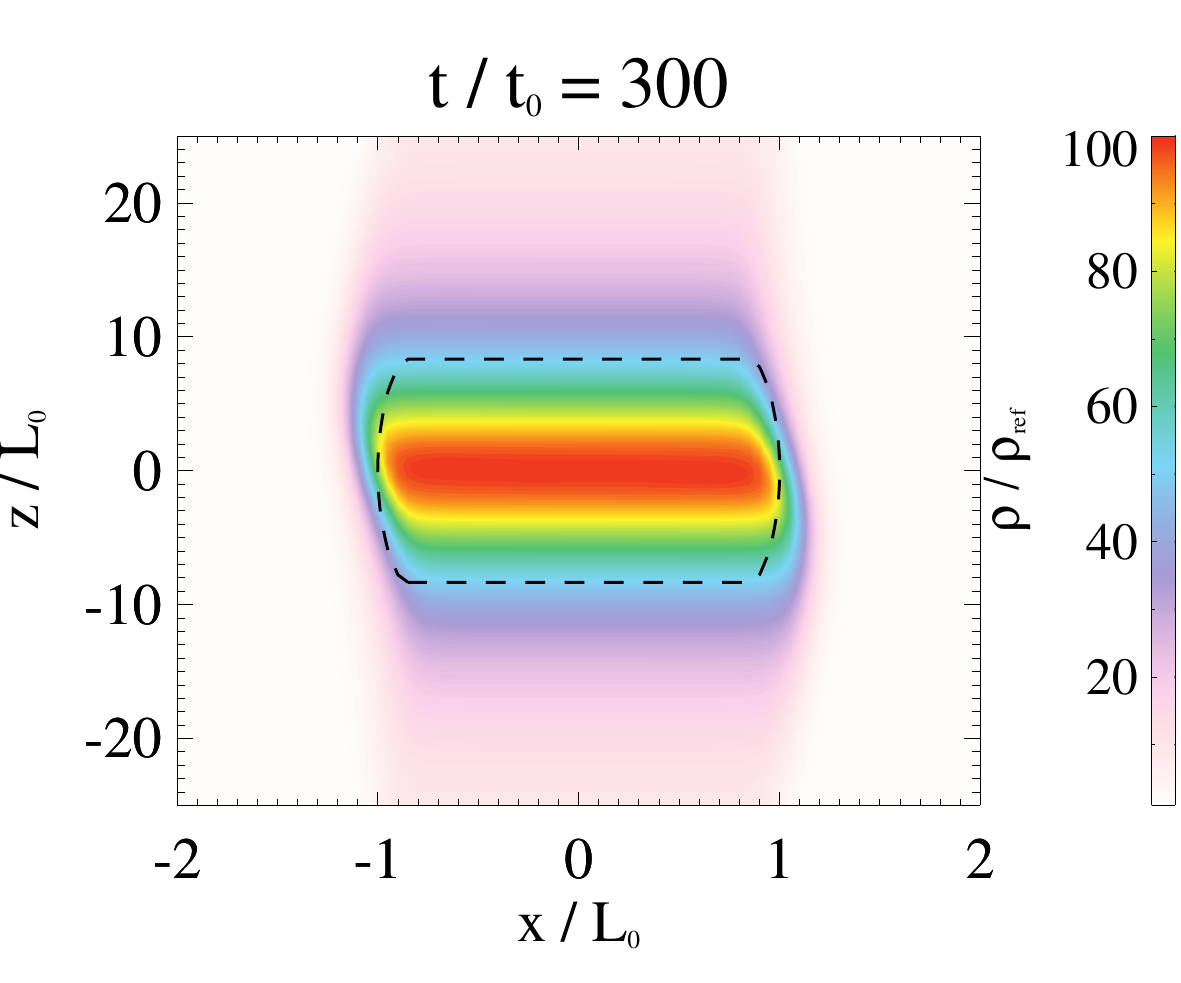}
    	\includegraphics[width=4cm,height=4cm]{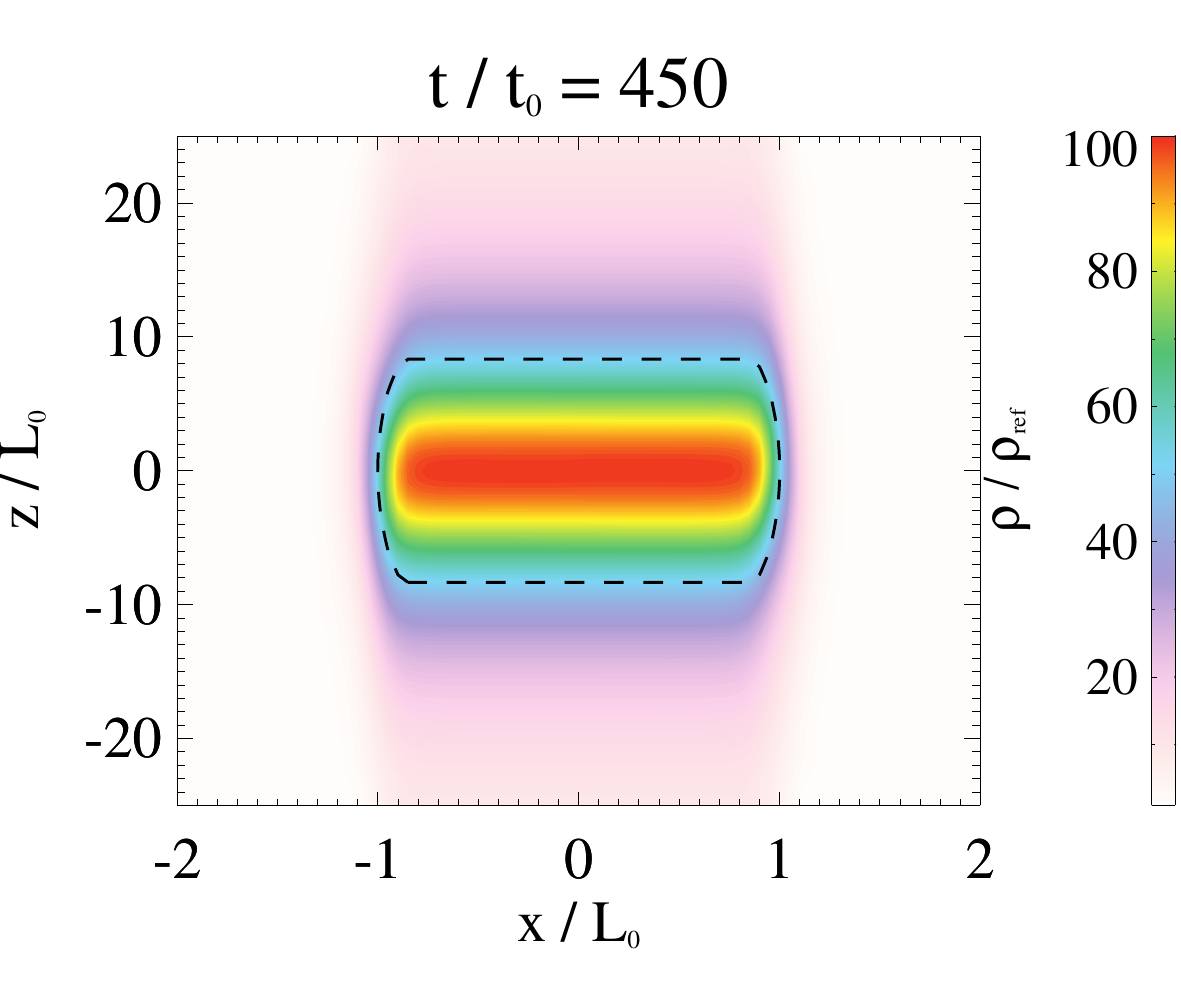}
    	\includegraphics[width=4cm,height=4cm]{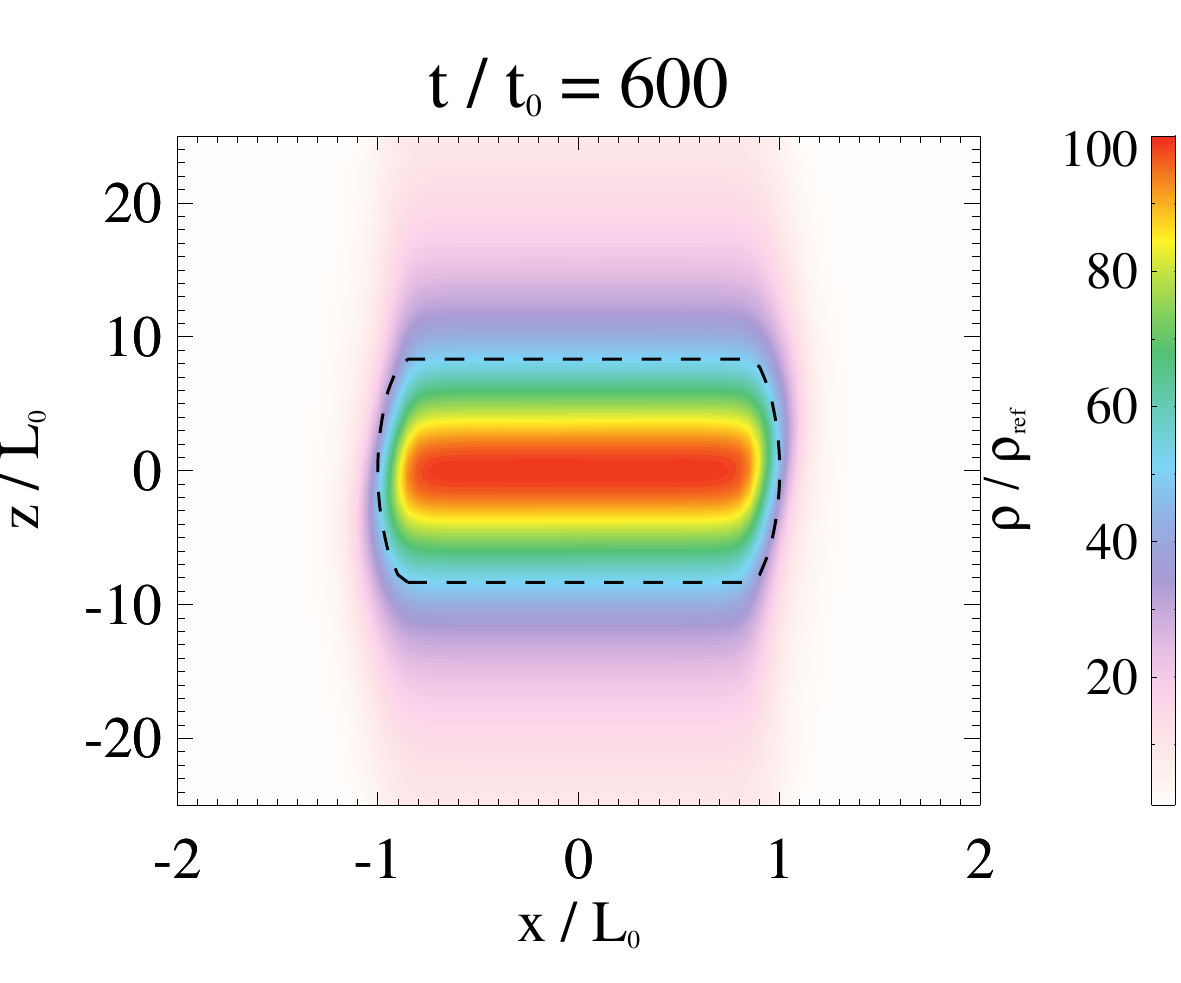}
    	\caption{Snaphots of a simulation with $\chi = 10$ and $n = 2$. Top: density colour maps at $z=-L_{z}/4 = -12.5 L_{0}$. Bottom: density colour maps at $y = 0$. Dashed lines represent the density contours $\rho \approx 15 \rho_{\rm{ref}}$ (top panels) and $\rho = 50 \rho_{\rm{ref}}$ (bottom) at $t = 0$.}
    	\label{fig:rho_snapshots_nk2}
    \end{figure*}

	We start by describing a simulation of the fundamental longitudinal mode of oscillation in an inhomogeneous thread with a longitudinal density ratio of $\chi = 10$. This transverse oscillation is triggered by applying the initial perturbation given by Eq. (\ref{eq:pert}) with $n = 1$. For this example, the amplitude of the perturbation is $V_{0} = 0.05 c_{\rm{A,in}}$. 
		
	The perturbation causes a lateral displacement of the tube that follows a sinusoidal motion. The displacement is maximum at the central position of the tube and zero at its ends. This behaviour is illustrated by the density snapshots displayed in Fig. \ref{fig:rho_snapshots}. In addition to the oscillation around the equilibrium position, the horizontal cuts of density at the height $z = 0$ (top panels of Fig. \ref{fig:rho_snapshots}) show that the shape of the originally cylindrical thread is deformed as time advances. This is a consequence of the triggering of the KHI \citep{2008ApJ...687L.115T,2014ApJ...787L..22A,2015A&A...582A.117M} and the non-linear generation of fluting modes \citep{2014SoPh..289.1999R,2016A&A...595A..81M,2018ApJ...853...35T}, which are modes with azimuthal wavenumber $m \geq 2$ \citep{2005LRSP....2....3N} localised around the boundary of the tube \citep{2017ApJ...850..114S}.
	
	Now, we turn our attention to the case of the first longitudinal overtone, with $n = 2$. For this simulation we used the same value of $V_{0}$ as in the previous case. Fig. \ref{fig:rho_snapshots_nk2} shows that the first overtone has a node at $z = 0$: there is no lateral displacement at the position where the tube is denser. Conversely, it has two anti-nodes, one at each half of the tube, which oscillate laterally in opposite directions. We see that the amplitude of the displacement for this mode is smaller than that for the fundamental mode although the same value of $V_{0}$ was used, and, consequently, the boundary of the tube is less distorted.
	
\begin{figure}
		\centering
		\resizebox{\hsize}{!}{\includegraphics[width=0.8\hsize]{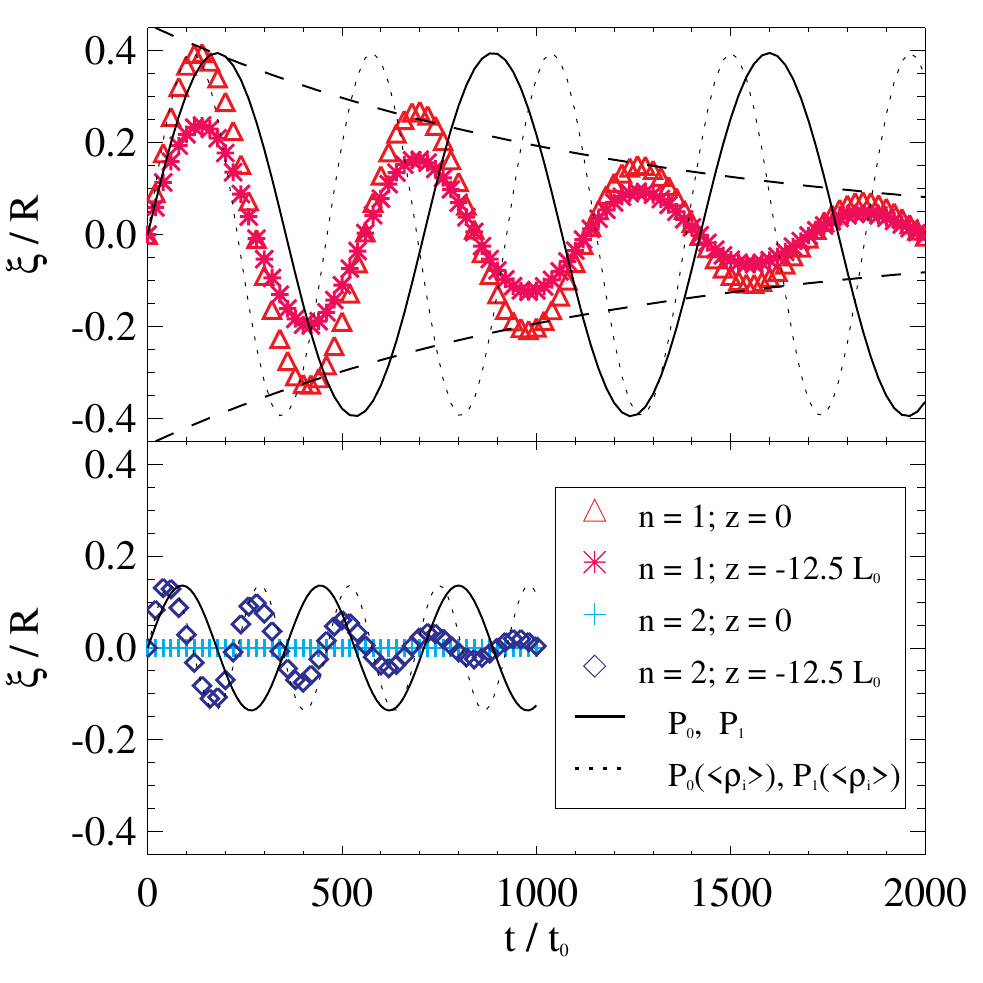}}
		\caption{Displacement of the tube axis at different longitudinal positions from simulations of modes $n = 1$ (top) and $n = 2$ (bottom). The solid and dotted lines represent oscillations with periods given by Eq. (\ref{eq:period}) for tubes of uniform density equal to $\rho_{\rm{i,0}}$ and $\langle \rho_{\rm{i}} \rangle$, respectively. The dashed lines represent the exponential decay of the displacement of the fundamental mode.}
		\label{fig:disp_axis}
	\end{figure}

	In order to obtain a more precise description of the two simulated oscillations, we have computed the displacement of the centre of mass of the density cuts located at $z = 0$ and $z = -L_{z}/4 = -12.5 L_{0}$, respectively. The results are presented in Fig. \ref{fig:disp_axis}. The top panel confirms that for the mode $n = 1$ the different longitudinal positions of the tube oscillate in phase and it shows that the amplitude of the oscillation decreases with time. The main reason for this attenuation is the linear process of resonant absorption, which causes a transfer of energy from the global kink oscillation to localised Alfvén modes in the transverse inhomogeneous layer \citep{1978ApJ...226..650I,2002ApJ...577..475R,2008ApJ...682L.141A,2010ApJ...722.1778S}. The non-linear excitation of fluting modes \citep[see e.g.,][]{2014SoPh..289.1999R,2017ApJ...850..114S} and the triggering of the KHI \citep{2018ApJ...853...35T,2021ApJ...910...58V} also contribute to the damping of the oscillation.
	
	The bottom panel of Fig. \ref{fig:disp_axis} shows that for the first overtone there is no lateral displacement at $z = 0$. The displacement at $z = -L_{z}/4$ follows a similar behaviour than that of the fundamental mode but with smaller amplitude and shorter period.
	
	We include in Fig. \ref{fig:disp_axis} black solid lines that represent oscillations of periods $P_{0}$ and $P_{1}$, respectively, which are given by Eq. (\ref{eq:period}). These periods correspond to kink oscillations in longitudinally uniform thin tubes with density equal to $\rho_{\rm{i,0}}$. The dotted black lines represent oscillations of tubes with longitudinally uniform density given by the average density of the Lorentzian profile, that is, $\rho_{\rm{i}}(z) = \langle \rho_{\rm{i}} \rangle$. We see that the thread with $\chi = 10$ oscillates faster than the tube with $\rho_{\rm{i}}(z) = \rho_{\rm{i,0}}$, but slower than the tube with $\rho_{\rm{i}}(z) = \langle \rho_{\rm{i}} \rangle$, in agreement with the findings of \citet{2015A&A...575A.123S}. The oscillation periods of the tube with the Lorentzian profile are $P(n=1) \approx 568 t_{0}$ and $P(n=2) \approx  226 t_{0}$. Therefore, the period ratio follows the relation $P(n=1) / P(n=2) > 2$, in agreement with the findings of \citet{2010ApJ...725.1742D}, \citet{2011A&A...533A..60A}, and \citet{2011A&A...531A.167S} for prominence threads.
	
	The dashed lines in the top panel of Fig. \ref{fig:disp_axis} show that the attenuation is well described by an exponential decay during the whole course of the simulation. This behaviour seems to differ from the results for longitudinally uniform threads obtained by \citet{2012A&A...539A..37P}, \citet{ 2013A&A...555A..27R}, \citet{2016A&A...595A..81M}, and \citet{2018ApJ...853...35T}, who found that during the first periods of the oscillation the damping is better fitted by a Gaussian profile. \citet{2013A&A...555A..27R} and \citet{2016A&A...595A..81M} found that the damping changes from a Gaussian to an exponential profile at a switch time given by
	\begin{equation} \label{eq:switch_time}
		t_{\rm{s}} = \frac{\rho_{\rm{i,0}}/\rho_{\rm{ex}}+1}{\rho_{\rm{i,0}}/\rho_{\rm{ex}}-1}P.
	\end{equation}
	
	 The switch time decreases with the density contrast between the internal and external plasmas. For the simulations we are analysing here (with $\rho_{\rm{i,0}} = 100 \rho_{\rm{ex}}$) the switch time is approximately equal to one period. Therefore, the behaviour represented in the top panel of Fig. \ref{fig:disp_axis} may agree with the findings for longitudinally uniform tubes: the apparent discrepancy may be a consequence of the chosen value for the ratio $\rho_{\rm{i,0}} / \rho_{\rm{ex}}$ instead of a consequence of the longitudinal variation of density given by the parameter $\chi$.
	
\subsection{Parametric study} \label{sec:parametric}
	In the following sections we vary the value of the parameter $\chi$ (which implies a variation of the average density of the threads) and study the general behaviour of the periods, damping times and amplitudes of the transverse oscillations.
	 
\subsubsection{Fundamental longitudinal mode} \label{sec:fundamental}
\begin{figure*}
    	\centering
    	\includegraphics[width=17cm,height=11cm]{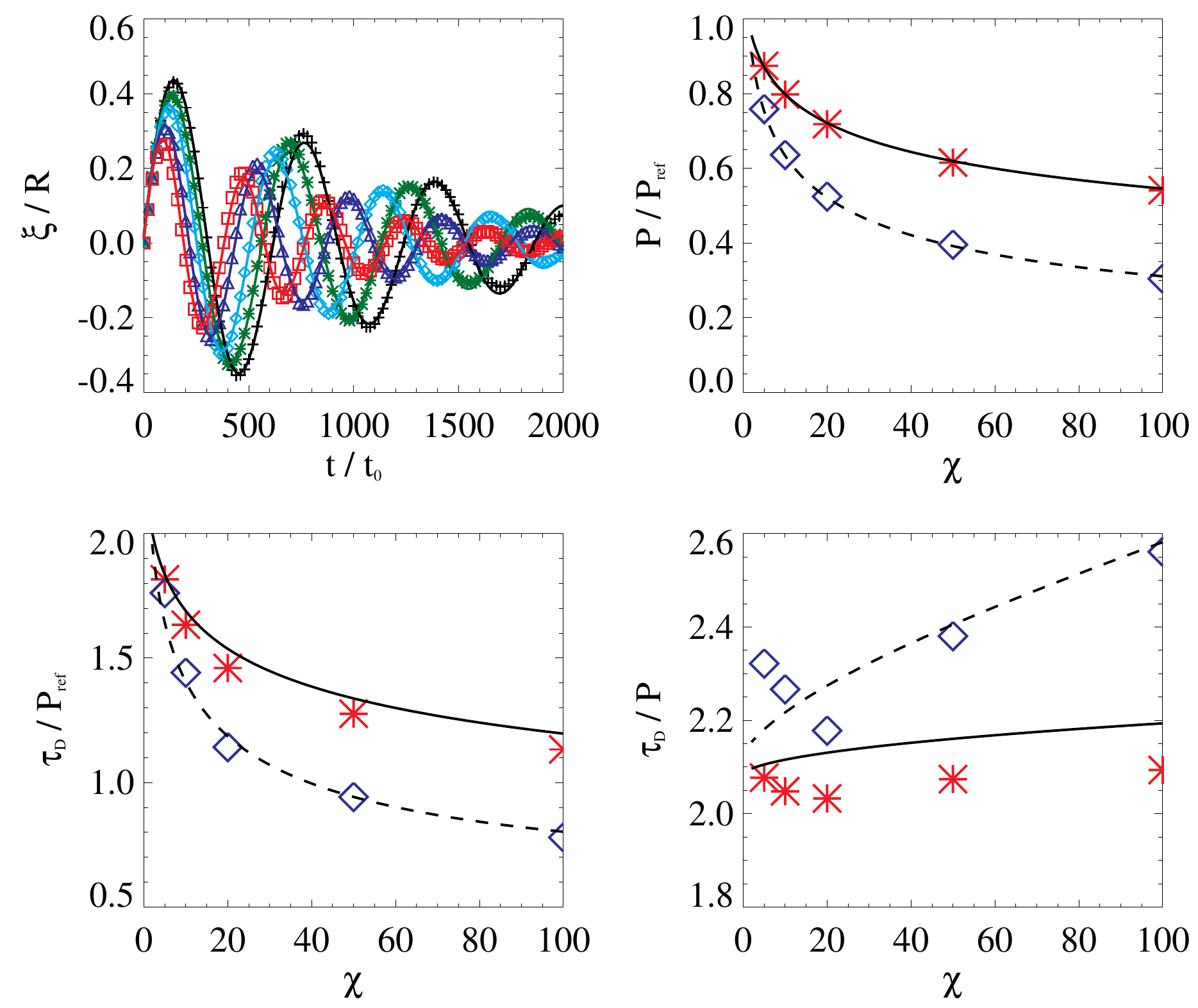}
    	\caption{Top left: displacement of the centre of mass as a function of time from simulations with $n = 1$ and with $\chi = 5$ (black crosses), $\chi = 10$ (green asterisks), $\chi = 20$ (light blue diamonds), $\chi = 50$ (dark blue triangles), and $\chi = 100$ (red squares). Top right: oscillation period as a function of the density contrast, $\chi$. Bottom left: damping time as a function of $\chi$. Bottom right: ratio of damping time to period as a function of $\chi$. On the top left panel, solid lines represent fits of the numerical results. On the remaining panels, the black solid and dashed lines correspond to the results obtained by solving the 2D eigenvalue problem for $n = 1$ and $n = 2$, respectively. Symbols represent the results from the simulations. For $n = 1$, $P_{\rm{ref}} = P_{0}$; for $n = 2$, $P_{\rm{ref}} = P_{1}$.}
    	\label{fig:fits}
    \end{figure*}

	The top left panel of Fig. \ref{fig:fits} presents the results from a series of simulations in which the initial perturbation corresponds to the fundamental mode of oscillation, $n = 1$, and its amplitude is given by $V_{0} = 0.05 c_{\rm{A,in}}$. It shows the transverse displacement of the centre of mass of the tube as a function of time for the density profiles plotted in Fig. \ref{fig:profiles}: black crosses, green asterisks, light blue diamonds, dark blue triangles, and red squares correspond to the longitudinal density ratios $\chi = 5$, $\chi = 10$, $\chi = 20$, $\chi = 50$, and $\chi = 100$, respectively. We see in this panel that the amplitude, period and the damping time of oscillation depend on the chosen longitudinal density ratio. To extract these oscillation parameters from the data of the simulations, we fitted each line to an exponentially decaying sinusoidal function,
	\begin{equation} \label{eq:fit_exp}
		\frac{\xi}{R} = A_{0} \sin \left(\frac{2\pi}{P}t\right) \exp \left(-\frac{t}{\tau_{\rm{D}}}\right),
	\end{equation}
	where $A_{0} \equiv \left(\xi/R\right)_{0}$ is the dimensionless amplitude of the oscillation, $P$ is the period, and $\tau_{\rm{D}}$ is the exponential damping time. Due to non-linear effects related to the initial large amplitudes of the perturbations, some of the oscillations would be better fitted by a Gaussian damping profile, as shown by \citet{2016A&A...595A..81M}. Nevertheless, we applied the exponential fitting function to every oscillation to be able to compare the results from the simulations with the solutions from the eigenvalue problem and with results from previous studies.

	The red asterisks on the top right, bottom left and bottom right panels of Fig. \ref{fig:fits} display the period, damping time and damping ratio, respectively, as functions of the longitudinal density ratio, $\chi$. We see in the top right panel that larger density ratios correspond to shorter periods, with the case of $\chi = 100$ having a period that is $\sim 0.55$ times the period of the uniform tube ($\chi = 1$). This is the same behaviour as the one presented in Fig. 3(a) of \citet{2015A&A...575A.123S}. The bottom left panel shows that the damping times also decrease as the density ratio is increased. Finally, the bottom right panel shows that the dependence of the ratio between the damping time and the period on the longitudinal density ratio is not so strong as for the previous parameters but a small increase is found as $\chi$ is increased.
	
	The above described behaviour of the damping time to period ratio is different from that of threads with no density variation along the axis. \citet{2002ApJ...577..475R}, \citet{2008ApJ...682L.141A} and \citet{2009ApJ...707..662S,2010ApJ...722.1778S} found that in a longitudinally homogeneous thread with a radial transition layer described by Eq. (\ref{eq:rho_layer}), the ratio of damping time to period of the kink oscillations in the thin tube and thin boundary regime is
	\begin{equation} \label{eq:damp_ratio}
		\frac{\tau_{\rm{D}}}{P} \approx \frac{2}{\pi}\frac{1}{l/R}\left(\frac{\rho_{\rm{i,0}}+\rho_{\rm{ex}}}{\rho_{\rm{i,0}}-\rho_{\rm{ex}}}\right).
	\end{equation}
	A similar result was obtained by \citet{2005A&A...430.1109A} and \citet{2005A&A...441..361A} in their studies of coronal loop oscillations although they considered the effect of longitudinal density stratification.	The reason for this different behaviour is that in the present work the density contrast between the internal and external layers varies with the parameter $\chi$, while in the above mentioned studies the density contrast was a constant \citep{2006A&A...457.1059D}.
	 
	The model we presented in Sect. \ref{sec:model} is more closely related to that used by \citet{2011A&A...533A..60A}, who represented a prominence thread as a tube divided in three longitudinally uniform regions with different values of density, connected by small transition layers. \citet{2011A&A...533A..60A} found that the damping ratio slightly increases for very short threads. In our model, short threads are represented by large values of $\chi$.
	
	Using Eq. (\ref{eq:damp_ratio}) as a reference, the slight increase of $\tau_{\rm{D}} /P$ with $\chi$ can be roughly understood as follows. As shown by Eq. (\ref{eq:rho_av}), a larger $\chi$ corresponds to a smaller average density. If we use the variable $\langle \rho_{\rm{i}} \rangle$ instead of $\rho_{\rm{i,0}}$ in Eq. (\ref{eq:damp_ratio}), we find that $\tau_{\rm{D}} / P$ increases as the average density decreases.
	
	The solid lines in the top right and the bottom panels of Fig. \ref{fig:fits} show the results from the 2D eigenvalue problem. There is a remarkable agreement with the results from simulations for the case of the oscillation periods. We see some differences in the damping times, with the simulations yielding smaller values, which means that the oscillations are attenuated slightly faster. As mentioned in Sect. \ref{sec:ref_sims}, this could be caused by the non-linear generation of fluting modes \citep{2010PhPl...17h2108R} and the triggering of the KHI \citet{2018ApJ...853...35T,2021ApJ...910...58V}, which grows by extracting energy from the flow. The numerical diffusivity used in the code to prevent the appearance of numerical noise also adds another damping mechanism. None of these processes (either physical or numerical) are taken into account by the 2D eigenvalue problem.
	
\subsubsection{First longitudinal harmonic} \label{sec:1harmonic}
	In this section we use the same amplitude of the velocity perturbation as in the previous one but here the perturbation corresponds to the oscillation mode $n = 2$, that is, the first longitudinal harmonic. 
	
	This oscillation mode presents a node at the middle position of the tube, so its centre of mass remains static and cannot be used to analyse the displacement of the tube. For this mode, the maximum displacements occur at $z = -L_{z}/4$ and $z = L_{z}/4$. We have chosen the former position for the present analysis. We have computed the centre of mass contained in the horizontal plane at that position and applied Eq. (\ref{eq:fit_exp}) to fit its displacement. The results from this fitting procedure are represented by blue diamonds on the top right and bottom panels of Fig. \ref{fig:fits}.
	
	As for the fundamental mode, both the oscillation periods and the damping times reduce as the longitudinal density ratio is increased. However, in this case the dependence on that parameter is even more pronounced than for the case of the fundamental mode. This behaviour qualitatively agrees with the results of \citet{2005ApJ...624L..57A}, who found that the first longitudinal overtone is more affected by the longitudinal variations of density than the fundamental mode. Then, Fig. \ref{fig:fits} shows that the period when $\chi = 100$ is $\sim 0.3$ times the oscillation period of the tube with uniform density, and the damping time is reduced to less than a half. This means that the inhomogeneity in density has a larger effect on the periods than on the damping times. In addition, the ratio of damping time to period has a larger increase with $\chi$ than for the case with $n = 1$. This is related to the large decrease of density contrast between the internal and external plasmas at $z = -L_{z} / 4$  as the parameter $\chi$ increases.

\subsubsection{$P_{0} / P_{1}$ ratio} \label{sec:period}
	\begin{figure*}
		\centering
		\includegraphics[width=17cm,height=6cm]{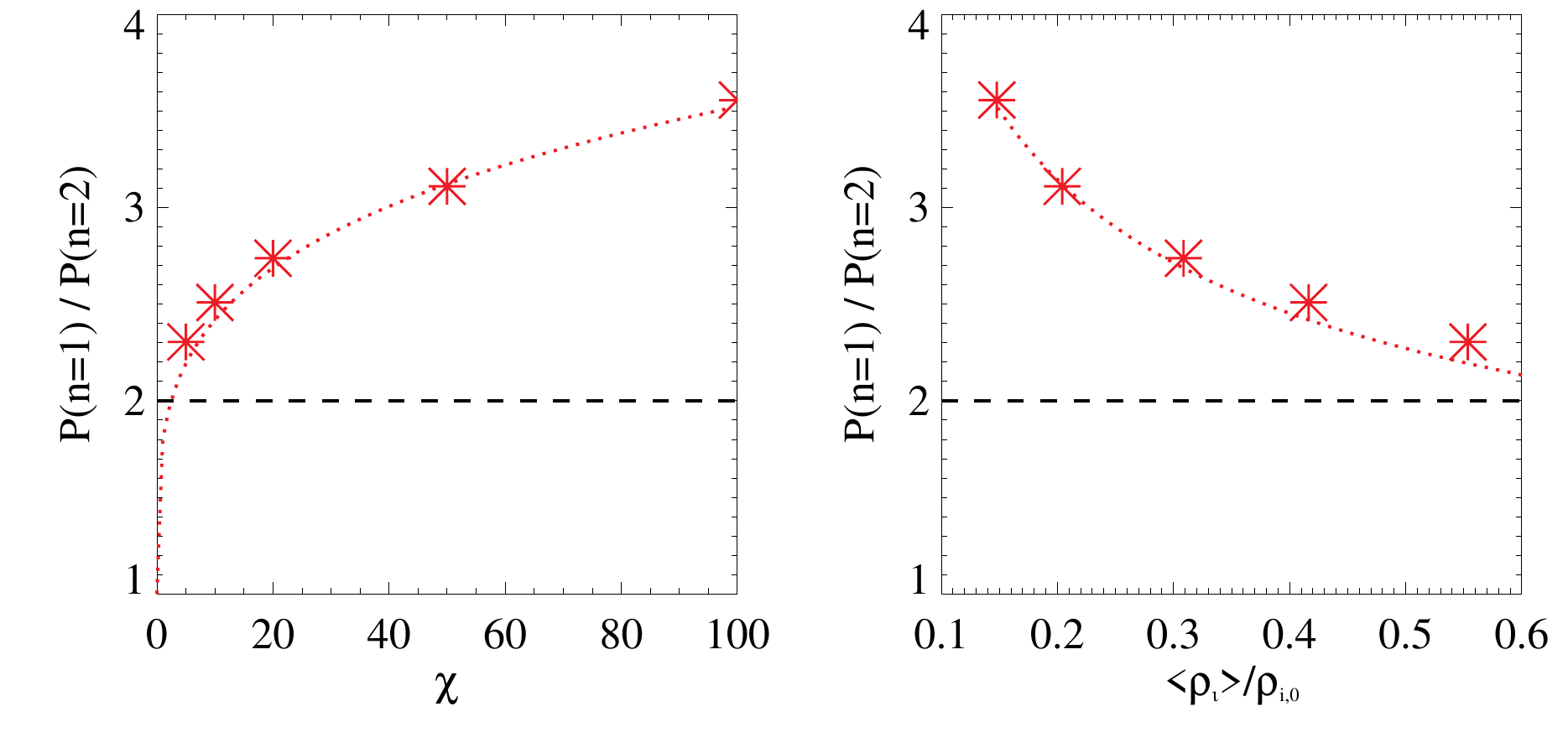}
		\caption{Left: period ratio as a function of density ratio. Right: period ratio as a function of the average density of the
			tube. Dashed lines correspond to the case of tubes with uniform density. The dotted lines represent the approximation
			given by Eq. (\ref{eq:approx}).}
		\label{fig:ratio}
	\end{figure*}

	\begin{figure*}
		\centering
		\includegraphics[width=17cm,height=6cm]{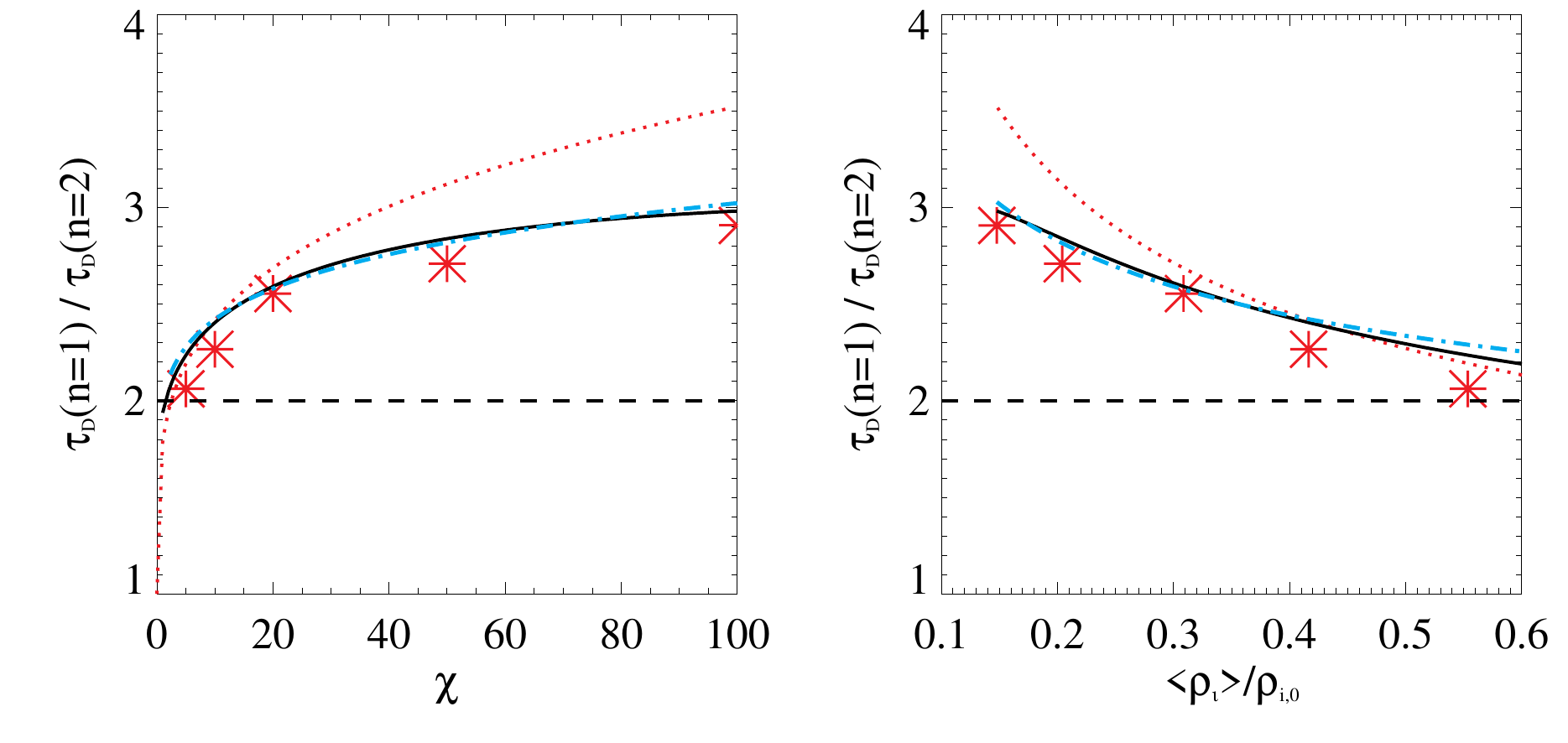}
		\caption{Damping time ratio as a function of density ratio (left) and as a function of the average density of the tube (right). Red asterisks represent the results from simulations. Solid black lines correspond to the results from the eigenvalue problem. Horizontal dashed lines represent the case of tubes with uniform density. The red dotted lines represent the approximations given by Eqs. (\ref{eq:approx}) and (\ref{eq:approx_2}). The blue dotted-dashed lines on the left and right panels represent the fits given by Eqs. (\ref{eq:damp_fit_chi}) and (\ref{eq:damp_fit_rhoav}), respectively.}
		\label{fig:ratio_damp}
	\end{figure*}

	Now, we perform a comparison of the periods of the two oscillation modes previously analysed. The results are shown in Fig. \ref{fig:ratio}. For the case of a tube with uniform density this ratio is $P_{0} / P_{1} = 2$ in the thin tube approximation \citep{1983SoPh...88..179E}. However, the value of the ratio varies when longitudinal inhomogeneities in density are taken into account. For the case of a prominence thread, the ratio is always larger than $2$ \citep{2010ApJ...725.1742D}. The left panel of Fig. \ref{fig:ratio} shows that the period ratio grows as $\chi$ increases.
	
	Up to this point, we have focused on the dependence of the oscillation parameters on the ratio of the central density to the footpoint density. However, we note that each one of the profiles shown in Fig. \ref{fig:profiles} represents a thread with a different total mass, so the comparison between each other and with the uniform density case is not so straightforward. A more appropriate procedure would be to consider the dependence on the total mass of the tube or, equivalently, on its average density \citep[see][]{2005ApJ...624L..57A}.
	
	The dependence of the period ratio on the average density is shown on the right panel of Fig. \ref{fig:ratio}. For a longitudinally uniform thread, the period ratio is independent on the average density. For a tube with a Lorentzian profile it increases as the average density decreases. As shown by \citet{2015A&A...575A.123S} this behaviour is not exclusive of the Lorentzian profile but it is shared by other non-uniform density distributions.
	
	The analytical approximation displayed on Fig. \ref{fig:ratio} is given by
	\begin{equation} \label{eq:approx}
		\frac{P(n=1)}{P(n=2)} \approx 1 + \left(\frac{4}{\pi^{2}} \chi\right)^{1/4},
	\end{equation}
	which was derived by \citet{2015A&A...575A.123S} after obtaining an empirical fit as a function of the average density,
	\begin{equation} \label{eq:approx_2}
		\frac{P(n=1)}{P(n=2)} \approx 1 + \left(\frac{\langle \rho_{\rm{i}}\rangle}{\rho_{\rm{i,0}}}\right)^{-1/2},
	\end{equation}
	using the relation given by Eq. (\ref{eq:rho_av}) and applying the limit $\chi \gg 1$. An analytical proof of the validity of this formula was provided by \citet{2016SoPh..291.1143R}.
	
	We see that the numerical results have a very good agreement with the analytical approximation for large values of the density ratio (or small average densities), while small differences appear in the opposite range. Prominence threads are expected to have large density ratios, so the use of this approximate formula to perform prominence seismology is justified \citep[see, e.g.,][]{2015A&A...578A.130A}.

\subsubsection{$\tau_{\rm{D,0}}/\tau_{\rm{D,1}}$ ratio} \label{sec:damping}

	According to Eq. (\ref{eq:damp_ratio}), the damping time to period ratio in a longitudinally homogeneous tube is independent from the oscillation mode $n$ as long as the longitudinal wavelength remains much longer than the radius of the tube, that is, in the thin tube limit. Therefore, the ratio between the damping times of the fundamental mode and the first harmonic is the same as the ratio between the periods, that is equal to $2$.
	
	However, it has been shown that when the tube has a Lorentzian profile in density, the period ratio approximately follows the relations given by Eqs. (\ref{eq:approx}) and (\ref{eq:approx_2}). We expect that these formulas are not applicable to the ratio of damping times, since the attenuation process is mainly affected by the transverse variation of density \citep[see e.g.,][]{2002ApJ...577..475R,2008ApJ...682L.141A,2009ApJ...707..662S} while the oscillation period is weakly affected by this variation but strongly depends on the longitudinal one \citep{2005A&A...430.1109A,2010ApJ...725.1742D,2015A&A...575A.123S}.
	
	Figure \ref{fig:ratio_damp} shows the damping times ratio as a function of $\chi$ (left) and on the average density (right). In addition to the results from the numerical simulations (red asterisks), this plot includes the solutions of the 2D eigenvalue problem (black solid line). We see that the damping ratio rises as the longitudinal density ratio is increased (or the average density is reduced) but not as fast as the period ratio (red dotted lines).
	
	With the goal of finding expressions similar to Eqs. (\ref{eq:approx}) and (\ref{eq:approx_2}) for the damping ratio, we fitted the solutions of the eigenvalue problem displayed in Fig. \ref{fig:ratio_damp}. We used the fitting function $f(X) = 1 + a_{1} X^{a_{2}}$, where $f(X)$ is the ratio of the damping times, $X$ is the independent variable ($\chi$ or the ratio $\langle \rho_{\rm{i}}\rangle/ \rho_{\rm{i,0}}$), and $a_{1}$ and $a_{2}$ are the coefficients of the fit. We got the following approximations:
	\begin{equation} \label{eq:damp_fit_chi}
		\frac{\tau_{\rm{D}}(n = 1)}{\tau_{\rm{D}}(n = 2)} \approx 1 + \chi^{0.15}
	\end{equation} 
	and
	\begin{equation} \label{eq:damp_fit_rhoav}
		\frac{\tau_{\rm{D}}(n = 1)}{\tau_{\rm{D}}(n = 2)} \approx 1 + 1.05 \left(\frac{\langle \rho_{\rm{i}}\rangle}{\rho_{\rm{i,0}}}\right)^{-0.34},
	\end{equation}
	which are shown in Fig. \ref{fig:ratio_damp} as light blue dotted-dashed lines. We see that Eqs. (\ref{eq:damp_fit_chi}) and (\ref{eq:damp_fit_rhoav}) yield a better fit of the numerical results for large values of the longitudinal density ratio or for small values of the average density, which are the expected ranges of values of prominence threads. 
	
	Taking into account the results displayed in Fig. \ref{fig:ratio_damp}, the ratio $\tau_{\rm{D,0}} / \tau_{\rm{D,1}}$ could be used as another seismology tool to compute the parameters $\chi$ or $\langle \rho_{\rm{i}}\rangle/ \rho_{\rm{i,0}}$ of threads and provide results which are independent from the ones obtained through the use of the period ratio. Nevertheless, the actual usefulness of this ratio and the approximations given by Eqs. (\ref{eq:damp_fit_chi}) and (\ref{eq:damp_fit_rhoav}) should be explored in a future work. \citet{2015A&A...575A.123S} and \citet{2016SoPh..291.1143R} demonstrated that the fundamental variable that determines the value of $P_{0} / P_{1}$ is the ratio $\langle \rho_{\rm{i}}\rangle/ \rho_{\rm{i,0}}$, with no dependence from the specific longitudinal density profile of the thread. A similar general behaviour has not yet been established for the ratio $\tau_{\rm{D,0}} / \tau_{\rm{D,1}}$, which might depend on other additional factors that would constrain the range of applicability of the approximations presented here.
		
\subsubsection{Amplitude of the oscillations} \label{sec:amplitudes}
	\begin{figure*}
		\centering
		\includegraphics[width=17cm,height=6cm]{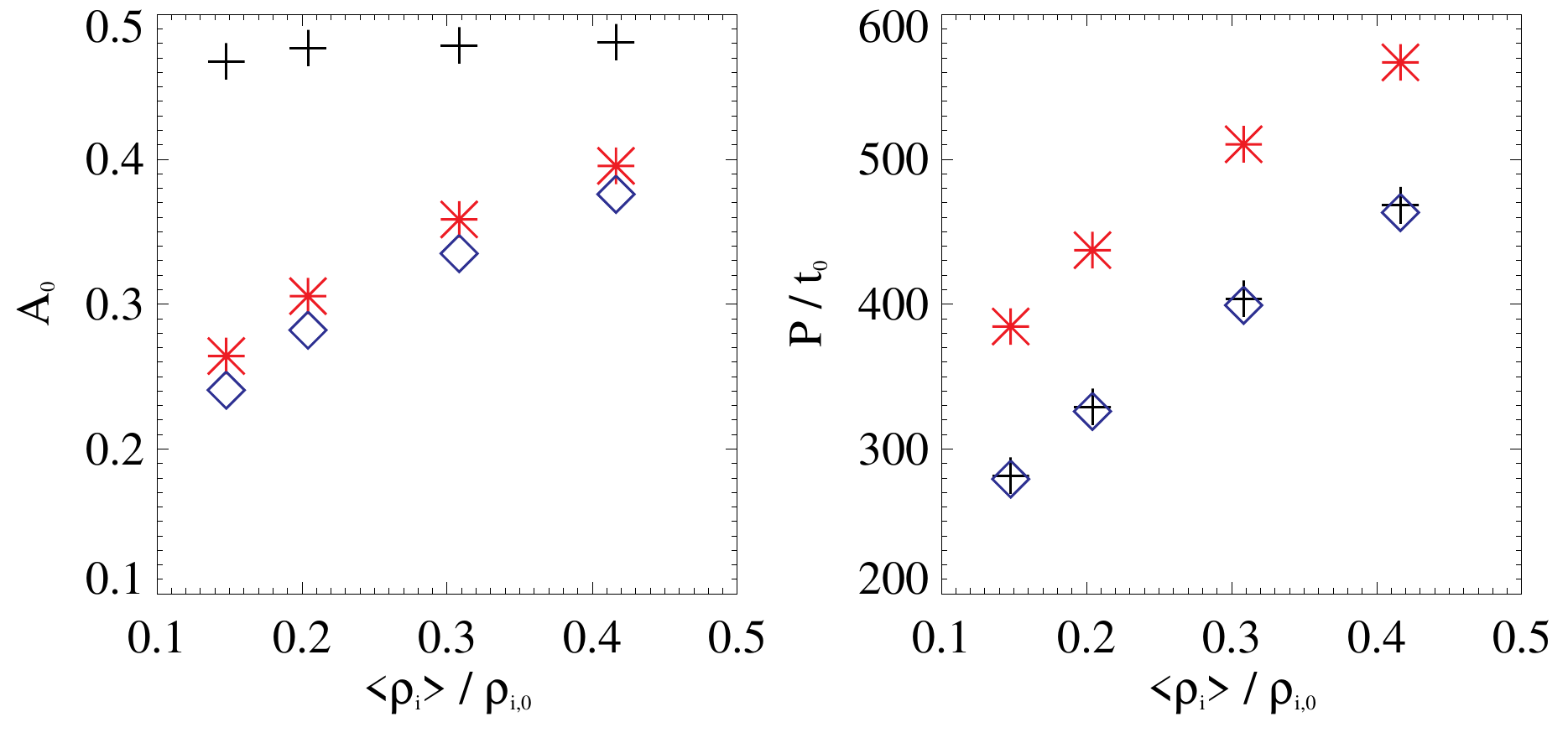}
		\caption{Displacement (left) and oscillation period (right) as functions of average density. Black crosses correspond to simulations of tubes with uniform density and an amplitude of the initial perturbation given by $V_{0} = 0.05 c_{\rm{A}}$. Red symbols represent the results with a Lorentzian profile and $V_{0} = 0.05 c_{\rm{A}}$. Blue diamonds represent simulations of tubes with uniform density but an initial kinetic energy that is the same as the cases with the Lorentzian profile.}
		\label{fig:disp1}
	\end{figure*}
	
	\begin{figure*}
		\centering
		\includegraphics[width=17cm,height=6cm]{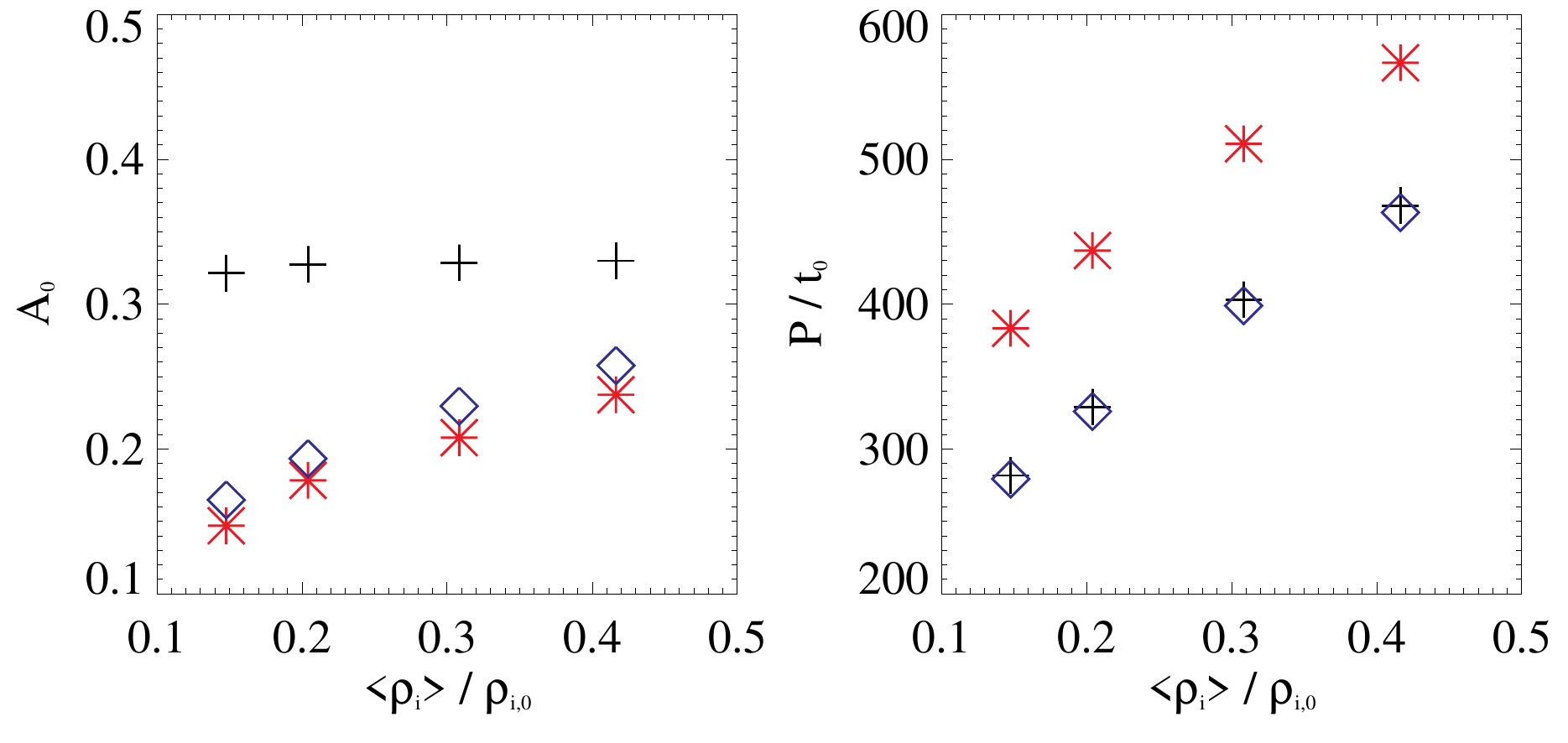}
		\caption{Same as Fig. \ref{fig:disp1} but for $z = -L_{z}/4$.}
		\label{fig:disp2}
	\end{figure*}

	Figure \ref{fig:disp1} shows how the average density of the tube affects the amplitudes of the transverse oscillations of the threads (left panel) and their periods (right panel).
	
	In the first place, we compare two series of simulations in which the amplitude of the initial perturbation is $V_{0} = 0.05 c_{\rm{A,in}}$ and the fundamental mode of oscillation, $n=1$, is considered: black crosses correspond to uniform tubes with the same density as the average density of the Lorentzian profiles, while red asterisks show the results for tubes with Lorentzian profiles. For a given average density, a tube with uniform density oscillates with a larger amplitude and shorter period (as already shown by Fig. \ref{fig:ratio}). In addition, we see that for the case of the Lorentzian profile the amplitude of the oscillation increases with the average density while for a uniform thread it remains almost constant.
	
	These results for the longitudinally homogeneous tube are in good agreement with the linear analysis for the transverse kink mode in the thin tube approximation performed by \citet{2014SoPh..289.1999R} and \citet{2018ApJ...853...35T}, who showed that the displacement of the tube is given by	
	\begin{equation} \label{eq:displ}
		A_{0}=\frac{L_{z}V_{0}}{\pi R c_{\rm{A,in}}}\sqrt{\frac{1+\rho_{\rm{ex}}/\rho_{\rm{i,0}}}{2}} = \frac{L_{z}\alpha}{\pi R}\sqrt{\frac{1+\rho_{\rm{ex}}/\rho_{\rm{i}}}{2}}.
	\end{equation}
	In the above formula we used the fact that in our simulations the amplitude of the initial perturbation is given by $V_{0} = \alpha c_{\rm{A,in}}$. Therefore, the displacement is almost a constant since $\rho_{\rm{i,0}} \gg \rho_{\rm{ex}}$ and the rest of parameters have been kept fixed for these series of simulations. The physical reason behind this behaviour is that all the simulations of this series start with the same total kinetic energy, independently from the average density of the tube. 
	
	We note that in a strict sense, Eq. (\ref{eq:displ}) is only applicable to the linear regime where $A_{0} \ll 1$, while our simulations present much larger values of the displacement. However, it seems that the predicted trend also appears in the non-linear regime (at least for the set of parameters explored in the present work).
	
	The behaviour of the simulations with a Lorentzian profile can be explained as follows. Although the value of $\chi$ varies throughout the series, the density at the central part of the tube and, consequently, the internal Alfvén speed remain fixed, which implies that the amplitude of the initial velocity perturbation is also fixed. Therefore, as the average density of the tube is increased (or $\chi$ is decreased) the total kinetic energy rises, which produces oscillations of larger amplitude.
	
	In addition, for a given value of the averaged density, the internal Alfvén speed of the tube with a uniform density is larger than the one with a Lorentzian profile, since $\rho_{\rm{i,0}}$ is smaller in the former case. This is the reason why all the simulations with uniform density show larger displacements.
	
	Next, we compare simulations with different density profiles but with the same initial kinetic energy. For that, we perform a new series of simulations of threads with uniform density but varying the amplitude of the initial perturbation, $V_{0}$, so their total kinetic energy matches that of the threads with a Lorentzian profile. The results of this new series are shown in Fig. \ref{fig:disp1} as blue diamonds. Surprisingly, we find on the left panel that the results of both series still differ, with the amplitudes for the cases with Lorentzian profiles being slightly larger.
	
	An explanation for the mentioned discrepancy could be that, although the total kinetic energy is the same, a larger fraction of it is present in the central region of the tube in the case of the Lorentzian profile in comparison with the case of the uniform profile. To check the validity of this explanation, we have computed the amplitude of the oscillation at a different position of the tube, namely $z = -0.25 L_{z}$. At that position, for a given value of $\langle \rho_{\rm{i}}\rangle / \rho_{\rm{i,0}}$ the uniform profile has a larger density (and larger kinetic energy) than the Lorentzian profile. Therefore, smaller amplitudes would be expected for the latter. The results of these computations are shown in Fig. \ref{fig:disp2}. It can be seen that the amplitudes for the cases with a uniform density are slightly larger, which agrees with the explanation provided.
	
	Finally, the right panels of Figs. \ref{fig:disp1} and \ref{fig:disp2} show that the results of the two series of simulations of threads with homogeneous density overlap. Thus, we find that in these simulations the periods of the oscillations are not affected by the amplitude of the initial perturbation, in agreement with the behaviour predicted by Eq. (\ref{eq:period}). However, we recall this formula is only strictly applicable to the linear regime, that is, for oscillations with amplitudes $A_{0} \ll 1$. Figures \ref{fig:disp1} and \ref{fig:disp2} show that the simulated oscillations have values of $A_{0}$ that go from $\sim 0.15$ to $\sim 0.5$, which means that they are out of the range of applicability of Eq. (\ref{eq:period}). According to \citet{2014SoPh..289.1999R}, non-linearity produces an increase of the oscillation frequency proportional to the square of the amplitude $A_{0}$. More precisely, the relation between the frequencies in the non-linear and linear regimes is given by $\omega_{\rm{nonl}} = \omega_{\rm{lin}} \left(1 + A_{0}^{2} \Upsilon \right)$, where $\Upsilon$ is a small parameter that depends on the ratio $\rho_{\rm{i}} / \rho_{\rm{ex}}$ and the geometry of the flux tube. Therefore, smaller periods are obtained as the amplitude is increased. Nevertheless, \citet{2014SoPh..289.1999R} also found that the non-linear frequency shift is usually small and it would become noticeable after a large number of periods. Hence, even larger amplitudes than the ones used here would be required to obtain a strong variation in the oscillation periods.

\section{Forward modelling of inhomogeneous flux tubes} \label{sec:fomo}
	\begin{figure*}
		\centering
		\includegraphics[width=4cm,height=4cm]{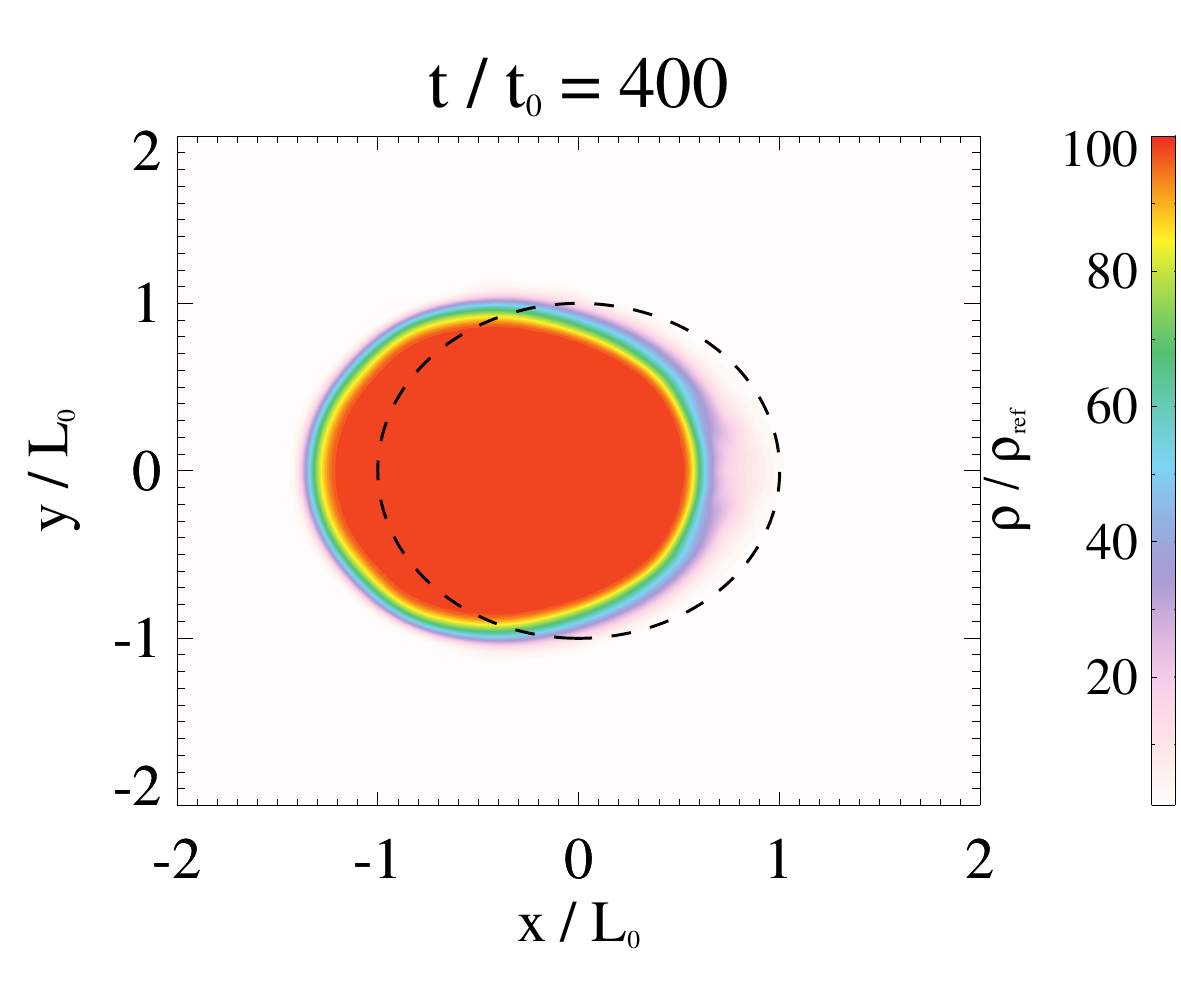}
		\includegraphics[width=4cm,height=4cm]{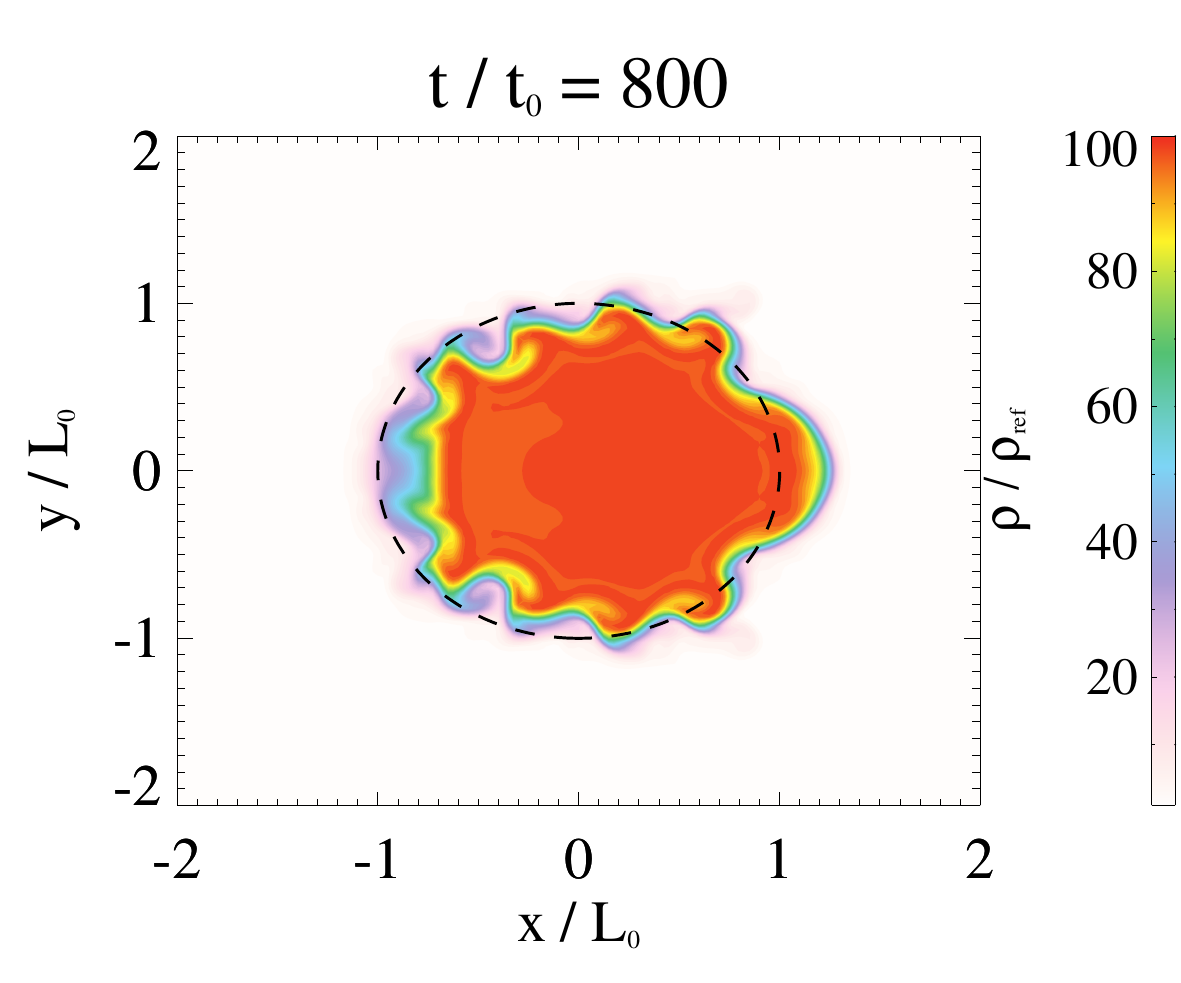}
		\includegraphics[width=4cm,height=4cm]{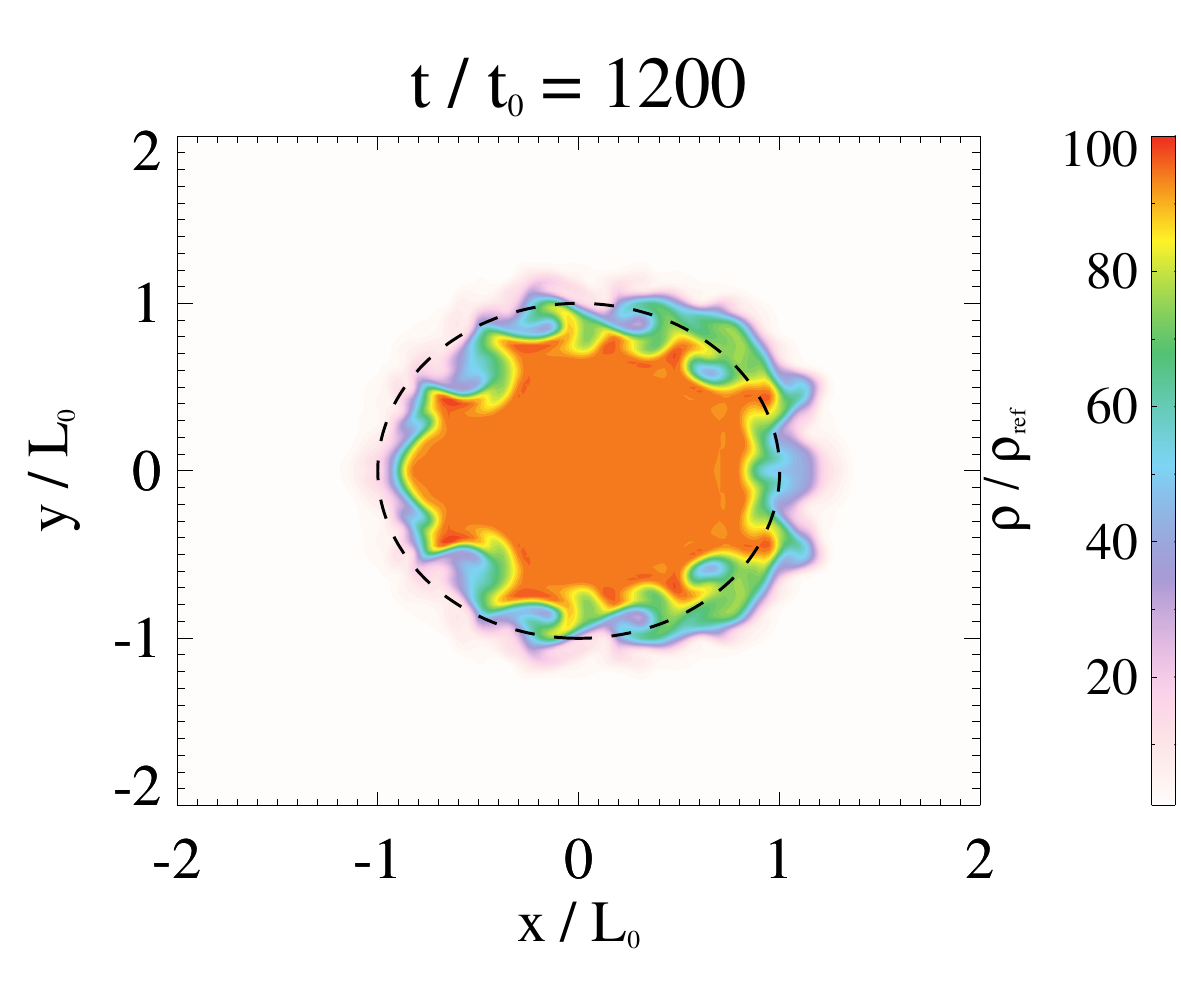}
		\includegraphics[width=4cm,height=4cm]{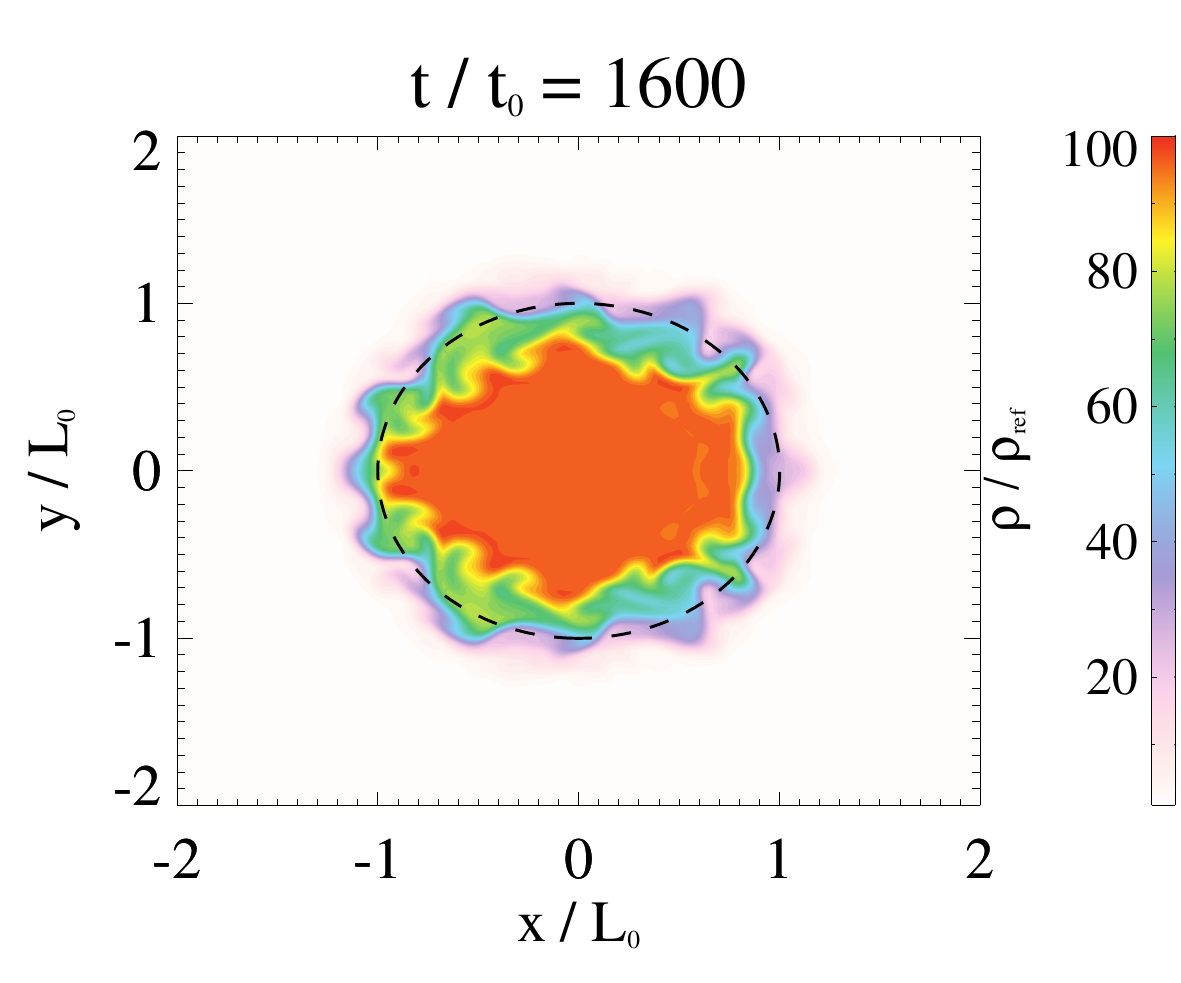} \\
		\includegraphics[width=4cm,height=4cm]{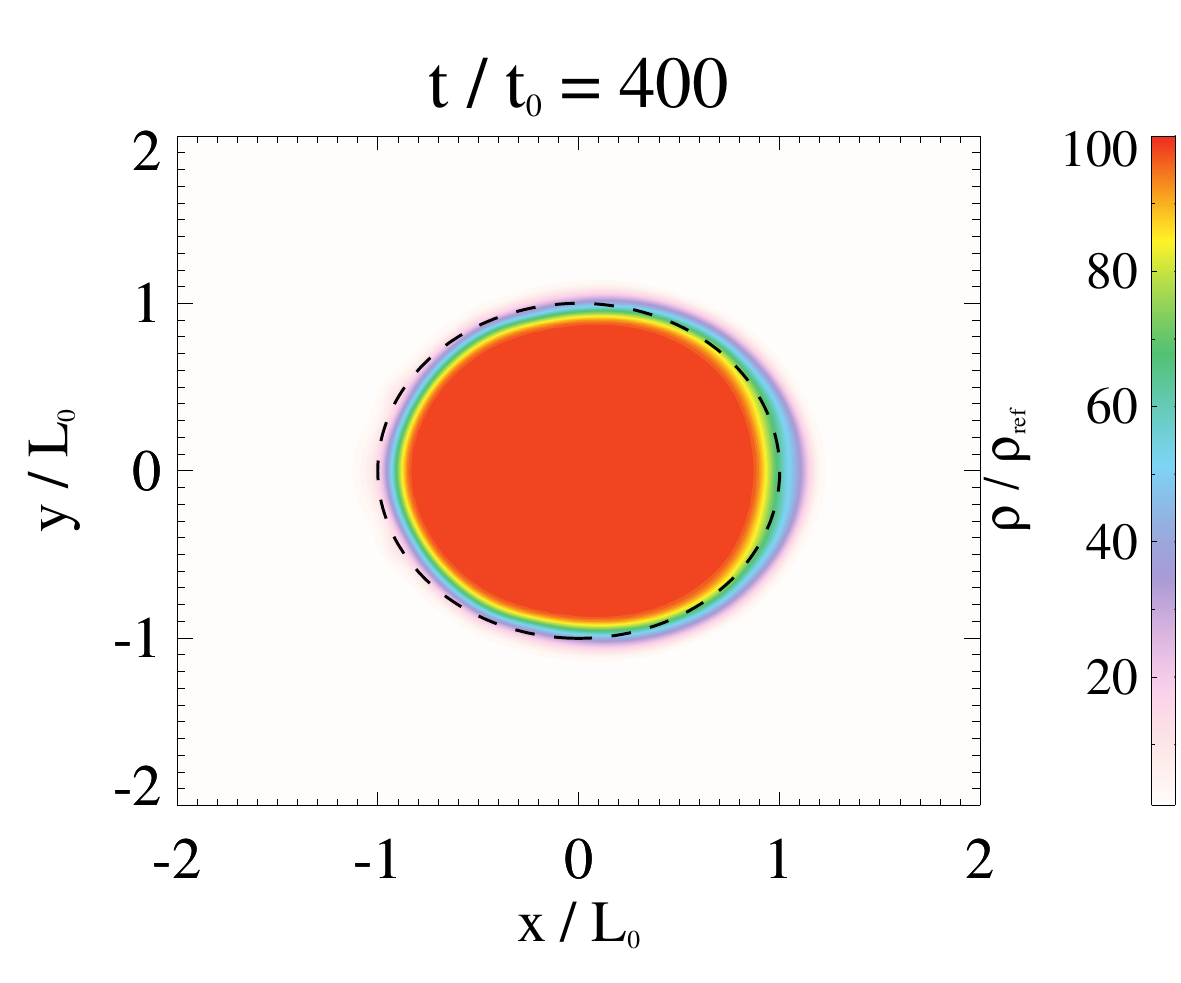}
		\includegraphics[width=4cm,height=4cm]{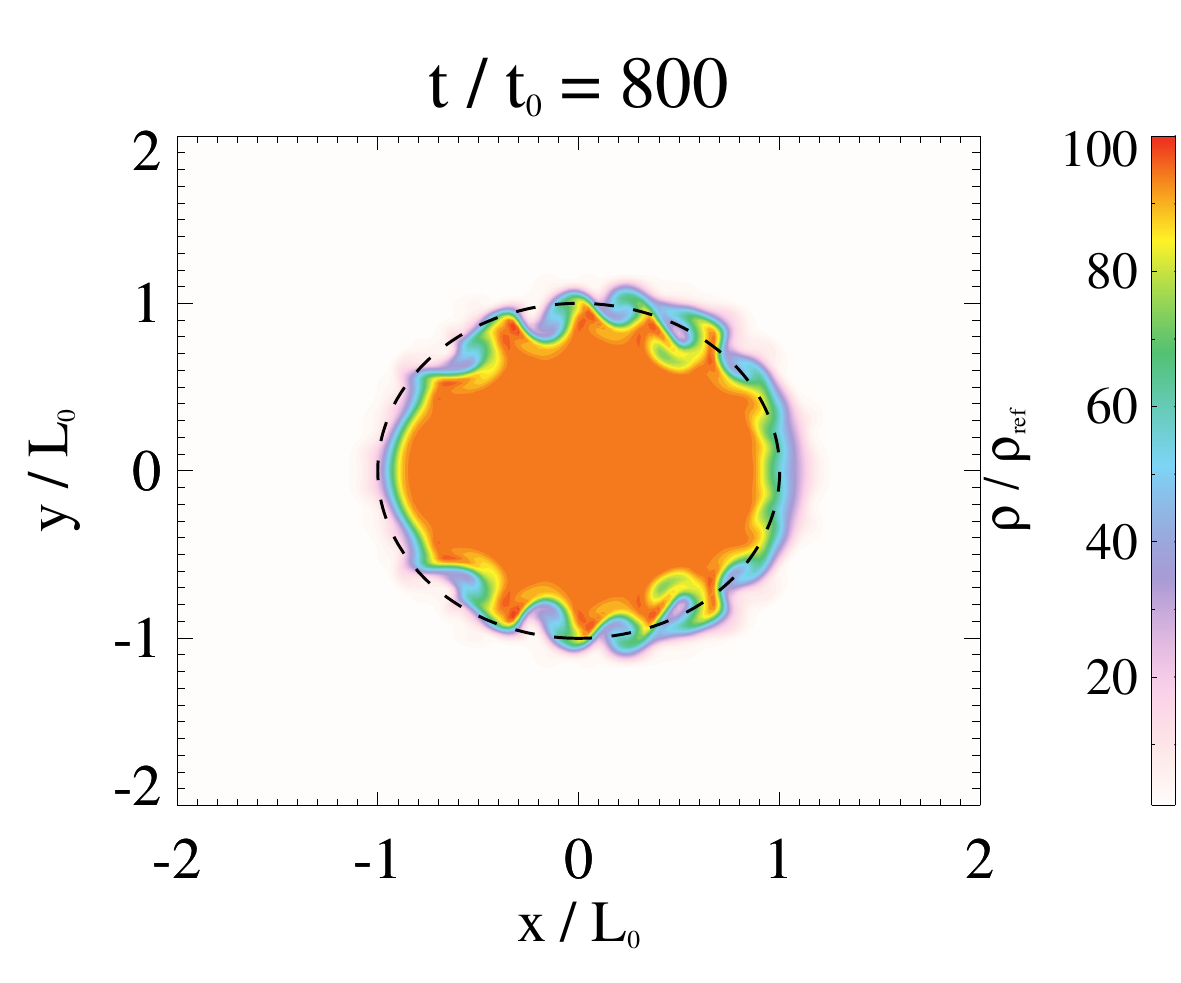}
		\includegraphics[width=4cm,height=4cm]{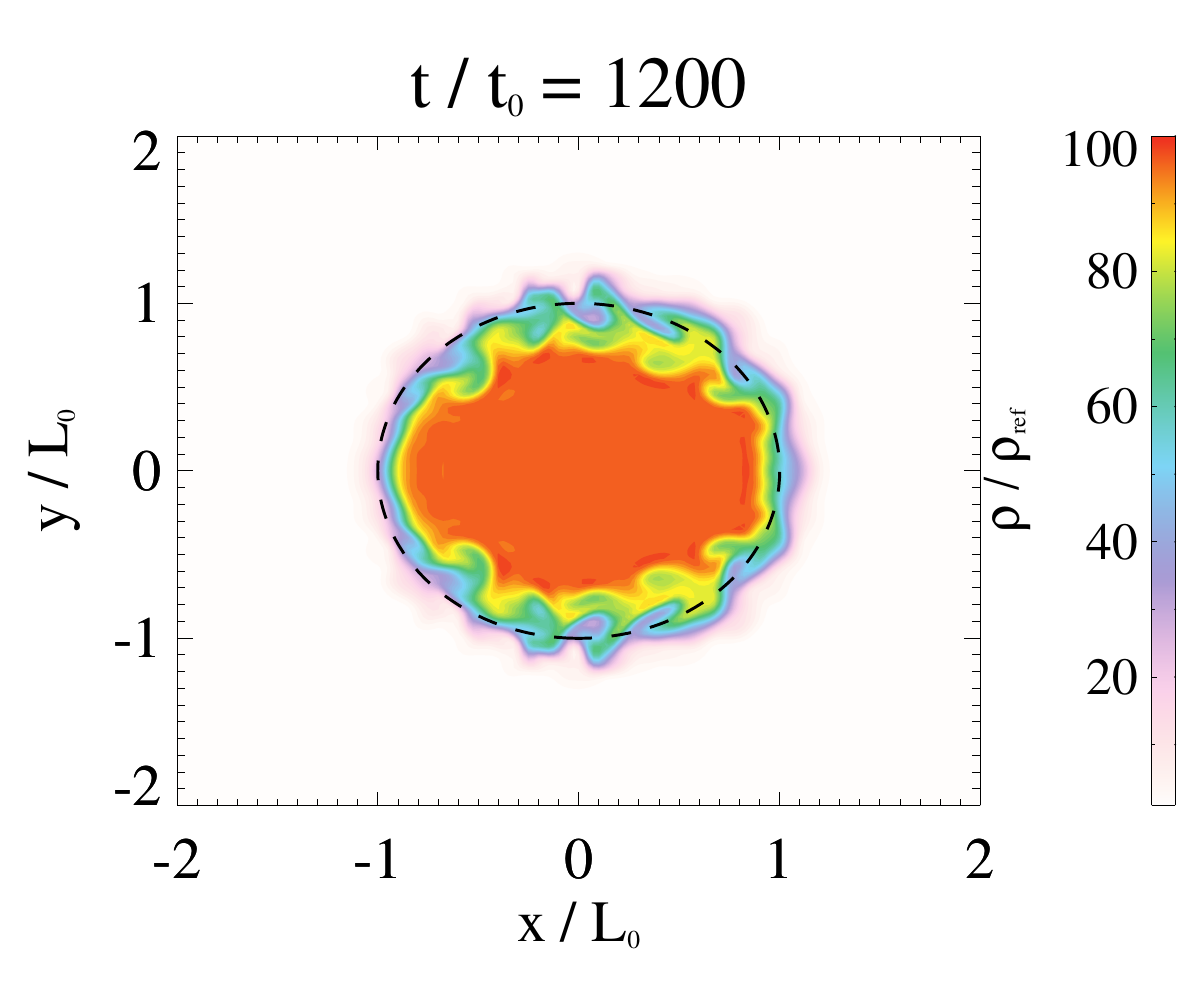}
		\includegraphics[width=4cm,height=4cm]{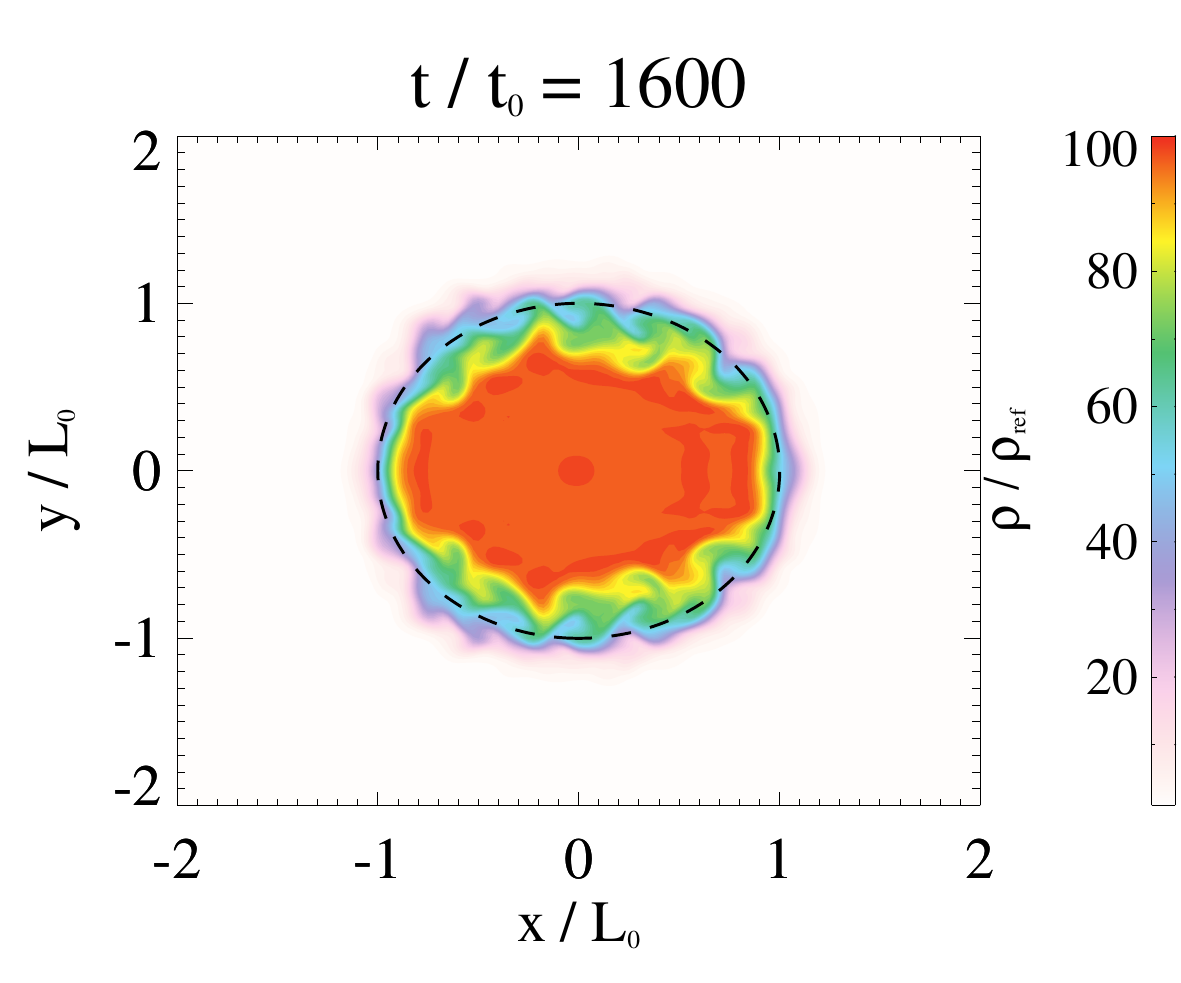}
		\caption{Density colour maps at $z=0$ from simulations with $V_{0} = 0.05 c_{\rm{A}}$, $n = 1$, and a numerical domain of $400 \times 400 \times 100$ points. Top panel: case with $\chi = 10$; bottom panels: case with $\chi = 100$. Dashed lines represent the density contour $\rho = 50 \rho_{\rm{ref}}$ at $t=0$.}
		\label{fig:khi_dens}
	\end{figure*}
	
	We perform forward modelling of our simulations to compute the synthetic intensity of the H$\alpha$ line emerging from the oscillating thread. We chose the H$\alpha$ spectral line for the present study because it is associated with the neutral component of the plasma (and, consequently, related to low temperatures) and it is typically used for imaging of the cool prominence fine structures \citep{2011SSRv..158..237L}.
	
	To obtain the synthetic emission of the H$\alpha$ line, we implemented the approximate method developed by \citet{2015A&A...579A..16H}. We used the reference values for a prominence at a height of $10 000$ km provided by Table 1 of \citet{2015A&A...579A..16H}, so the plasma in the interior of the thread has a temperature of $T = 10 000 \ \rm{K}$, a pressure of $P = 0.005 \ \rm{Pa}$ and an ionisation degree of $i = 0.7$. The lighter external plasma is fully ionised, $i = 1$, and initially has the same pressure as the internal plasma in order to fulfill the condition of mechanical equilibrium: since the internal and external densities follow the relation $\rho_{\rm{i,0}} = 100 \rho_{\rm{ex}}$, the initial temperature of the external medium is $T \approx 8.5 \times 10^{5} \ \rm{K}$. As time advances, at every step of the simulation the ionisation degree of each point of the numerical domain is computed by performing a bilinear interpolation from the data included in Table 1 of \citet{2015A&A...579A..16H}.
	
	In this section we increase the resolution of our simulations and use a numerical domain of $400 \times 400 \times 100$ points instead of the $200 \times 200 \times 100$ used for the previous section. Increasing the numerical resolution provides a more accurate description of the non-linear dynamics of the flux tube. It allows to better resolve smaller scales that were not captured by the low resolution simulations, as we show in more detail in the paragraphs below, and obtain better predictions of the results that can be expected from observations.
	
	Before presenting the H$\alpha$ synthetic profiles, for reference purposes we show in Fig. \ref{fig:khi_dens} density snapshots taken from simulations with $V_{0} = 0.05 c_{\rm{A}}$ and $n = 1$. The snapshots correspond to horizontal cuts at $z = 0$, where the displacement is maximum. The top row shows the results for the simulation with $\chi = 10$, while the bottom row corresponds to $\chi = 100$. We see that up to the time $t / t_{0} = 400$ the tube oscillates keeping its cylindrical shape almost unaltered. On the contrary, at later times its boundary is strongly distorted as the KHI vortexes develop. For detailed investigations on the onset and evolution of the KHI in magnetic flux tubes we refer the reader to the works of, e.g., \citet{2008ApJ...687L.115T}, \citet{2014ApJ...787L..22A}, \citet{2015A&A...582A.117M}, \citet{2016A&A...595A..81M}, \citet{2017ApJ...836..219A}, \citet{2018ApJ...853...35T} or \citet{2021A&A...648A..22D}.
	
	\begin{figure*}
		\centering
		\includegraphics[width=0.495\hsize]{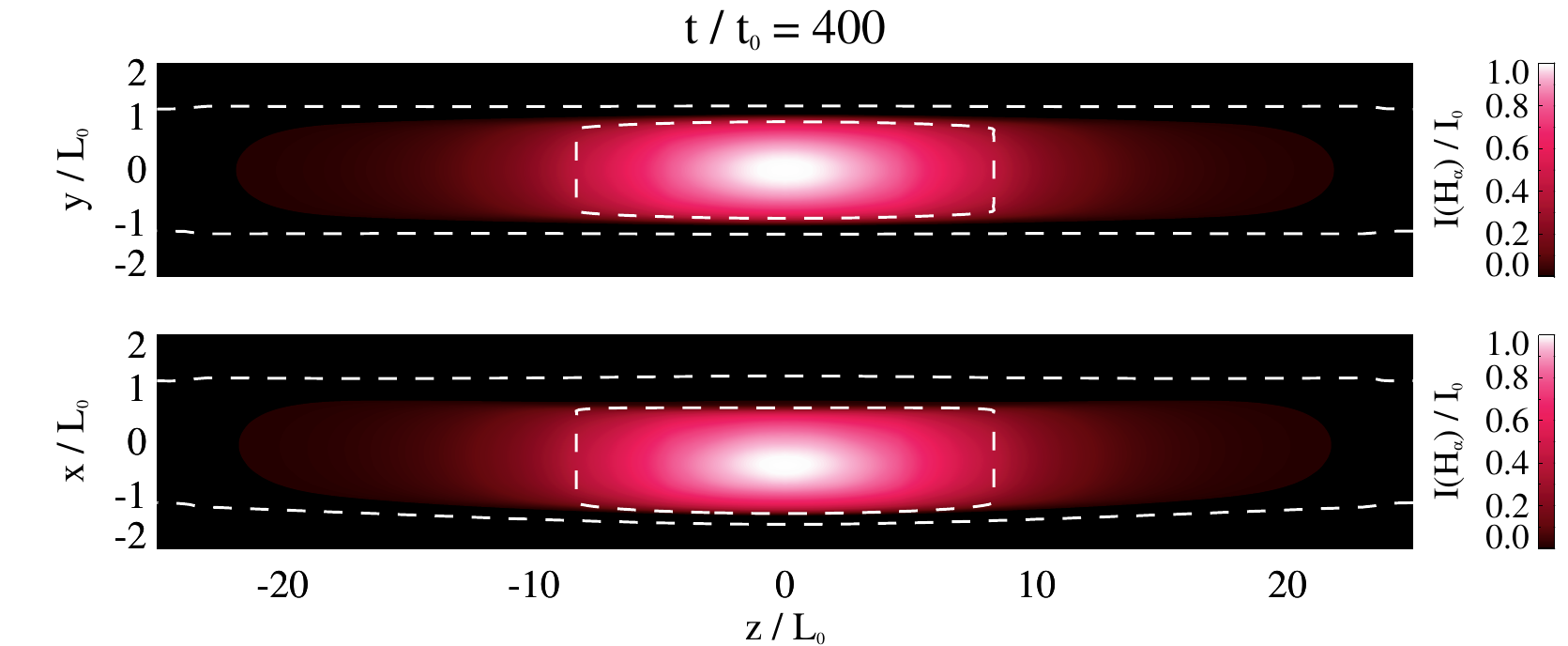}
		\includegraphics[width=0.495\hsize]{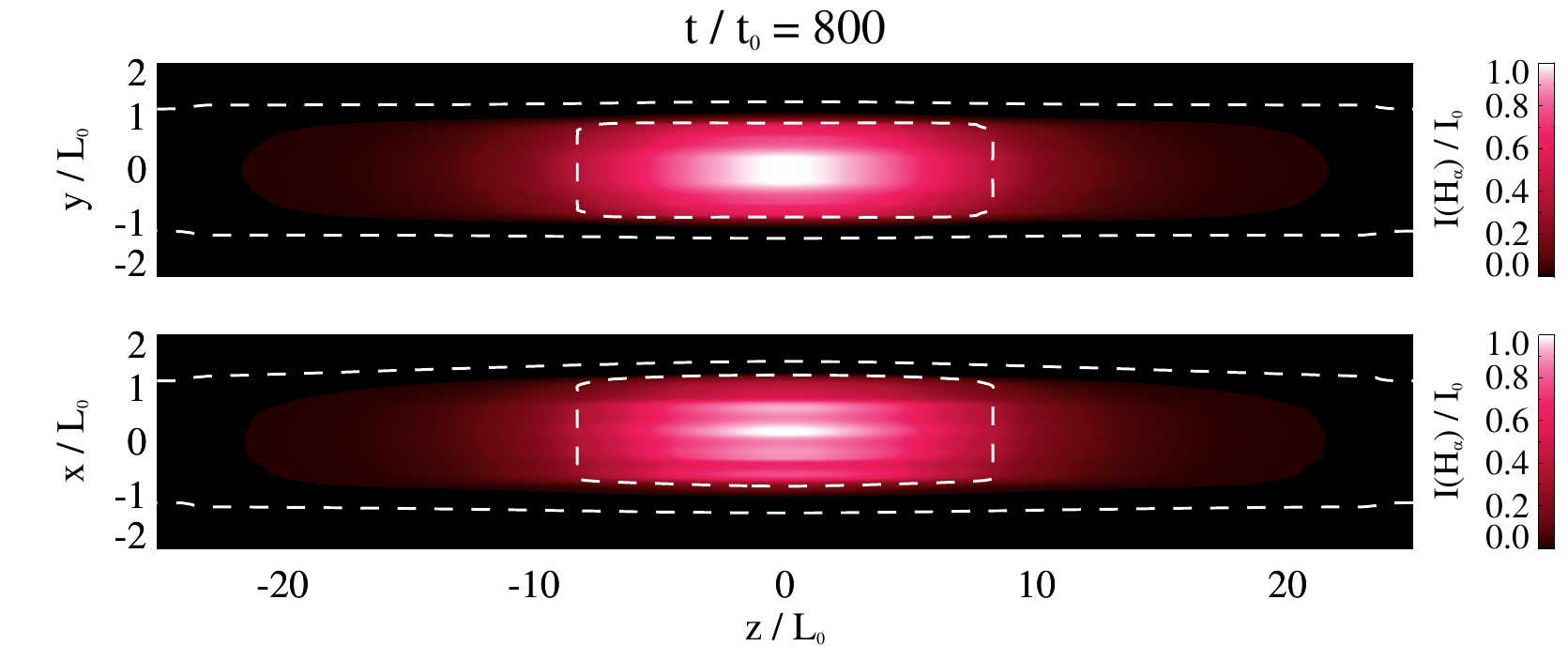} \\
		\includegraphics[width=0.495\hsize]{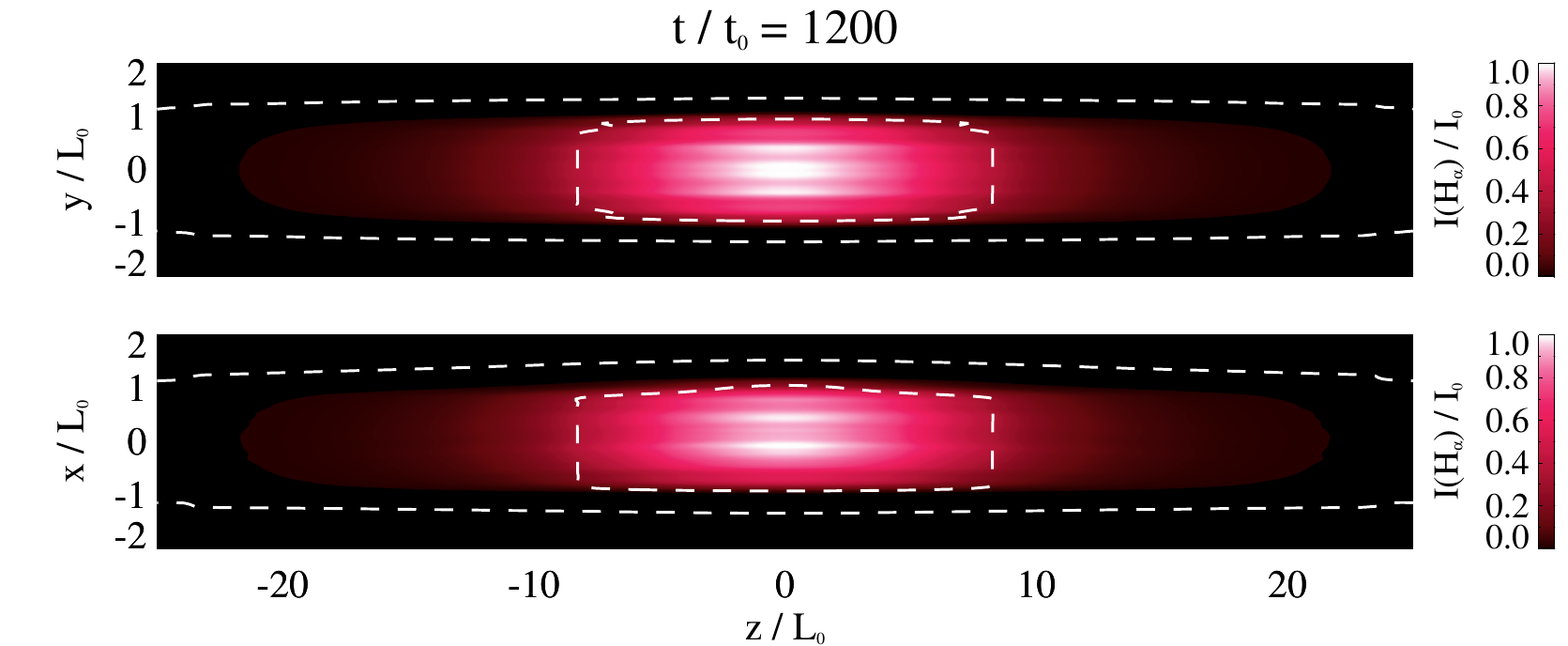}
		\includegraphics[width=0.495\hsize]{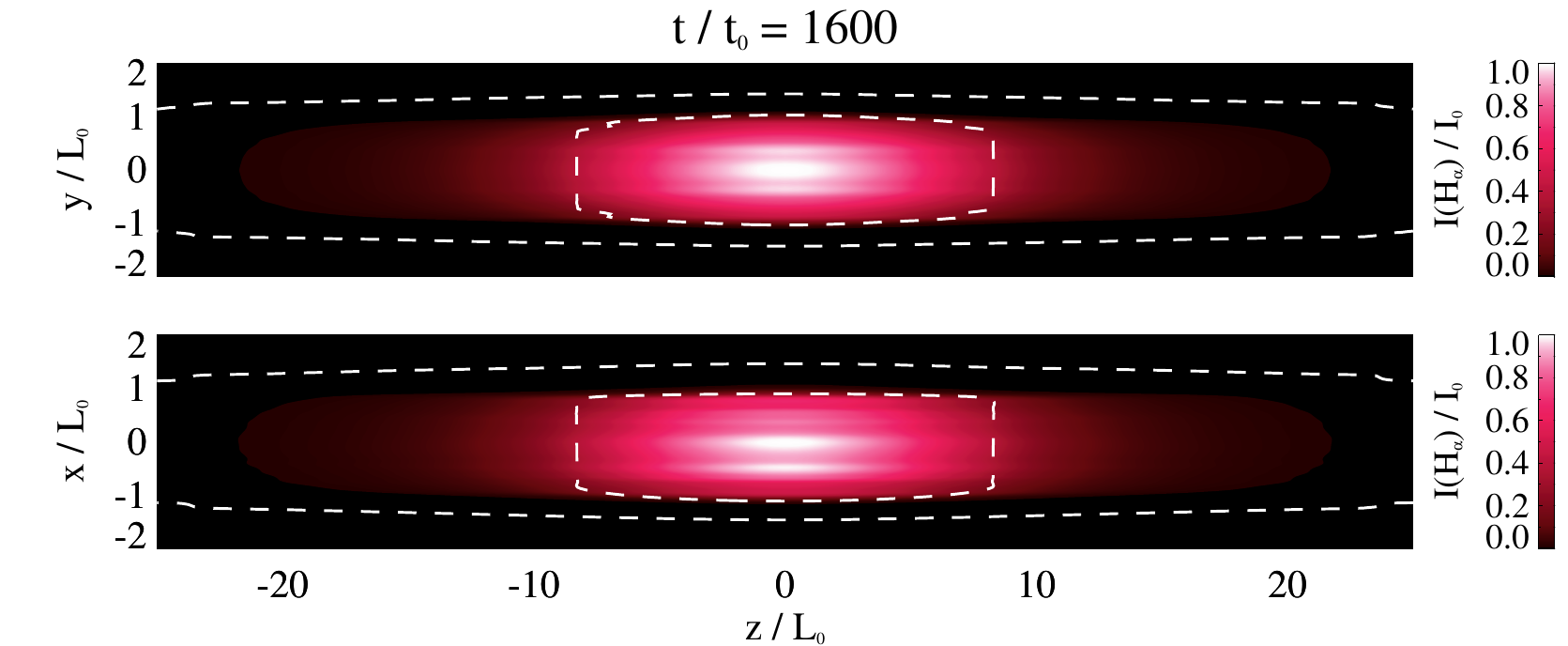}
		\caption{Intensity in the H$\alpha$ line at different times of the oscillation for the case with $\chi = 10$. In each snapshot the top panel corresponds to the $x_{\rm{LOS}}$, while the bottom panel corresponds to the $y_{\rm{LOS}}$. White dashed lines show density contours for $\rho= 1.05 \rho_{\rm{ex}}$ and $\rho = 50 \rho_{\rm{ex}}$. An animation of this figure is available online.}
		\label{fig:fomo_tube_dr10}
	\end{figure*}
	
	\begin{figure*}
		\centering
		\includegraphics[width=0.495\hsize]{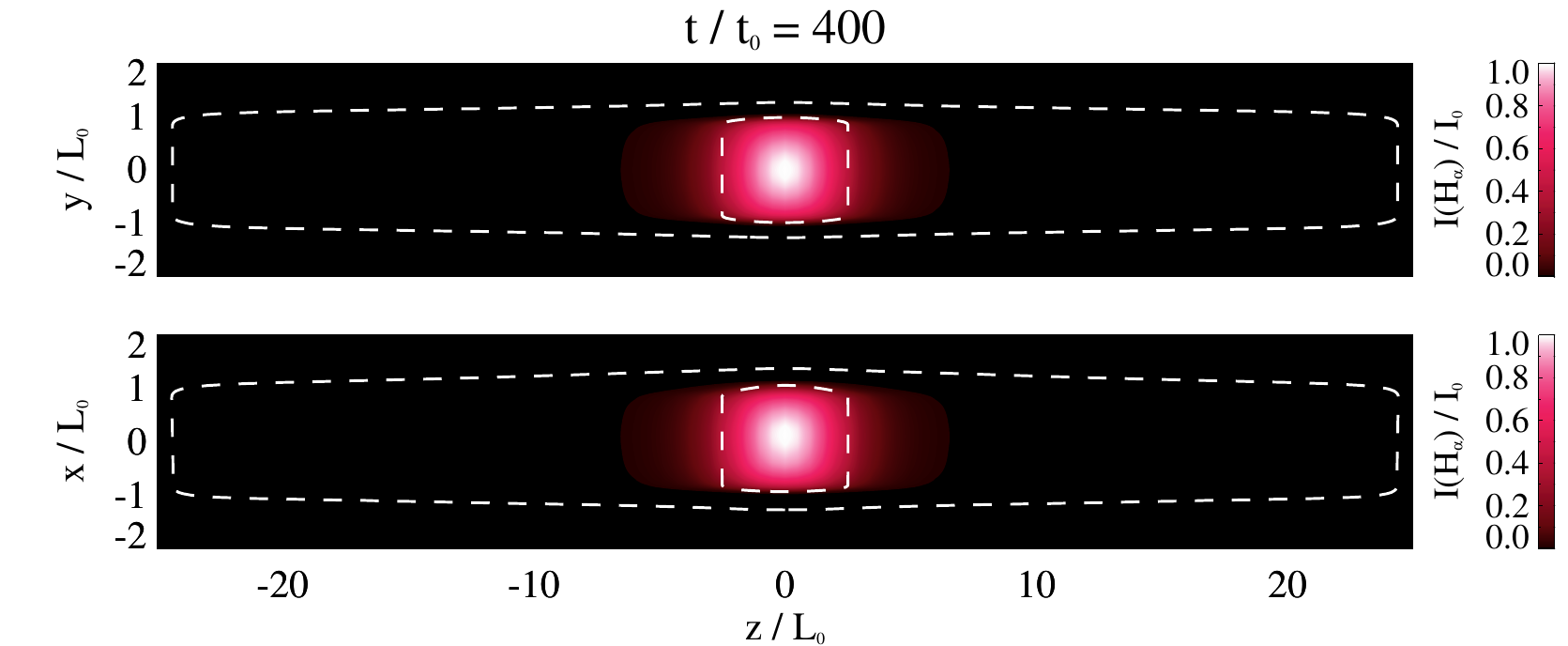}
		\includegraphics[width=0.495\hsize]{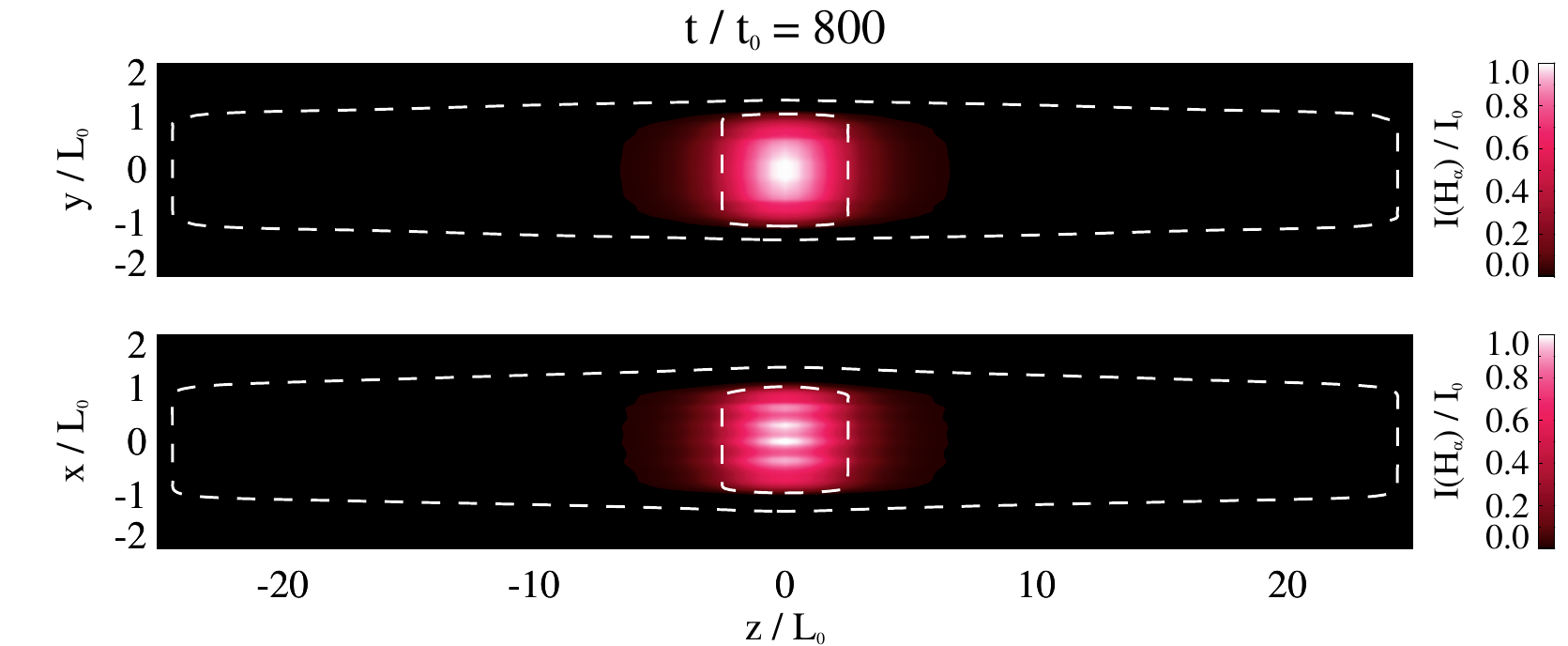} \\
		\includegraphics[width=0.495\hsize]{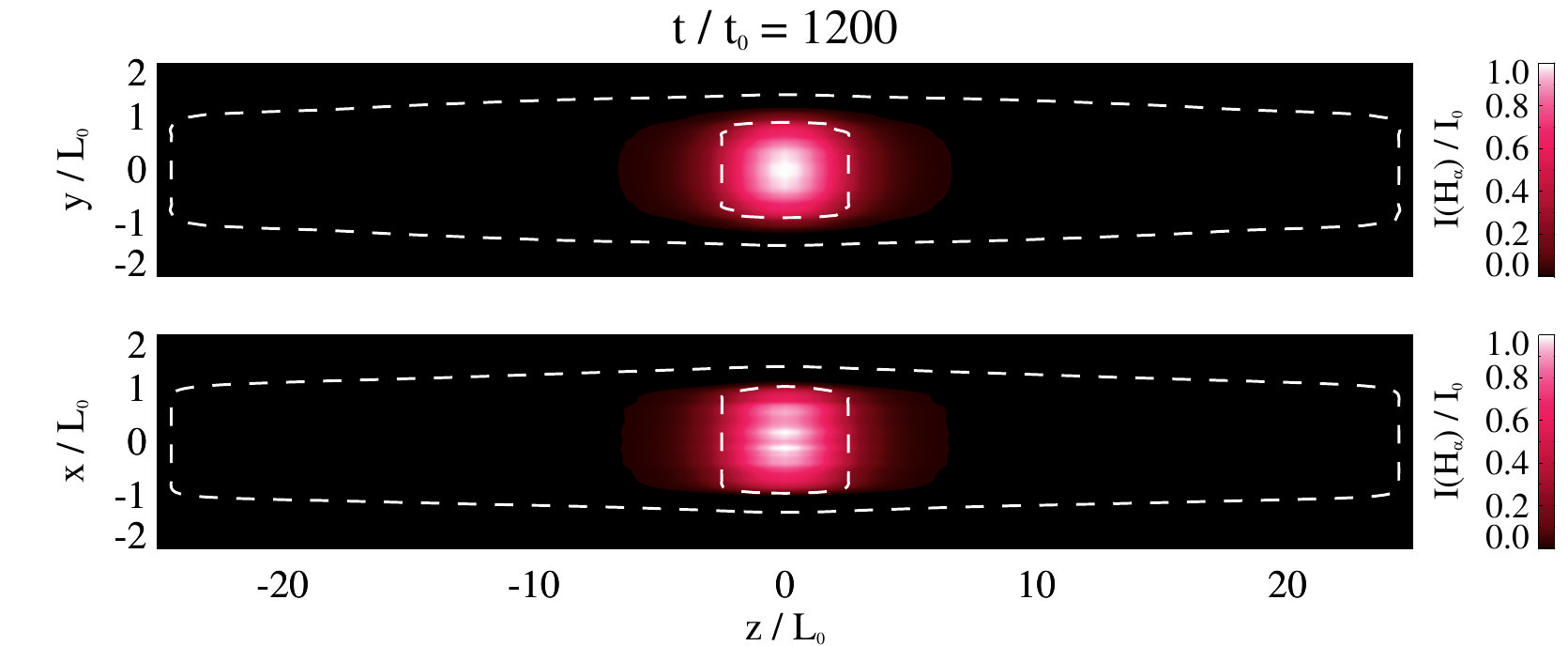}
		\includegraphics[width=0.495\hsize]{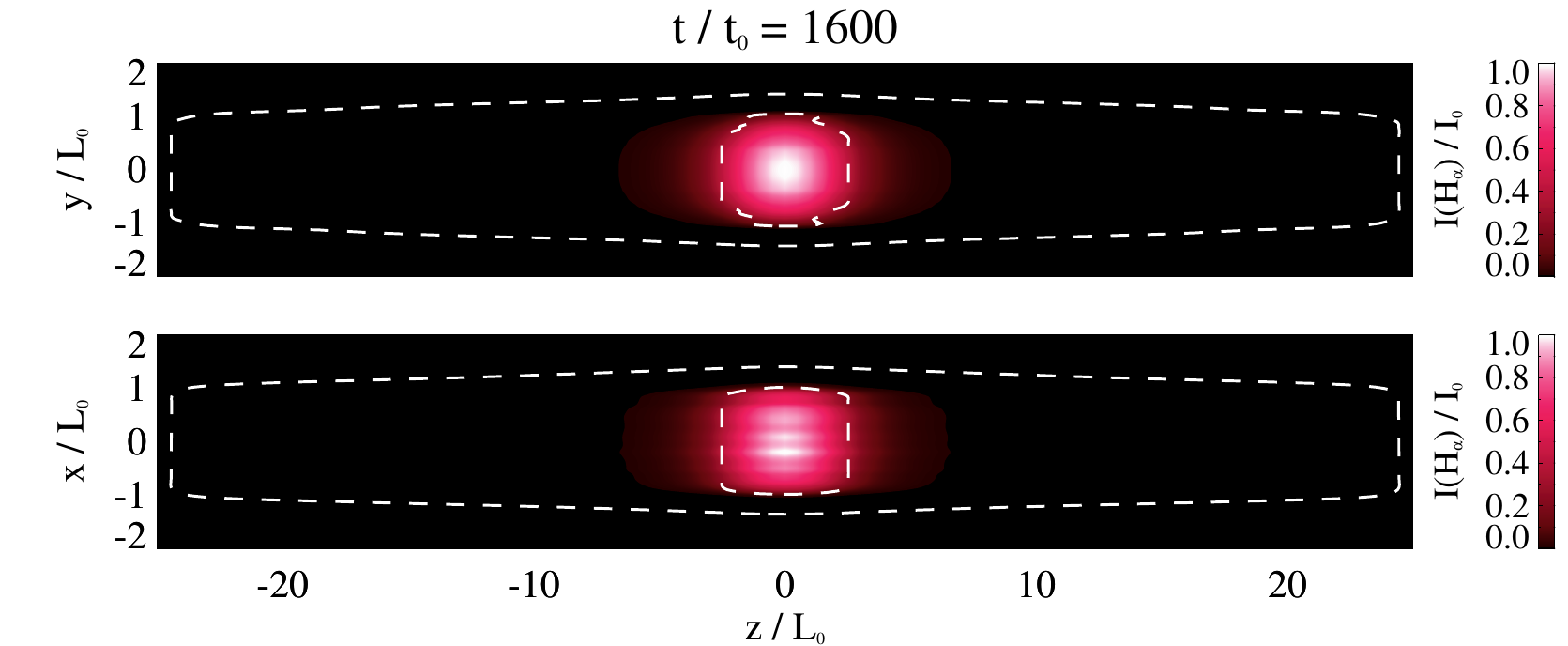}
		\caption{Same as Fig. \ref{fig:fomo_tube_dr10} but for the simulation with $\chi = 100$. An animation of this figure is available online.}
		\label{fig:fomo_tube_dr100}
	\end{figure*}
	
	\begin{figure*}
		\centering
		\includegraphics[width=0.495\hsize]{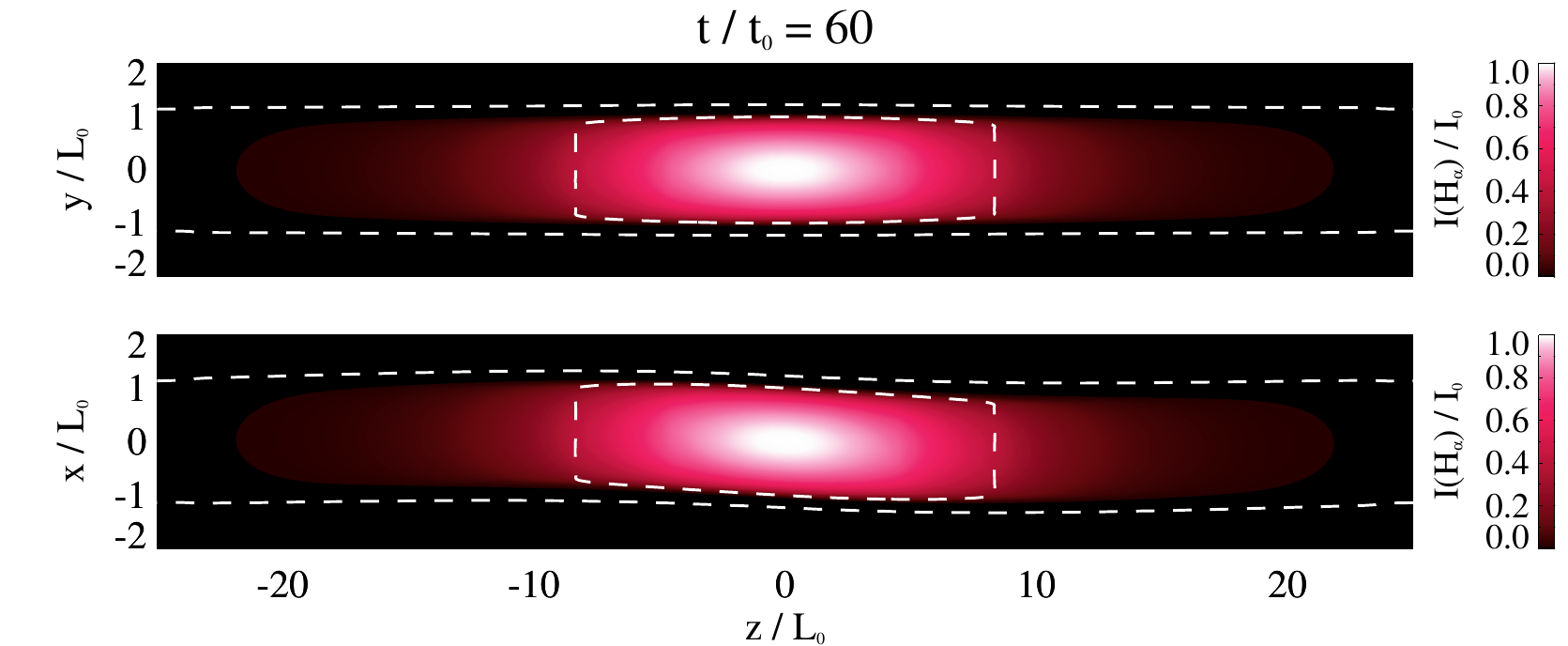}
		\includegraphics[width=0.495\hsize]{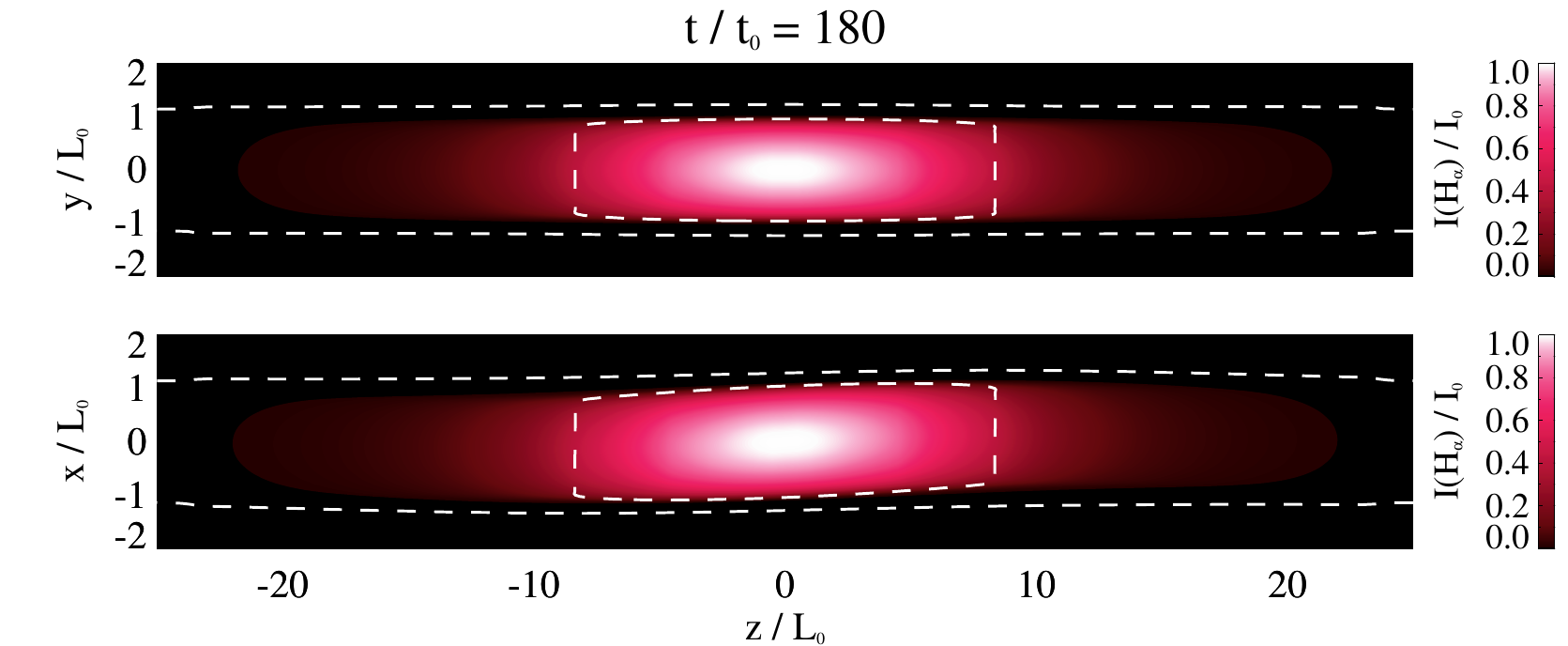} \\
		\includegraphics[width=0.495\hsize]{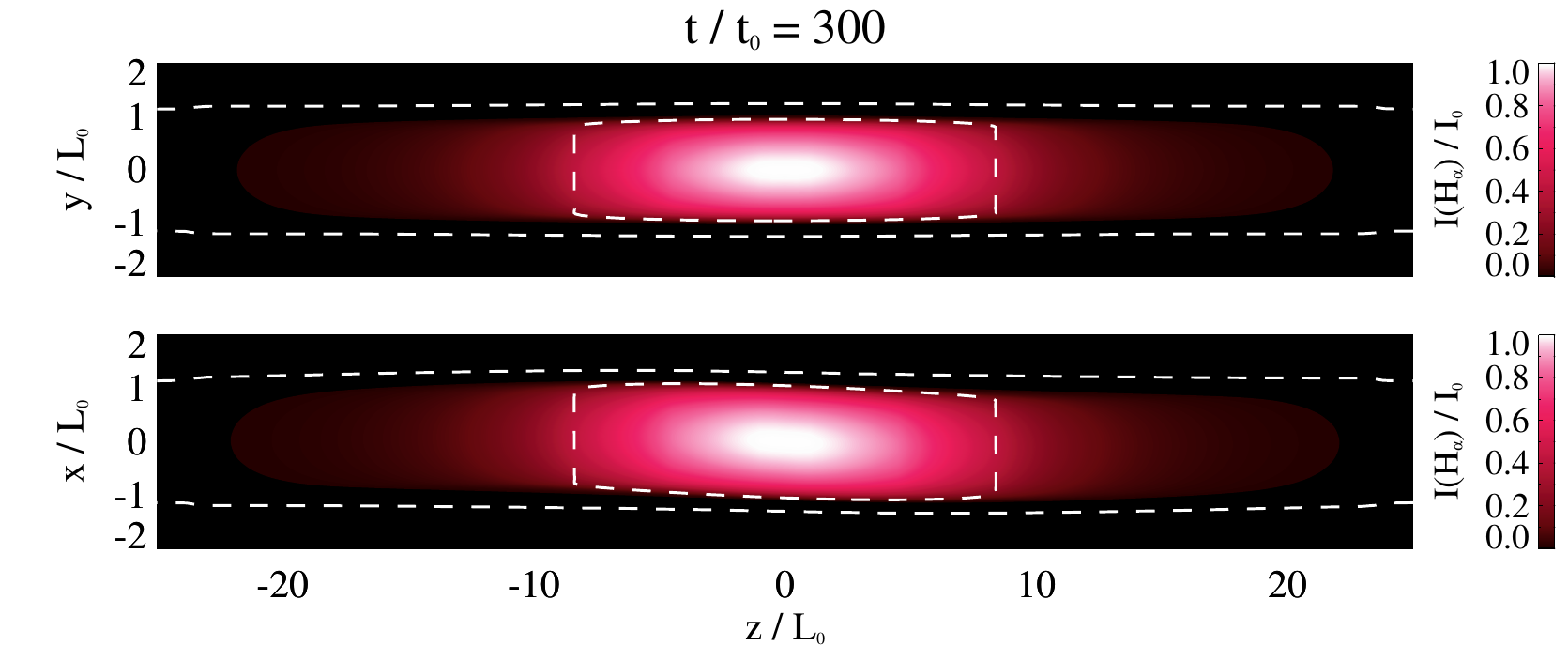}
		\includegraphics[width=0.495\hsize]{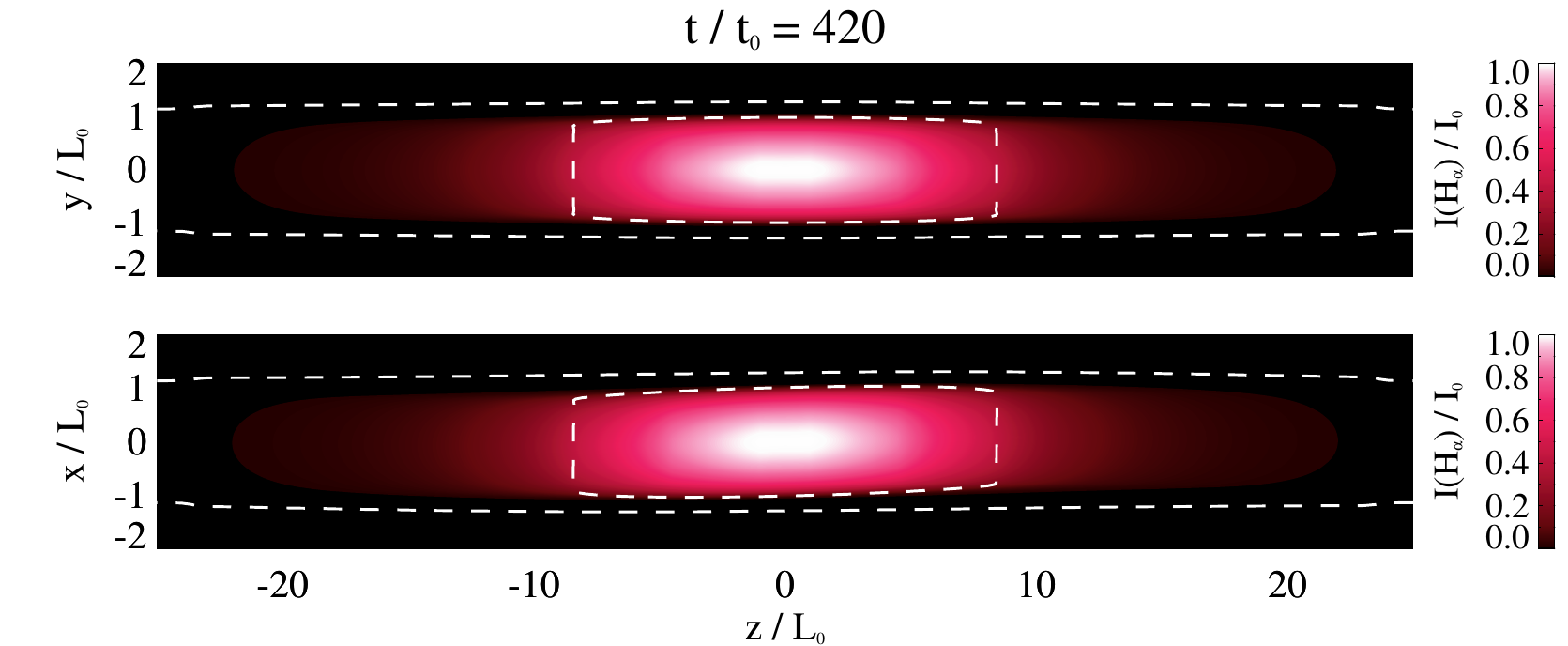}
		\caption{Intensity in the H$\alpha$ line at different times of the oscillation for the case with $\chi = 10$ and $n = 2$ (first harmonic). In each snapshot the top panel corresponds to the $x_{\rm{LOS}}$, while the bottom panel corresponds to the $y_{\rm{LOS}}$. White dashed lines show density contours for $\rho= 1.05 \rho_{\rm{ex}}$ and $\rho = 50 \rho_{\rm{ex}}$. An animation of this figure is available online.}
		\label{fig:fomo_tube_dr10_nkz2}
	\end{figure*}
		
	The top panels of Fig. \ref{fig:khi_dens} can be compared with those of Fig. \ref{fig:rho_snapshots}. They represent oscillations of threads with the same physical parameters (density profile, longitudinal oscillation mode, and amplitude of the perturbation) but differ in the numerical resolution employed in each simulation: the simulation depicted in Fig. \ref{fig:rho_snapshots} used a lower numerical resolution, while the simulation represented in Fig. \ref{fig:khi_dens} used a higher resolution. This comparison shows the importance of using a higher numerical resolution (a smaller cell size) to properly simulate the evolution of the KHI. It allows to resolve smaller scales, associated with larger azimuthal wavenumbers which have larger growth rates \citep{2010ApJ...712..875S}, and show more details of the development of the fine structure during the thread oscillation, as demonstrated by \citep{2015ApJ...809...72A}. These authors also found that improving the numerical resolution produces an increase of small-scale vortexes without altering the global oscillation of the tube.

	Then, in Fig. \ref{fig:fomo_tube_dr10} and \ref{fig:fomo_tube_dr100} we present the results of applying the approximate forward modelling to our simulations. We took into account the whole 3D simulation to compute the H$\alpha$ synthetic profile and show how the full longitudinal structure of the flux tube would be seen in the observations, revealing in that way the effect of the inhomogeneity in density. We considered two different lines of sight (LOS): the first one, denoted as $x_{\rm{LOS}}$, is parallel to the direction of oscillation, which goes along the $x$-axis; the second one, $y_{\rm{LOS}}$, is perpendicular to the direction of oscillation, that is, along the $y$-axis. Boths LOS are perpendicular to the longitudinal axis of the tube. In each of the panels of Figs. \ref{fig:fomo_tube_dr10} and \ref{fig:fomo_tube_dr100} the top and bottom panels correspond to the $x_{\rm{LOS}}$ and the $y_{\rm{LOS}}$, respectively. In addition, we have added density contours for the values $\rho = 1.05\rho_{\rm{ex}}$ and $\rho = 50 \rho_{\rm{ex}}$ as white dashed lines. The density contours are computed at the plane that is perpendicular to the line of sight and crosses the longitudinal axis of the tube, that is, the plane $x = 0$ for the $x_{\rm{LOS}}$ and the plane $y = 0$ for the $y_{\rm{LOS}}$. These lines serve to illustrate how the whole magnetic tube, and not only the dense central part, oscillates.
	
	We also applied the approximate forward modelling method to a simulation of the mode $n = 2$ and the longitudinal density ratio $\chi = 10$. The rest of parameters of the equilibrium state and the initial perturbation are the same as the ones used in the previous simulations. Fig. \ref{fig:fomo_tube_dr10_nkz2} presents several snapshots from this simulation. The density contour $\rho = 1.05 \rho_{\rm{ex}}$ approximately shows how the whole magnetic flux tube oscillates, with its maximum displacement occurring at $z = -12.5 L_{0}$ and $z = 12.5 L_{0}$.
	
	The comparison between Figs. \ref{fig:fomo_tube_dr10} -- \ref{fig:fomo_tube_dr10_nkz2} shows how important is from the observational point of view the effect of the longitudinal distribution of density in the thread. The central part of the tube appears much brighter than the remaining regions. The reason is that the intensity has an approximately quadratic dependence on the density, so the variations of density are very strongly reflected on the intensity profiles. It also reveals one of the main practical problems that the field of prominence seismology has to face when using tools like the period ratio $P_{0} / P_{1}$. On the one hand, the H$\alpha$ line provides information about the cool and dense central part of the flux tube and it allows to measure the fundamental mode of oscillation. On the other hand, the first harmonic has a node at $z=0$ while its maximum displacement takes place in the dark regions of the H$\alpha$ profile (see Fig. \ref{fig:fomo_tube_dr10_nkz2}). Therefore, another filter associated with the lighter and hotter plasma would be needed to measure the properties of this oscillation mode.
	
	\begin{figure*}
		\centering
		\includegraphics[width=0.85\hsize,keepaspectratio]{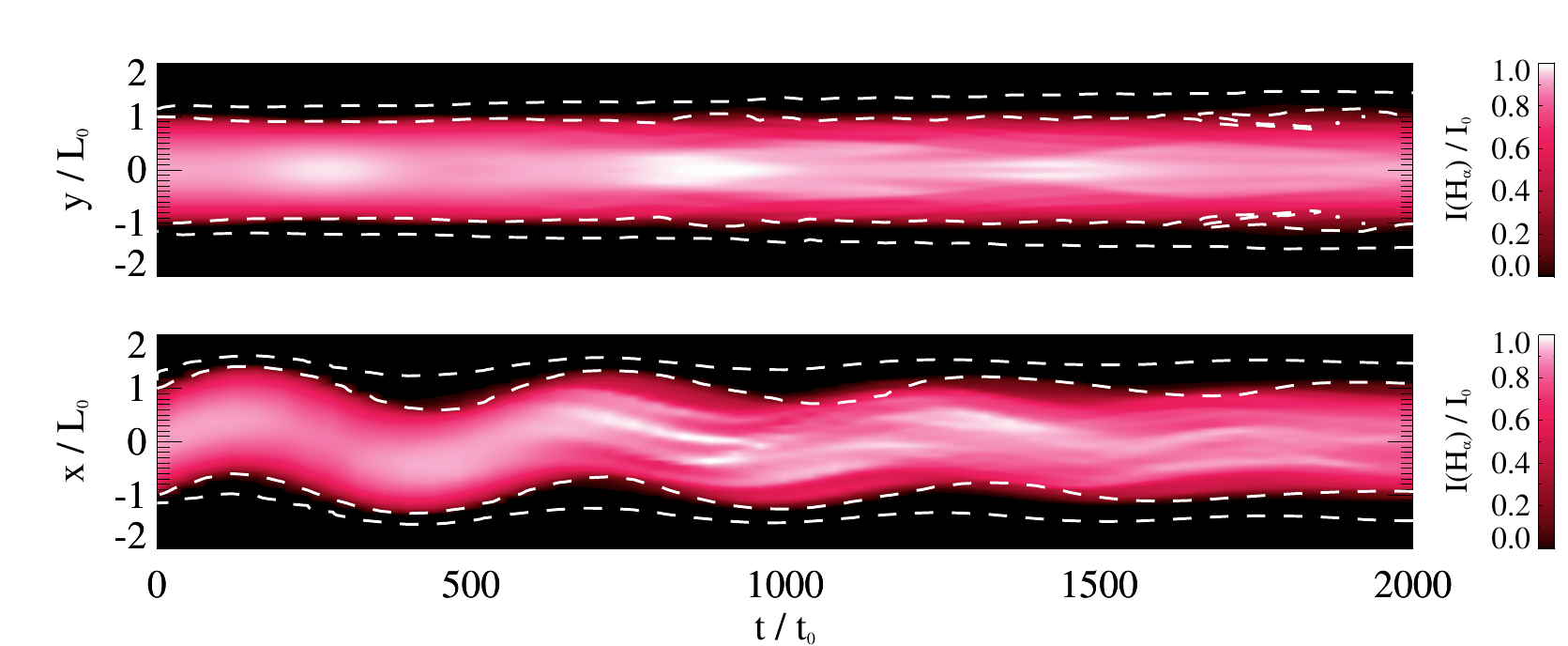}
		\caption{Intensity of H$\alpha$ line at $z=0$ as a function of time from a simulation with a density ratio of $\chi = 10$.}
		\label{fig:fomo_slice_dr10}
	\end{figure*}
	
	\begin{figure*}
		\centering
		\includegraphics[width=0.85\hsize,keepaspectratio]{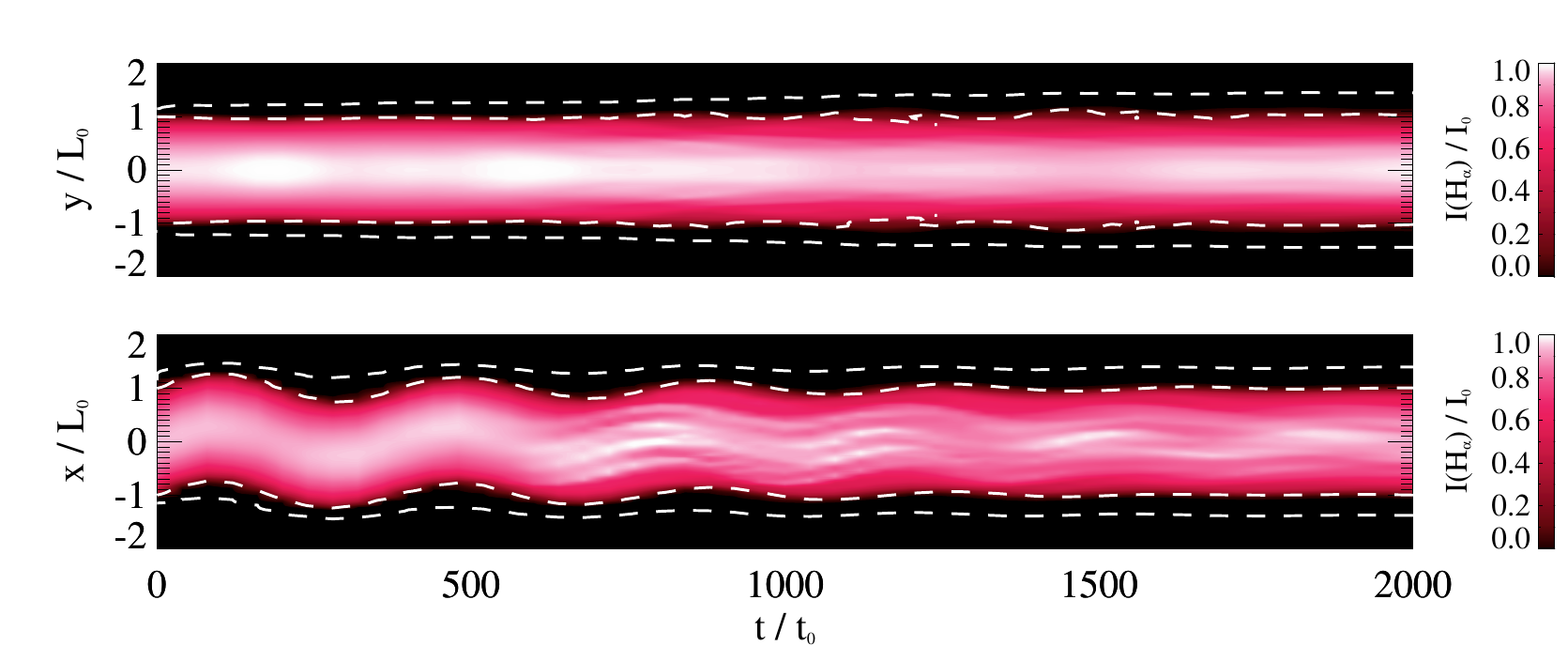}
		\caption{Intensity of H$\alpha$ line at $z=0$ as a function of time from a simulation with a density ratio of $\chi = 100$.}
		\label{fig:fomo_slice_dr100}
	\end{figure*}
	
	Another remarkable feature of the intensity profiles is the development of bright and dark fringes as time advances. These fringes are the observational signatures of the deformation of the tube boundary caused by the KHI. The stripes are more evident when the thread is observed along the LOS perpendicular to the direction of oscillation. However, for the case with $\chi = 10$, they can also be seen in the other line of sight. For $\chi = 100$, there are variations of brightness along the $x_{\rm{LOS}}$ but the stripes are not so easily discerned. The development of this fine strand-like structure as a consequence of the KHI was already described by \citet{2014ApJ...787L..22A,2016ApJ...830L..22A,2017ApJ...836..219A} for the case of coronal loops and by \citet{2015ApJ...809...72A} for the case of prominences.
	
	There are two main differences between the results we present in this work and the results shown by \citet{2014ApJ...787L..22A,2015ApJ...809...72A,2016ApJ...830L..22A,2017ApJ...836..219A}. In the first place, these authors performed forward modelling of spectral lines associated with higher temperatures while we used a line related to cool plasma. Therefore, our forward modelling results and theirs can be compared qualitatively but not quantitatively.
	The other difference is that \citet{2014ApJ...787L..22A,2015ApJ...809...72A,2016ApJ...830L..22A,2017ApJ...836..219A} considered a tube model with no longitudinal variation of density. They found that the KH vortexes are present at every longitudinal position of the tube. We expect the same to occur in the case with the longitudinal inhomogeneity, although they can only be seen in the intensity profiles near $z = 0$, where the plasma is denser. By inspecting density snapshots at different positions of the flux tube (which we do not present here), we checked that this is what indeed occurs. 

	In the last place, Figs. \ref{fig:fomo_slice_dr10} and \ref{fig:fomo_slice_dr100} show how a horizontal cut at $z = 0$ would be seen as a function of time for the cases with $\chi = 10$ and $\chi = 100$, respectively. Again, the parallel and perpendicular lines of sight are displayed in the top and bottom panels, respectively. The oscillatory motion of the thread is more clearly appreciated in the perpendicular line of sight, $y_{\rm{LOS}}$, which allows to easily trace the displacement of the central part of the tube, the development of the instability and the attenuation of the oscillation due to the resonant absorption. From these panels we can also estimate that the width of the most conspicuous bright and dark strands is on the order of $w \sim R / 5$ or $R/4$ (that is, from $200$ to $250 \ \text{km}$), which corresponds to an angular diameter of $\delta \sim 0.3''$. Therefore, these fine structures could be resolved by H$\alpha$ instruments such as CRisp Imaging SpectroPolarimeter \citep[CRISP][]{2008ApJ...689L..69S} installed at the Swedish Solar Telescope \citep[SST,][]{2003SPIE.4853..341S} or Visible Imaging Spectrometer \citep[VIS,][]{2010AN....331..636C} installed at the Goode Solar Telescope \citep[GST,][]{2012ASPC..463..357G}, which have spatial resolutions on the order of $\sim 0.1'' (\sim 70 \ \rm{km})$ \citep[see e.g.,][]{2019ApJ...880..143J,2020A&A...633A..11F}. However, we note that, as already mentioned in Section \ref{sec:model}, the value of $R$ used in this work is slightly larger than what has been observed for prominence threads. Typically, the radius of a thread ranges from $50 \ \rm{km}$ to $500 \ \rm{km}$ \citep{2007Sci...318.1577O,2008AdSpR..42..803L,2011SSRv..158..237L}, which would give a range of $10$ to $125$ km for the bright and dark fringes. Consequently, those instruments would not be able to resolve the KH-strands of thin threads, but they would allow to observe the fine structure of thicker threads. More recent instruments, such as the Visible Broadband Imager \citep[VBI,][]{2021SoPh..296..145W}, installed at the Daniel K. Inouye Solar Telescope \citep[DKIST,][]{2020SoPh..295..172R}, should be able to detect strands with a width of the order of $\sim 25 \ \rm{km}$. 
	
	On the other hand, when observed along the direction of oscillation, the amplitude of this motion cannot be directly determined from these time-distance diagrams. Nonetheless, some evidences of the oscillation can still be found: apart from the bright and dark fringes that can be associated with the KHI, there are also periodic variations of brightness. These variations are related to the contractions and expansions that appear in the tube before the KH vortexes develop. As shown by \citet{2017ApJ...836..219A} and \citet{2018ApJ...853...35T}, the width of the tube periodically increases in the $x$-direction and decreases in the $y$-direction, and viceversa. A contraction in one direction produces a decrease of the H$\alpha$ intensity along the parallel line of sight but an increase in the perpendicular line of sight, while expansions have the opposite effect. This behaviour can be checked by comparing the top panels of Figs. \ref{fig:fomo_slice_dr10} and \ref{fig:fomo_slice_dr100} with their corresponding bottom panels.
	
	By comparing Figs. \ref{fig:fomo_slice_dr10} and \ref{fig:fomo_slice_dr100} it can be seen that the thread with a lower density ratio oscillates with a larger amplitude and a longer period, in agreement with the results presented in Section \ref{sec:fundamental}. In addition, it presents larger variations of brightness in its $x_{\rm{LOS}}$ profile, which is a consequence of having a larger amplitude of the oscillation. In both cases, the fringes associated with the non-linear stage of the KHI start to be clearly seen at around the same time-step, $t / t_{0} \approx 550$, which is after $1$ oscillation period for $\chi = 10$ and about $1.5$ periods for $\chi = 100$. It is also interesting to note how the density contours follow the oscillation of the bright region of the tube during the first periods but they become out of phase as time advances. This is a consequence of the deformations caused by the KHI and the way the contours and the H$\alpha$ intensity are computed: the former uses the values of density at only one plane, while the latter is the result of integrating along the whole line of sight (which crosses different planes with different values of density).

\section{Summary and future work} \label{sec:concl}
	In the present work, we performed a numerical study of the longitudinal fundamental and first harmonic modes of transverse oscillations in inhomogeneous threads. We confirmed the findings of \citet{2015A&A...575A.123S} that, if the central density of the tube is kept fixed, the oscillation period is reduced as the ratio between the densities at the centre of the tube and at its ends, $\chi$, is increased. We also confirmed that the period ratio of the two modes, $P_{0} / P_{1}$, increases as the longitudinal density ratio is increased, or as the average density of the tube is decreased.
	
	Then, we analysed the damping times of the oscillations. The attenuation is mainly due to the process of resonant absorption \citep[see e.g.,][]{1978ApJ...226..650I,2002ApJ...577..475R}. There is also a small contribution due to the non-linear generation of modes with high azimuthal wavenumber \citep{2014SoPh..289.1999R,2017ApJ...850..114S}, and the development of the KHI \citep{2018ApJ...853...35T,2021ApJ...910...58V}. We found that the damping times also decrease as the longitudinal density ratio of the tube is increased. However, the damping time to period ratio increases with $\chi$, which means that the increase of $\chi$ causes a relatively larger reduction of the oscillation periods in comparison with the damping times. This result qualitatively agrees with those of \citet{2011A&A...533A..60A}, who considered a longitudinal density structuring and found a slight increase of the damping ratio for very short threads (which in our model would be represented by an increasing value of $\chi$).
	
	In addition, a more pronounced effect of the increase of $\chi$ on the damping ratio is found for the case of the first harmonic. The reason is that the efficiency of resonant absorption depends on the density contrast between the internal and external densities of the flux tube and the thickness of the radial transition layer \citep{2002ApJ...577..475R,2008ApJ...682L.141A,2009ApJ...707..662S}. In the present work the density contrast varies with height and with the longitudinal density ratio $\chi$. At the height of the maximum displacement of the first harmonic there are larger variations of the density contrast with $\chi$ than at the height where the fundamental mode has its maximum displacement.
	
	Taking into account the results described in the previous paragraphs, we proposed to use the ratio $\tau_{\rm{D,0}} / \tau_{\rm{D,1}}$ in the same way the period ratio $P_{0} / P_{1}$ is used for prominence seismology. From the solutions of the 2D eigenvalue problem we obtained two approximate expressions, Eqs. (\ref{eq:damp_fit_chi}) and (\ref{eq:damp_fit_rhoav}), for the dependence of $\tau_{\rm{D,0}} / \tau_{\rm{D,1}}$ on the parameters $\chi$ and $\langle \rho_{\rm{i}}\rangle/ \rho_{\rm{i,0}}$, respectively. However, the utility of this parameter for extracting information from the observations should be analysed in more detail in a future work. For instance, in contrast with the period ratio, which only depends on the relation between the average density and the density at the central part of the tube, as shown by \citet{2015A&A...575A.123S} and \citet{2016SoPh..291.1143R}, the ratio $\tau_{\rm{D,0}} / \tau_{\rm{D,1}}$ might depend on the specific longitudinal profile of density and its application might not be as straightforward. Nevertheless, we expect that this longitudinal dependence would be less relevant than its dependence on the profile of the radial transition layer.
	
	The last part of the present work consisted of performing forward modelling from our 3D simulations. We computed synthetic H$\alpha$ profiles using the approximate method developed by \citet{2015A&A...579A..16H}. In this way, we showed the effect that the Lorentzian profile of density would have in the observations: only the central denser part of the thread would be visible in the intensity profiles, while most of the tube would appear dark in comparison. This is an important issue for the field of prominence seismology. The H$\alpha$ profiles allow to detect and measure with some ease the characteristics of the fundamental oscillation mode, since its maximum displacement happens at the bright part of the thread. On the contrary, the first longitudinal harmonic has a node in that bright central region and the maximum displacement occurs in the dark regions, making extremely difficult to measure the properties of this mode. We recall that, to the best of our knowledge, no unequivocal detection of the first harmonic in prominence threads has been reported to date. This issue can be resolved by combining the H$\alpha$ observations with measures in spectral lines associated with the emission from lighter and hotter plasma.
	
	Apart from the periods and damping times of the oscillations and the density distribution of the thread, the intensity profiles also present some observational signatures of non-linear effects that take place as the tube oscillates. The large amplitudes of the initial perturbations used in our simulations cause the appearance of strong velocity shears that trigger the KHI \citep{1984A&A...131..283B,2008ApJ...687L.115T}. This instability generates large deformations of the external layers of the tube, as shown by \citet{2014ApJ...787L..22A}, \citet{2015A&A...582A.117M}, \citet{2016A&A...595A..81M}, \citet{2017ApJ...836..219A} or \citet{2018ApJ...853...35T}. These deformations are reflected in the H$\alpha$ profiles by longitudinal bright and dark stripes. This is the same mechanism that can generate the strand-like structure seen in coronal loops \citep{2014ApJ...787L..22A,2016ApJ...830L..22A,2017ApJ...836..219A} and prominences \citep{2015ApJ...809...72A}. We have estimated that, for a typical prominence thread, the size of this fine structure varies in the range from $10$ to $125 \ \rm{km}$. Therefore, it should be possible to observe these fine strands in thick threads using instruments such as SST/CRISP \citep{2003SPIE.4853..341S,2008ApJ...689L..69S} and GST/VIS \citep{2010AN....331..636C,2012ASPC..463..357G}, which have an H$\alpha$ spatial resolution of $\sim 70 \ \rm{km}$, or DKIST/VBI \citep{2020SoPh..295..172R,2021SoPh..296..145W} (with a resolution of $\sim 25 \ \rm{km}$).
	
	Here, we described the dynamics of magnetic flux tubes through the ideal MHD equations. However, prominence threads are made of partially ionised plasma. Thus, for a more realistic analysis, the effects of the interaction between the ionised and neutral components of the plasma should be taken into account. \citet{2009ApJ...707..662S,2014IAUS..300...48S} showed that ambipolar diffusion has a negligible impact on the attenuation of transverse oscillations in comparison with the mechanism of resonant absorption. Nevertheless, the ion-neutral interaction may have an important effect on non-linear processes, such as the heating of the plasma, specially as the KHI develops and smaller and smaller scales are generated, in which the energy dissipation due to ion-neutral collisons is more efficient \citep{2005A&A...442.1091L,2011A&A...529A..82Z,2013ApJ...767..171S,2019A&A...630A..79P}. In addition, this energy dissipation may oppose the action of the ponderomotive force, as shown by \citet{2018ApJ...856...16M} and \citet{2020A&A...641A..48B} for the case of non-linear Alfvén waves.
	
	Other effects that influence the oscillations in flux tubes but we did not consider are the presence of mass flows, which have already been studied using uniform or piecewise constant models \citep{2008ApJ...678L.153T,2011A&A...531A.167S,2014SoPh..289..167E,2020MNRAS.496...67B,2021arXiv210102064S} but not with longitudinally inhomogeneous density profiles, or the twist, curvature and cross section of the tube, which have been analysed for coronal loops \citep{2008A&A...486.1015V,2008ApJ...687L..45V,2008ApJ...686..694R,2009A&A...502..315M,2012ApJ...757..186K} but not for prominence threads. In addition, \citet{2015A&A...578A.104M} showed that since the plasma is partially ionised in prominence threads, they are always unstable in the presence of longitudinal shear flows, which trigger the KHI. The inclusion in our model of various of the aforementioned effects is left for future works.
	
\begin{acknowledgements}
	D.M. and E.K. acknowledge support from the European Research Council through the Consolidator Grant ERC-2017-CoG-771310-PI2FA. D.M. also acknowledges support from the Spanish Ministry of Science and Innovation through the grant CEX2019-000920-S of the Severo Ochoa Program. This publication is part of the R+D+i project PID2020-112791GB-I00, financed by MCIN/AEI/10.13039/501100011033.
\end{acknowledgements}

\bibliographystyle{aa}
\bibliography{ft1}	

\end{document}